\documentclass[preprint,12pt,authoryear]{elsarticle}

\usepackage{amssymb}
\usepackage{amsthm}

\usepackage{enumitem}
\usepackage{amsmath}
\usepackage{longtable}
\usepackage{makecell}
\usepackage{multirow}
\usepackage{adjustbox}
\usepackage{booktabs}
\usepackage{graphicx}
\usepackage{float}
\usepackage{xcolor}
\usepackage{mathtools}
\usepackage{etoolbox}
\usepackage{setspace}
\usepackage{subfigure}
\usepackage{units}
\usepackage{comment}
\usepackage[ruled,vlined]{algorithm2e}
\usepackage[hidelinks]{hyperref}
\usepackage{threeparttable}

\newtheorem*{assumption*}{\assumptionnumber}
\providecommand{\assumptionnumber}{}
\makeatletter

\definecolor{red(ncs)}{rgb}{0.0, 0.0, 0.0}

\definecolor{redd(ncs)}{rgb}{0.0, 0.0, 0.0}

\usepackage[margin=3.5cm]{geometry}

\newcommand{\rd}[1]{\textcolor{red(ncs)}{#1}}
\newcommand{\rdd}[1]{\textcolor{redd(ncs)}{#1}}

\newcommand{\algrule}[1][.2pt]{\par\vskip.5\baselineskip\hrule height #1\par\vskip.5\baselineskip}

\journal{}

\begin{document}

\begin{frontmatter}



\title{2-Level Reinforcement Learning for Ships on Inland Waterways\rdd{: Path Planning and Following}}


\author[a]{Martin Waltz\corref{CorrespondingAuthor}}
\ead{martin.waltz@tu-dresden.de}
\author[a]{Niklas Paulig}
\ead{niklas.paulig@tu-dresden.de}
\author[a,b]{Ostap Okhrin}
\ead{ostap.okhrin@tu-dresden.de}

\affiliation[a]{organization={Technische Universität Dresden, Chair of Econometrics and Statistics, esp. in the Transport Sector, Wuerzburger Str. 35},
            city={Dresden},
            postcode={01062}, 
            country={Germany}}
\affiliation[b]{organization={Center for Scalable Data Analytics and Artificial Intelligence (ScaDS.AI)}, city={Dresden/Leipzig}, country={Germany}}

\cortext[CorrespondingAuthor]{Corresponding author}
\begin{abstract}
This paper proposes a realistic modularized framework for controlling autonomous surface vehicles (ASVs) on inland waterways (IWs) based on deep reinforcement learning (DRL). The framework \rdd{improves operational safety and} comprises two levels: a high-level local path planning (LPP) unit and a low-level path following (PF) unit, each consisting of a DRL agent. The LPP agent is responsible for planning a path under consideration of \rdd{dynamic vessels, closing a gap in the current research landscape.~In addition, the LPP agent adequately considers traffic rules and the geometry of the waterway. We thereby introduce a novel application of a spatial-temporal recurrent neural network architecture to continuous action spaces. The LPP agent outperforms} a state-of-the-art artificial potential field \rdd{(APF)} method by increasing the minimum distance to other vessels by 65\% on average. The PF agent performs low-level actuator control while accounting for shallow water influences and the environmental forces winds, waves, and currents. Compared with a proportional-integral-derivative \rdd{(PID)} controller, the PF agent yields only 61\% of the mean cross-track error \rdd{(MCTE)} while significantly reducing control effort \rdd{(CE)} in terms of the required absolute rudder angle.~Lastly, both agents are jointly validated in simulation, employing the lower Elbe in northern Germany as an example case and using real automatic identification system (\rdd{AIS}) trajectories to model the behavior of other ships.
\end{abstract}




\end{frontmatter}


\sloppy
\section{Introduction}
Inland waterway (IW) transport is widely regarded as an energy-efficient and low-emitting mode of transportation in terms of greenhouse gases compared to road or rail transportation. Additionally, IW transport offers a significant freight capacity, making it an integral part of a sustainable transport system \citep{rohacs2007role, de2022inland}. Traditionally, the control of inland vessels is performed by human operators, and recent research began to investigate the use of autonomous surface vehicles (ASVs) for such inland operations \citep{gan2022ship, vanneste2022safety}. For a comprehensive understanding of ASV systems, we refer to \cite{liu2016unmanned}, \cite{fossen2021handbook}, and \cite{negenborn2023autonomous}.

One crucial factor affecting the economic potential of shipping companies is the costs associated with crew members \citep{al2017developing}. These costs can be significantly reduced since ASVs require little to no on-board personnel \citep{vagale2021path}. Furthermore, although IW operations already exhibit relatively low accident rates compared to other transportation modes \citep{hofbauer2020external}, the human factor can still pose a significant threat to operational safety. For instance, according to \cite{bavckalov2023lessons}, human failures were responsible for 58\% and 19\% of accidents in Austria and Serbia, respectively, from the early 2000s to 2017.~\rdd{As emphasized by \cite{gan2022ship}, IW operations are particularly challenging} for human operators due to waterway geometry and potential high traffic densities, further reinforcing the potential benefits of employing ASVs in inland operations.

There are different approaches to categorizing the level of autonomy of autonomous ships \citep{ringbom2019regulating}. In this paper, we adopt the definition from \citet[p.~1296]{vagale2021path}, which defines an ASV as \emph{"a vessel capable of making decisions and operating independently, without human guidance, navigation, and control"}. An ASV has to continuously generate a path which it can subsequently follow. The literature distinguishes between global (GPP) and local path planning \rdd{(LPP)}, where GPP addresses the static problem of defining a plan for the entire voyage ignoring kinematic and dynamic constraints, while LPP is an ongoing process based on real-time information to generate a feasible local path \citep{siegwart2011introduction, vagale2021path}. Path following \rdd{(PF)} describes the task of following a pre-determined (local) path without considering the visitation time of a particular waypoint \citep{fossen2021handbook}. As stated in the beginning, in this paper, we focus on LPP and PF. We refer to the controlled ASV as the \emph{own ship} and to its surrounding ships as \emph{target ships}.

In recent years, the \rdd{vessel traffic} literature has started to embrace the advancements in artificial intelligence \citep{munim2020big, ZHUGE2023}. One notable approach is deep reinforcement learning (DRL, \citealt{matsuo2022deep}), which combines reinforcement learning (RL, \citealt{sutton2018reinforcement, szepesvari2010algorithms}) with deep neural networks \rd{\citep{goodfellow2016deep}}. DRL has showcased remarkable success across challenging application domains \citep{vinyals2019grandmaster, ibarz2021train, bellemare2020autonomous, silver2018general, soler2024reinforcement, kitchat2024deep}, including \rdd{vessel} control tasks \citep{heiberg2022risk, hart2023vessel}. In RL, an agent acquires knowledge by engaging in iterative interactions with an environment, where it learns to make decisions through trial-and-error. For instance, in the context of LPP, the agent represents the own ship and is required to adapt its heading based on the presence of nearby target ships and the geometric characteristics of the waterway. By defining an appropriate feedback mechanism, known as a reward, the agent receives penalties for causing collisions or running aground. 

\rd{In contrast to many traditional control methods, DRL does not require prior knowledge or a complete model of the environment. Instead, this essential information is inherently embedded in the experiences the agent accumulates through training. Moreover, due to the use of deep neural networks as powerful function approximators, DRL can generalize across environmental situations, facilitating adaptation to previously unencountered scenarios \citep{kang2019generalization}.}

\rdd{While there have been applications of DRL to LPP and PF tasks on open waters, which are reviewed in Section \ref{sec:background}, research on IWs is strongly limited. Regarding the LPP task on IWs, to the best of our knowledge, only \cite{vanneste2022safety} have explored the application of DRL. However, their study is restricted to static obstacles and does not address the dynamic collision avoidance (COLAV) problem. Our study fills this gap by thoroughly considering the unique challenges associated with IW operations, including dealing with dynamically moving target ships, navigating through narrow waterways, and accounting for the effects of water depth on vessel dynamics. For the PF task on IWs, the closest work we identified is \cite{paulig2023robust}, who applied the Deep $Q$-Network (DQN, \citealt{mnih2015human}) modification proposed in \cite{waltz2022two} to achieve strong PF performances on various paths under different current influences. We extend their work by also considering the impact of waves and winds on the vessel, thereby further enhancing the comprehensiveness and robustness of the control system.}

In summary, our study makes the following contributions:
\begin{itemize}
    \item \rdd{We introduce the first holistic control architecture for ASVs on IWs based on DRL.} Our framework comprises separate agents dedicated to the LPP and PF tasks. \rdd{The architecture accounts for dynamically moving target ships,} environmental disturbances such as winds, waves, and currents, while also adhering to existing traffic rules and considering the geometry of the waterway.
    \item To accommodate continuous action spaces, we transfer the spatial-temporal recurrent network architecture \rd{from} \cite{waltz2022spatial} to actor-critic frameworks, and employ it for the LPP agent.
    \item To assess the effectiveness of our approach, we conduct extensive testing on both agents in a variety of challenging scenarios that are representative of the most difficult practical occurrences. In particular, we focus on advanced overtaking maneuvers and situations involving strong environmental forces. Furthermore, we validate the entire architecture by using real trajectories obtained from the automatic identification system (AIS).
    \item \rd{We numerically compare the performance of the two agents with strong baselines in the form of an artificial potential field (APF) \citep{liu2023colregs, wang2019obstacle} method for the LPP task and a PID controller \citep{paramesh2021unified} for the PF task. The performance metrics used for the comparison build on \cite{jadhav2023collision}.}
    \item To ensure reproducibility and facilitate further research, we have made the source code for this paper publicly accessible \rd{in \cite{TUDRL}}. In addition, we have open-sourced our trajectory extraction pipeline from AIS data \rd{in \cite{pytsa}}. By providing these resources, we aim to encourage the inclusion of real AIS data in the validation process of ASV control systems.
\end{itemize}
This paper is structured as follows: Section \ref{sec:background} provides additional information about traffic rules, sensor systems, and \rdd{PF} and planning algorithms. The newly proposed architecture is visualized and described in \rd{S}ection \ref{sec:architecture}, while \rd{S}ection \ref{sec:theory} details the relevant theory used in this work. Sections \ref{sec:LPP_module} and \ref{sec:PF_module} contain detailed descriptions of the LPP and PF modules, respectively. The results and validation scenarios are shown in \rd{S}ection \ref{sec:results_validation}, and \rd{S}ection \ref{sec:conclusion} concludes the paper.

\section{Background and related work}\label{sec:background}
\sloppy
\subsection{Traffic rules}\label{subsec:traffic_rules}
While path planning is an extensively studied problem in robotics \citep{siciliano2008springer}, \rdd{path planning for vessels} presents unique challenges due to the incorporation of traffic rules specific to each waterway. The International Regulations for Preventing Collisions at Sea (COLREGs, \citealt{COLREGs1972}) form the basis of these rules, governing the required behavior of ships during encounters. Additionally, national regulations exist that further define the traffic rules for specific waters.~\rdd{Although our architecture is generally applicable for any IW, as a use case, we focus in this paper on the lower part of the Elbe river in northern Germany. The relevant regulation in this area} is the \emph{Seeschifffahrtsstraßen-Ordnung} of the \cite{SeeSchStrO1998}. Of particular importance to us are two specific rules within this regulation, which are also representative of \rdd{IW} regulations of other countries. First, overtaking should be conducted on the portside of the vessel being overtaken. Second, the vessel being overtaken should facilitate the overtaking maneuver as much as possible. For the convenience of the reader, the exact wording of the regulation \rd{in German together with its translation into English} is provided in \ref{app:traffic_rules}.

\subsection{ASV sensors}
\rdd{ASVs rely on navigational information on their current state}, including position, velocity, accelerations, and environmental forces such as current and wind speed. These quantities stem from various sensors such as radar, LIDAR, sonar, visual sensors, infrared sensors, inertial measurement units, or the global positioning system \citep{liu2016unmanned}. However, not every seagoing vessel is necessarily equipped with each of these sensors, and the sensor data might be noisy or erroneous and requires advanced state estimation techniques \citep{lefeber2003tracking, motwani2013interval}. Moreover, data about target ships is received via AIS, mandatory equipment since the end of 2004 for all cargo ships of certain sizes and all passenger ships \citep{IMO_AIS}. \rdd{Furthermore, information about the waterway geometry and depth are provided via Electronic Navigational Charts \citep{blindheim2021electronic}, which are digital maps usually available to seafarers.}

\subsection{\rd{Path following algorithms}}
The PF task describes the derivation of low-level control commands from a given local path. First, the controller needs to derive a guidance law from the received set of waypoints \rdd{\citep{breivik2009guidance, fossen2021handbook}. In this study, we use the vector-field guidance (VFG, \citealt{nelson2006vector}) detailed in Section \ref{subsec:VFG}}. \rdd{Second, the controller converts the directional awareness established through the guidance law to} low-level actuator command\rdd{s} that minimize spatial and angular deviation from the desired path and the guidance signal. The actuator\rdd{s} are typically propeller\rdd{s} or rudder\rdd{s} to control the vessel's speed and course, respectively. In this paper, we focus on using rudder adjustments as the primary actuator command. Steering changes are preferred over speed adjustments due to fuel considerations and better visual and radar observability of \rd{the} course changes by other ships \citep{wang2017ship}.

Various control algorithms can be employed to translate the guidance signal into an actuator command. Arguably\rd{,} one of the most popular techniques is the classic proportional-integral-derivative (PID) controller \citep{paramesh2021unified}, which will also serve as a benchmark for our PF module. Further \rdd{conventional} methodologies used in the literature include $H_\infty$ control \citep{donha1998}, linear quadratic Gaussian control \citep{sharma2012autopilot}, model predictive control \citep{annamalai2015robust}, sliding motion control \citep{liu2018ship}, backstepping control \citep{zhang2017improved}, \rd{$\mathcal{L}_1$ adaptive control \citep{breu2011}, or dynamic surface control \citep{wan2020improved}, to mention a few. For a comprehensive comparison of \rdd{PF} control methods for ASVs, we refer to the recent survey of \cite{xu2023review}.}

\rdd{Moreover, recent research started to investigate the suitability of DRL for PF of ASVs, although primarily concentrated on open water situations while overlooking the joint impact of currents, winds, and waves. An important contribution in this realm is \cite{woo2019deep}, where the Deep Deterministic Policy Gradient (DDPG, \citealt{lillicrap2015continuous}) algorithm was used to successfully control a small-scale vessel to follow a linear path in a real-world experiment. \cite{zhao2020path} outline a smoothly convergent modification of the DQN algorithm to perform PF on polygonal and sinusoidal paths. Another relevant related work is \cite{martinsen2018curved}, in which the DDPG algorithm was used for curved PF of three different vessel classes: a mariner, a container, and a tanker. \cite{gonzalez2020usv} combine the DDPG algorithm with an adaptive sliding mode control strategy \citep{castaneda2021continuous} to let an ASV follow straight and zig-zag-like paths. Further contributions to DRL-based PF of ASVs include \cite{wang2023path}, \cite{zhao2021usv}, and \cite{peng2023model}.}

\sloppy
\subsection{Path planning algorithms}
\cite{tam2009review} provide historical perspectives on the development of path planning algorithms \rdd{for open waters}, while recent reviews are available in \cite{Vagale2021}, \cite{ozturk2022review}, and \cite{yu2023review}.~The most important methodological approaches in the field include evolutionary algorithms \citep{tam2010path}, velocity obstacle methods \citep{kuwata2013safe}, model predictive control \citep{johansen2016ship}, APF methods \citep{lyu2019colregs}, rapidly-exploring random trees \rdd{(RRT, \citealt{zhang2019path})}, dynamic-window approaches \citep{serigstad2018hybrid}, fast marching methods \citep{liu2015path}, \rdd{whale optimization \citep{han2022multi}}, and particle swarm optimization \rdd{(PSO, \citealt{ding2018energy, tutsoy2024minimum})}.

\rdd{In addition, there have been recent attempts to perform planning on open waters based on DRL. The first major study on this topic was \cite{cheng2018concise}, in which a concise DRL algorithm for static obstacle avoidance was proposed. \cite{heiberg2022risk} designs a DRL agent using Proximal Policy Optimization \citep{schulman2017proximal} while relying on advanced collision risk assessment theory. \cite{xu2022path} apply the DDPG with a modified experience replay mechanism to tackle a path planning and COLAV task. A similar proposal has been made in \cite{zhai2022intelligent}, where the authors build on the DQN to construct a COLAV algorithm. Finally, \cite{waltz2022spatial} outline a spatial-temporal recurrent network architecture for the DQN in maritime LPP. The architecture effectively handles an arbitrary number of target ships and is robust to partial observability. Further contributions for DRL-based approaches to maritime path planning include \cite{chun2021deep}, \cite{li2021path}, \cite{meyer2020colreg}, and \cite{guo2020autonomous}.}

\rdd{In contrast to the extensive literature on algorithms for ASVs on open waters, there is a notable scarcity of work done on vessels operating on IWs, even in the realm of classical algorithms.} \cite{zhang2023path} tackle this gap by employing an anisotropic fast marching algorithm for ASVs, with a particular emphasis on navigating through restricted areas near bridges. \cite{chen2016path} compare the performance of the heuristic search algorithm A* \citep{hart1968formal} and its various extensions for GPP in Dutch IWs. Similar to the concepts introduced in \cite{lyu2018fast}, \cite{gan2022ship} develop a planning algorithm for inland rivers that incorporates the safety potential field theory to account for static and dynamic obstacles and waterway bank walls. Finally, \cite{cao2022inland} detail a modification of the RRT algorithm to perform path planning on the Yangtze River Channel in Zhenjiang.

\section{Proposed architecture}\label{sec:architecture}
\begin{figure}[ht]
    \centering
    \includegraphics[width=\textwidth]{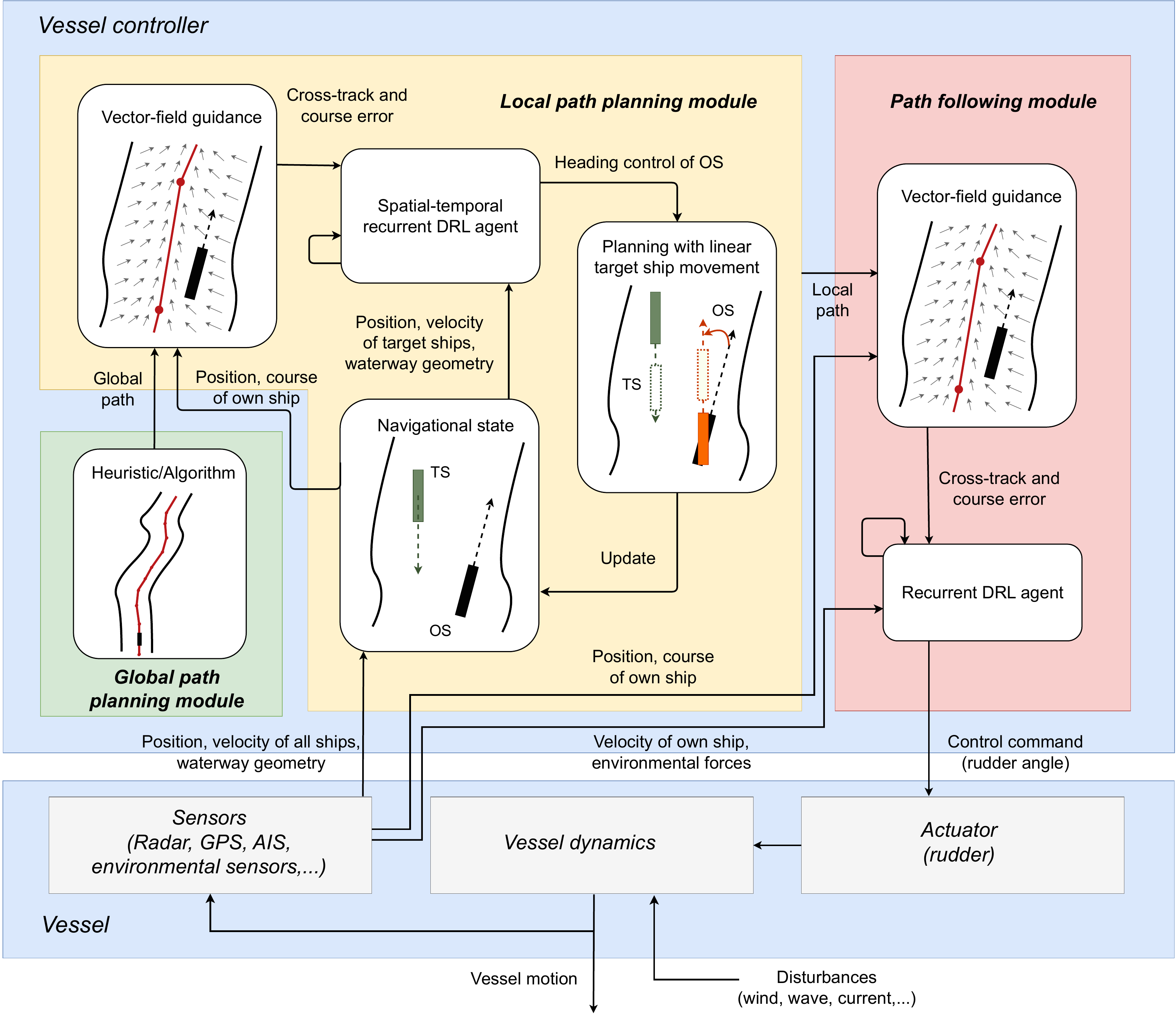}
    \caption{The proposed architecture for an ASV based on DRL, with the visualization being inspired by \cite{chen2016path}.}
    \label{fig:architecture}
\end{figure}

Figure \ref{fig:architecture} visualizes the proposed architecture for ASV control on \rdd{IWs}. The vessel controller comprises three key components: a GPP module, an LPP module, and a PF module. Two DRL agents are involved: a high-level agent at the core of the LPP module \rd{(in yellow)} and a low-level agent for the PF task \rd{(in red)}. The computational flow of the architecture is as follows: The global path generated by the GPP module \rd{(in green)} serves as input for the LPP module, which in turn produces a local path. The PF module processes this local path and generates an actuator control command. Note that we leave the algorithm or heuristic employed in the GPP module unspecified, although A* or its extensions are reasonable choices \citep{chen2016path, singh2018constrained}.

\begin{figure}[ht]
\hfill
\subfigure[Linear local path]{\includegraphics[width=6.5cm]{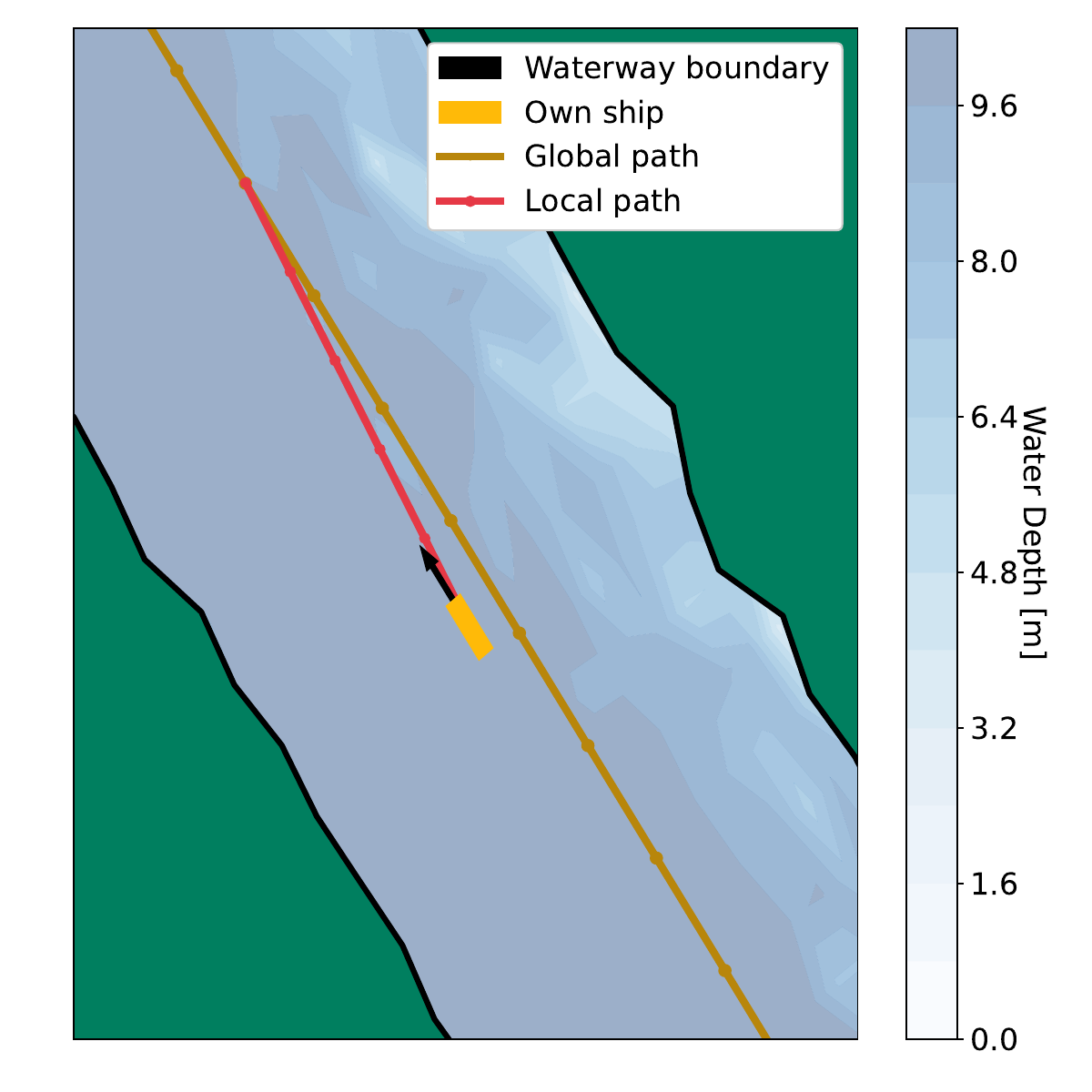}}
\hfill
\subfigure[Overtaking maneuver]{\includegraphics[width=6.5cm]{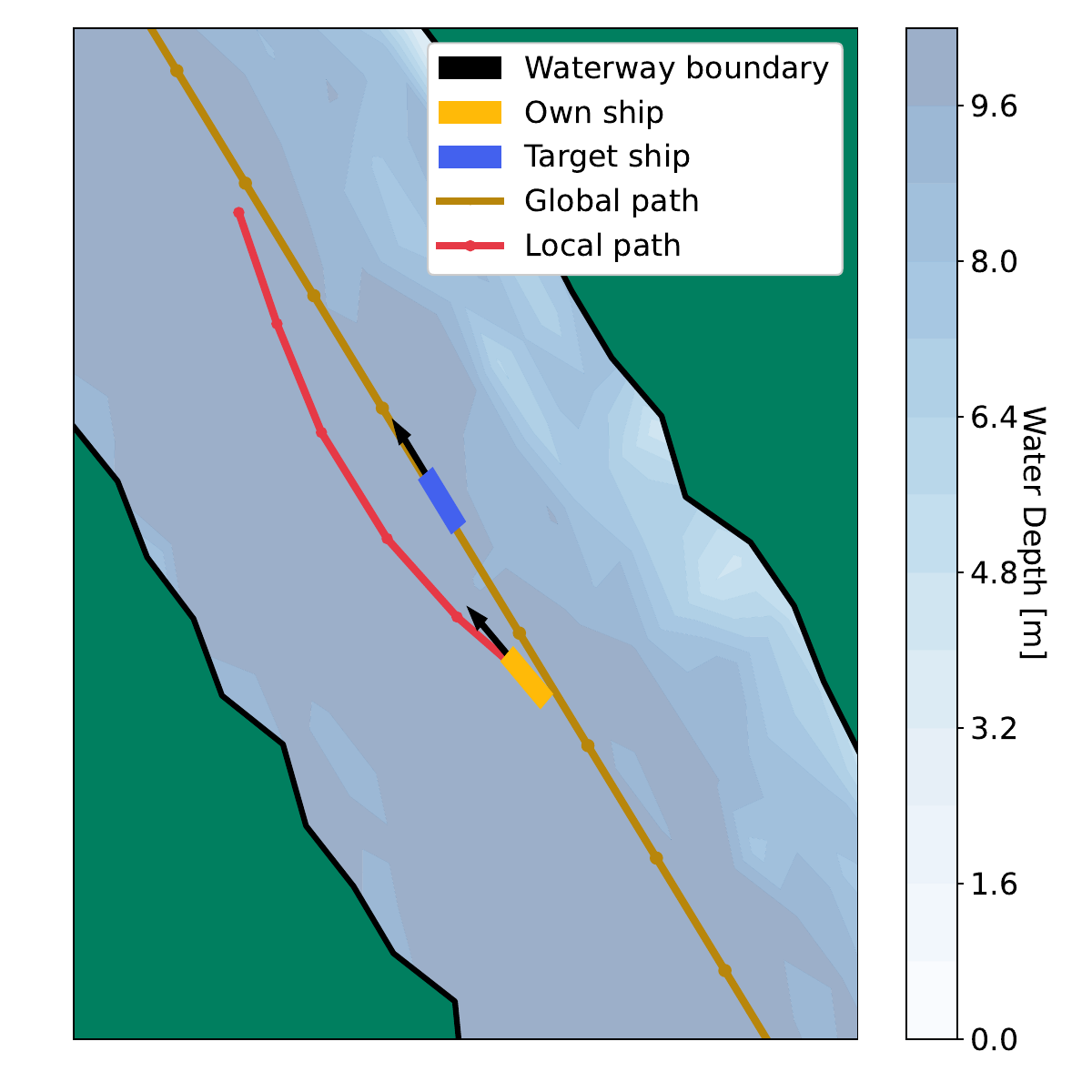}}
\hfill
\caption{\rd{The simplification of the \rdd{LPP} procedure if no target ships are present (left), and an overtaking maneuver (right).}}
\label{fig:GlobalLocalIllustration}
\end{figure}

Describing the procedure in more detail, the LPP module is activated every $t_{\rm replan}$ seconds to generate a new local plan based on existing information. In our simulation, we set $t_{\rm replan} = \unit[30]{s}$, although this parameter can be freely chosen depending on the geometrical difficulty of the waterway segment or the traffic density. To optimize computation time, the local path is replanned only when target ships are near the own ship. Otherwise, a simple linear local path can be constructed to return to the global path. In this case, it suffices to select a waypoint of the global path and linearly connect it to the own ship's position. Figure \ref{fig:GlobalLocalIllustration} (a) illustrates this procedure, while Figure \ref{fig:GlobalLocalIllustration} (b) visualizes the functionality of the framework in an overtaking scenario. In the latter case, the own ship recognizes the target ship in front and plans a safe local path accordingly.

When the LPP unit is activated, the DRL agent within it processes information from two sources: a cross-track and course error signal, indicating deviation from the global path, and navigational data from the sensor suite. The sensor data includes the positions and velocities of target ships as well as information about the waterway's geometry. Based on this information, the DRL agent generates a heading change command for the own ship, and the positions and velocities of all vessels are updated according to the vessel's model detailed in Section \ref{sec:theory}. The own ship's position is stored as a waypoint of the local path, and the error signals to the global path are updated. This process is repeated until a sufficient number of waypoints has been generated. We emphasize that the LPP procedure always occurs in simulation, even when deployed on a real ship, as path planning does not involve any actuator control.

For simplicity, control commands for the target ships are not incorporated during these planning iterations. Therefore, the LPP agent assumes that all target ships move linearly and maintain a constant course and speed, regardless of their actual behavior. In practice, this simplification can be replaced with a more sophisticated intent estimation or trajectory prediction unit \citep{huang2020ship, ma2021intent}, which can be seamlessly integrated into the architecture.

Finally, the PF module \rdd{is activated every $t_{\rm control}$ seconds, where it} receives the local path and calculates the cross-track and course error using the VFG approach. Combined with velocity information and observations of environmental forces, an actuator command is generated to navigate the vessel along the desired path. \rdd{In our simulation, we set the realistic value of $t_{\rm control} = \unit[5]{s}$. In addition, we include Algorithm \ref{algo:Architecture} to provide a concise summary of the overall computational flow.}

\begin{algorithm}[htp]
\begin{small}
\setstretch{1.10}
\DontPrintSemicolon
\SetAlgoLined
\SetKwComment{Comment}{\# }{}
\vspace{0.2cm}
\textbf{\rdd{Pre-Departure:}}\\
 \rdd{plan global path using any heuristic or algorithm}\\
 \algrule
 \textbf{\rdd{Procedure:}}\\
 \While{\rdd{not arrived}}{
    \uIf{\rdd{time to replan locally (every $t_{\rm replan}$ seconds)}}{
    \uIf{\rdd{no target ships around}}{
        \rdd{plan linear path back to global path}
    }\Else{
        \rdd{plan local path using LPP agent (Section \ref{sec:LPP_module})}
  }
  }
  \vspace{0.25cm}
 \uIf{\rdd{time to control actuators (every $t_{\rm control}$ seconds)}}{
    \rdd{adjust rudder angle using PF agent (Section \ref{sec:PF_module})}
 }
 }
\caption{\rdd{2-Level ASV control using DRL}}
\label{algo:Architecture}
\end{small}
\end{algorithm}

\section{Theory}\label{sec:theory}
\subsection{Vessel dynamics}\label{sec:vessel_dynamics}
\begin{figure}[!htb]
    \centering
    \includegraphics[width=.5\textwidth]{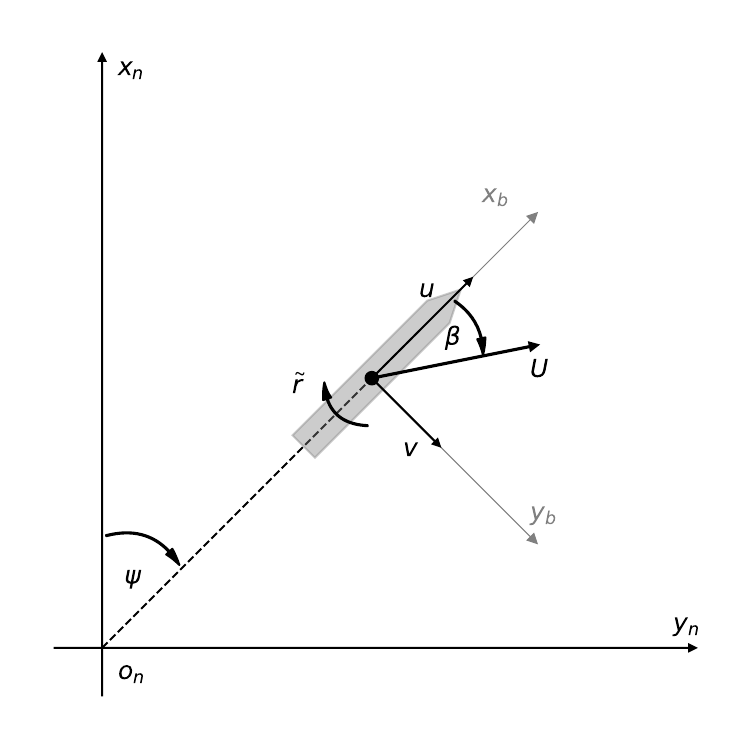}
    \caption{Coordinate systems considered in this work; see \cite{yasukawa2015introduction}.}
    \label{fig:coord_systems}
\end{figure}

In this study, we employ the 3-degree-of-freedom Maneuvering Modeling Group (MMG) model introduced by \cite{yasukawa2015introduction} to simulate the dynamic behaviour of a 1:5 scale replica of the \rdd{commonly used} KVLCC2 tanker. The principal particulars of the tanker can be found in \cite{paulig2023robust}. Two coordinate systems are considered in our research, which are illustrated in Figure \ref{fig:coord_systems}. The first system, denoted as $\{n\}$, follows the \emph{North-East-Down} convention. It describes the navigational status of the vessel using the vector $\eta = (x_n, y_n, \psi)^\top$. The variables $x_n$ and $y_n$ represent the north and east positions of the vessel, respectively, relative to the origin point $o_n$. The variable $\psi$ is the heading angle of the ship and represents the angle between the $x_n$-axis and the $x_b$-axis of the second coordinate system, denoted as $\{b\}$. The $\{b\}$ reference frame is a \emph{body-fixed} frame that is centered at the midship position of the vessel. Within this coordinate system, the longitudinal axis is denoted as $x_b$, and the transversal axis is denoted as $y_b$. The velocity of the vessel is described by the vector $\nu = (u, v, \Tilde{r})^\top$, consisting of the surge velocity $u$, sway velocity $v$, and yaw rate $\Tilde{r}$. Following \cite{fossen2021handbook}, the total speed of the vessel is $U = \sqrt{u^2 + v^2}$ and the course angle is $\chi = \psi + \arctan(v/u)$.

Unlike the majority of prior studies in the field of ASVs \citep{heiberg2022risk, fan2022novel, zhai2022intelligent}, our research incorporates environmental forces, namely winds, waves, and currents, into our simulations. These forces play a critical role in real-life operations. Consequently, the vessel's movement is governed by the following set of equations:
\begin{equation}\label{eq:MMG_model}
\begin{aligned}
    (m + m_{x_b}) \dot{u} - (m + m_{y_b}) v \Tilde{r} - x_G m \Tilde{r}^2 &= X = X_H + X_R + X_P + X_{WI} + X_{WA}, \\
    (m + m_{y_b}) \dot{v} + (m + m_{x_b})u\Tilde{r} + x_G m \dot{\Tilde{r}} &= Y = Y_H + Y_R + Y_{WI} + Y_{WA}, \\
    (I_{zG} + x_{G}^{2} m + J_z) \dot{\Tilde{r}} + x_G m (\dot{v} + u\Tilde{r}) &= N_m = N_H + N_R + N_{WI} + N_{WA},
\end{aligned}
\end{equation}
where $X$ \rd{is} the surge force, $Y$ denotes the lateral force, and $N_m$ signifies the yaw moment around the midship. The first order derivative of a variable $x$ with respect to time is denoted $\dot{x}$. The variables $m$, $m_{x_b}$, and $m_{y_b}$ represent the mass of the ASV, the added masses in the $x_b$ and $y_b$ directions, respectively. Additionally, $x_G$ \rd{is} the longitudinal coordinate of the center of gravity in $\{b\}$, $I_{zG}$ represents the moment of inertia, and $J_z$ \rd{denotes} the added moment of inertia. The set of equations (\ref{eq:MMG_model}) encompasses five distinct force components: hull (H), rudder (R), propeller (P), wind (WI), and wave (WA). It is important to note that the propeller component only affects the surge force \rd{$X$}. The equations for the hull, rudder, and propeller components are derived in \cite{yasukawa2015introduction}, which also provides relevant hydrodynamic derivatives and further parameter values. Importantly, the rudder components $X_R$, $Y_R$, and $N_R$ depend on the rudder angle $\delta$, which is used to control the ship.

To compute the wind forces depending on the wind speed $V_{wi}$ and wind angle $\beta_{wi}$, we use the wind coefficient approximation for symmetrical ships described in \citet[Chapter 10.1]{fossen2021handbook}. Regarding wave forces and moments, we follow \cite{taimuri20206} and \cite{sakamoto1986minimisation} to compute the respective quantities depending on the wave amplitude $\zeta_{wa}$, wave angle $\beta_{wa}$, wave period $T_{wa}$, and wave length $\lambda_{wa}$. The most significant environmental force for vessels operating on IWs are currents, represented by the current speed $V_c$ and the current angle $\beta_c$ in the global frame $\{n\}$. Following \citet[Chapter 6.7]{fossen2021handbook}, these are not considered via an additional force or moment term in (\ref{eq:MMG_model}) but by directly specifying the velocity vector of the vessel relative to the currents. Consequently, the total speed of the ship in the presence of currents becomes $\sqrt{(u-u_c)^2 + (v-v_c)^2}$, where $u_c$ and $v_c$ are the longitudinal and lateral components of the currents in $\{b\}$, respectively.

Furthermore, another crucial yet often neglected environmental characteristic for vessels operating on IWs is the influence of the water depth $H$ on the vessel dynamics. \rdd{W}e follow \cite{taimuri20206} and consider the effects of shallow waters on wake fraction, thrust deduction, and flow-straightening coefficient as proposed by \cite{amin2010generalised}, while the hydrodynamic derivatives in shallow waters are approximated following \cite{kijima1990prediction} and \cite{ankudinov1990manoeuvring}.

In our simulation, we discretize the dynamics in (\ref{eq:MMG_model}) using a step size of 5 seconds\rdd{, aligning with the control frequency $t_{\rm control}$ of the PF unit,} and use the ballistic method of \cite{treiber2015comparing} to update the velocity and the position vector of the vessel. Throughout the paper, we refer to a variable $x$ at time $t$ as $x_t$. \rdd{We provide an overview of the used abbreviations and notation in \ref{app:Nomenclature} for the convenience of the reader.}

\subsection{Vector-field guidance}\label{subsec:VFG}
\begin{figure}[!htb]
    \centering
    \includegraphics[width=0.5\textwidth]{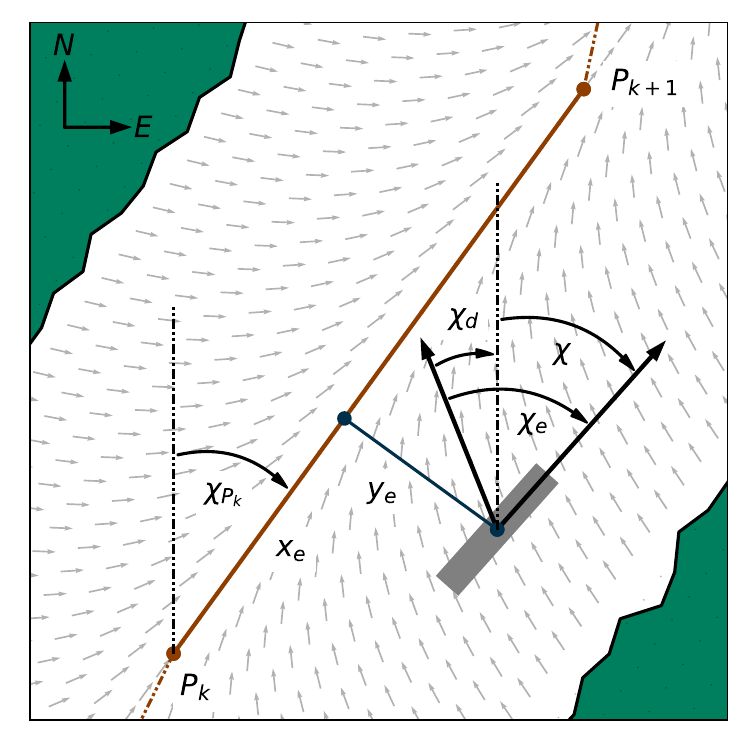}
    \caption{Visualization of the induced vector field; inspired by \cite{paulig2023robust}.}
    \label{fig:VFG}
\end{figure}

Both our LPP and PF units use VFG signals to ensure accurate path tracking. The concept of VFG originated from unmanned aerial vehicle control \citep{nelson2006vector} and is \rdd{visualized} in Figure \ref{fig:VFG}. The objective is to establish a vector field that guides the vessel back to its intended path, with the steepness of the field determined by a proportional gain parameter, denoted as $k > 0$. When given two consecutive waypoints, $P_k$ and $P_{k+1}$, the angle from $P_k$ to $P_{k+1}$ in the global frame $\{n\}$ is denoted as $\chi_{P_k}$. The vessel's cross-track and along-track errors are denoted as $y_e$ and $x_e$ respectively. Computation details for these quantities can be found in \citet[Chapter 12.4]{fossen2021handbook}.

\rdd{T}he VFG method defines the desired course of the vessel as $\chi_d = \chi_{P_k} - \chi^\infty \frac{2}{\pi} \arctan(k \cdot y_e)$, where $\chi^\infty \in \left(0, \frac{\pi}{2}\right]$. In this study, we set $\chi^\infty = \frac{\pi}{2}$, effectively reducing the guidance mechanism to a proportional line-of-sight guidance law \citep{fossen2014uniform}. The formulation introduces a course error, $\chi_e = \chi_d - \chi$, which represents the deviation between the desired course and the actual course of the ship, $\chi$. \rd{To address sudden changes in the desired course when a waypoint is passed, we adopt the approach proposed by \cite{paulig2023robust} and redefine the desired course to allow for a weighted transition between the current and subsequent path segments, resulting in smoother course adjustments. In particular, we set:}
\begin{equation}
    \chi_d = \left\{\left[1-\frac{x_e}{d(P_k, P_{k+1})}\right]\chi_{P_k} + \frac{x_e}{d(P_k, P_{k+1})}\chi_{P_{k+1}}\right\}  - \arctan(k \cdot y_e),
\end{equation}
where $d(P_k, P_{k+1})$ represents the Euclidean distance between waypoints $P_k$ and $P_{k+1}$.

\subsection{Collision risk assessment}
Estimating the collision risk with nearby target ships is a core task \rdd{of ASVs} \citep{ozturk2019individual}. In the literature, two concepts hold special importance. The first concept, the \emph{ship domain}, was originally introduced by \cite{toyoda1971marine} and specifies a safe area around a vessel that should not be entered by any other ship. While various shapes of ship domains have been explored in the literature \citep{szlapczynski2017review}, we adopt a symmetric domain configuration that allows for one ship length of space in front of the USV's bow, and one ship width to the USV's starboard, stern, and port side. Furthermore, we define a collision event when the midship position of a target ship is at or inside our own ship's ship domain.

The second concept is the \emph{closest point of approach} (CPA) \citep{lenart1983collision}, which includes the measures distance (DCPA) and time (TCPA) to the \rdd{CPA}. The CPA quantifies the criticality of a scenario by assuming that both the own ship and the target ship maintain their speed and course. Several recent studies have built upon these two concepts and defined specific collision risk metrics to assess the severity of a situation \citep{mou2010study, ha2021quantitative, waltz2022spatial}.

\subsection{Reinforcement learning}
Reinforcement learning is a fundamental pillar of artificial intelligence that involves an agent learning through iterative trial-and-error interactions with an environment \citep{sutton2018reinforcement}. In this paradigm, the agent's interactions with the environment are formalized as a Markov decision process (MDP) \citep{puterman2014markov}, represented by a tuple $(\mathcal{S}, \mathcal{A}, \mathcal{P}, \mathcal{R}, \gamma)$. Here, $\mathcal{S}$ denotes the state space of the system, $\mathcal{A}$ represents the action space, $\mathcal{P}: \mathcal{S} \times \mathcal{A} \times \mathcal{S} \rightarrow [0,1]$ is the state transition probability function, $\mathcal{R}: \mathcal{S} \times \mathcal{A} \rightarrow \mathbb{R}$ is a bounded reward function, and $\gamma \in [0,1)$ is a discount factor that balances the importance of immediate and future rewards. \rd{The sum of discounted rewards is called \emph{return}. The agent's objective within this framework is to optimize for a policy $\pi: \mathcal{S} \times \mathcal{A} \rightarrow [0,1]$, a mapping from states to probability distributions over actions, that maximizes the expected return.}

\rd{The methods to achieve this task are commonly divided into \emph{value-based} and \emph{policy gradient} algorithms \citep{sutton2018reinforcement}. Value-based methods typically learn a policy-dependent action-value for each state-action pair $(s,a)$, denoted $Q^{\pi}(s,a)$, which represents the expected return when being in state $s$, executing action $a$, and following policy $\pi$ afterward. Crucially, under fulfillment of regularity conditions, an optimal policy $\pi^*$ can be derived by behaving greedily to the optimal action-value function $Q^*$, where the latter fulfills $Q^*(s,a) = \max_{\pi}Q^{\pi}(s,a)$ for all $s \in \mathcal{S}$ and $a \in \mathcal{A}$ \citep{puterman2014markov}. Policy gradient algorithms, on the other hand, directly optimize a parametrized policy $\pi^{\theta}$ with respect to a performance measure such as the expected return under policy $\pi^{\theta}$ over the initial state distribution. The set $\theta$ is the parameter set of the policy $\pi^{\theta}$, for example, the biases and weights of a neural network. Lastly, several state-of-the-art algorithms belong to the class of \emph{actor-critic} algorithms, which are policy gradient algorithms that use action-value estimates during the gradient step. In particular, the policy is referred to as the \emph{actor}, and the action-value function is the \emph{critic} \citep{fujimoto2018addressing}.}

\subsection{\rd{RL algorithm LSTM-TD3}}
In practical applications, fully-observed systems, where the agent has access to the complete state vector, are uncommon due to sensor noise, delays, or other disturbances \citep{meng2021memory}. Recognizing this reality, we extend the MDP formalism to encompass partially-observable \rdd{MDPs} \citep{kaelbling1998planning} by introducing an observation space $\mathcal{O}$ and an observation function $\mathcal{Z}: \mathcal{S} \times \mathcal{A} \times \mathcal{O} \rightarrow [0,1]$. Consequently, at each time step $t$, the system resides in a state $s_t \in \mathcal{S}$, the agent receives an observation $o_t \in \mathcal{O}$ generated according to $\mathcal{Z}$, and then selects an action $a_t \in \mathcal{A}$. The system transitions to the next state $s_{t+1} \in \mathcal{S}$ based on $\mathcal{P}$ and provides a reward $r_t$ determined by $\mathcal{R}$. \rd{During implementation, the state input $s \in \mathcal{S}$ for an action-value function and policy is replaced by an observation $o \in \mathcal{O}$.} \rd{For example,} in our PF module, the observation vector includes the deviation to the local path and the currently acting environmental forces, while the action is a change in the rudder angle.

In this study, we use the Long Short-Term Memory-based Twin Delayed Deep Deterministic Policy Gradient (LSTM-TD3) algorithm proposed by \cite{meng2021memory} for both the \rd{LPP and PF agents. Building upon prior work by \cite{lillicrap2015continuous} and \cite{fujimoto2018addressing}, the LSTM-TD3 is an actor-critic algorithm maintaining an actor $\mu$ with parameter set $\theta_{\mu}$ and two critics $Q_1$ and $Q_2$ with parameter sets $\theta_{1}$ and $\theta_{2}$, respectively. Crucially, all networks incorporate LSTM layers \citep{hochreiter1997long}, which process the information of several times steps and offer a robust solution to address the challenges posed by partial observability. The optimization of the critics uses a DQN-style update \citep{mnih2015human} with the clipped double $Q$-Learning target proposed in \cite{fujimoto2018addressing}, while the actor is optimized based on the deterministic policy gradient theorem of \cite{silver2014deterministic}.}

While the algorithm is the same, the architecture of the \rdd{neural} networks differ for our LPP and PF agents and will be explained in detail in the upcoming Sections \ref{sec:LPP_module} and \ref{sec:PF_module}.

\sloppy
\section{Local path planning module}\label{sec:LPP_module}
\subsection{Configuration of the RL agent}
\subsubsection{Observation space}
\rd{The LPP agent constitutes the core of the LPP module, as illustrated in Figure \ref{fig:architecture}. Its responsibility is} to generate a reliable local plan while taking into account surrounding target ships, traffic rules, and the waterway geometry. This module does not consider environmental forces such as winds, waves, and currents, as those fall under the responsibility of the PF unit.

The observation for the LPP agent at time $t$, denoted as $o_{t}^{\rm LPP}$, is defined \rd{by stacking three vector components}: $o_{\rm OS,t}^{\rm LPP}$, which summarizes information about the own ship; $o_{\rm IW,t}^{\rm LPP}$, which describes the navigational area; and $o_{\rm TS,t}^{\rm LPP}$, which delivers information about the surrounding target ships. Specifically, we have:
\begin{equation}
o_{t}^{\rm LPP} = \left(\left(o_{\text{OS},t}^{\rm LPP}\right)^\top, \left(o_{\text{IW},t}^{\rm LPP}\right)^\top, \left(o_{\text{TS},t}^{\rm LPP}\right)^\top \right)^\top.
\end{equation}

\noindent\textbf{Own ship observation.} For the observation about the own ship, $o_{\rm OS,t}^{\rm LPP}$, the following features are included:
\begin{equation}
o_{\text{OS},t}^{\rm LPP} = \left(\frac{U_{\text{OS},t}}{U_{\rm scale}}, \frac{[\psi_{\text{OS},t} - {\chi_{P_k,t}]_{-\pi}^{\pi}}}{\pi}, \frac{[\chi^{\rm global}_{e,t}]_{-\pi}^{\pi}}{\pi}, \frac{y^{\rm global}_{e,t}}{y_{\rm scale}} \right)^{\top},\end{equation}
where $U_{\text{OS},t}$ represents the speed of the own ship at time $t$, $\psi_{\text{OS},t}$ is the heading of the own ship, $\chi_{P_k,t}$ is the angle between the two active waypoints of the own ship (as shown in Figure \ref{fig:VFG}), $y^{\rm global}_{e,t}$ is the cross-track error on the global path, and $\chi^{\rm global}_{e,t}$ is the course error derived from the VFG method. The superscript $global$ indicates that these features are specific to the LPP agent and are computed with respect to the global path. The VFG gain parameter is set to $k^{\rm LPP} = 0.001$. The function $[\cdot]_{a}^{a+2\pi}: \mathbb{R} \rightarrow [a, a+2\pi)$ is used to transform an angle to a desired domain; see \cite{waltz2022spatial}. Additionally, we have scaling parameters $U_{\rm scale} = \unit[3]{m/s}$ and $y_{\rm scale} = \unit[64]{m}$, where 64 meters corresponds to one length between perpendiculars ($L_{pp}$) of the downscaled KVLCC2 tanker.

\begin{figure}[h]
    \centering
    \includegraphics[width=0.7\textwidth]{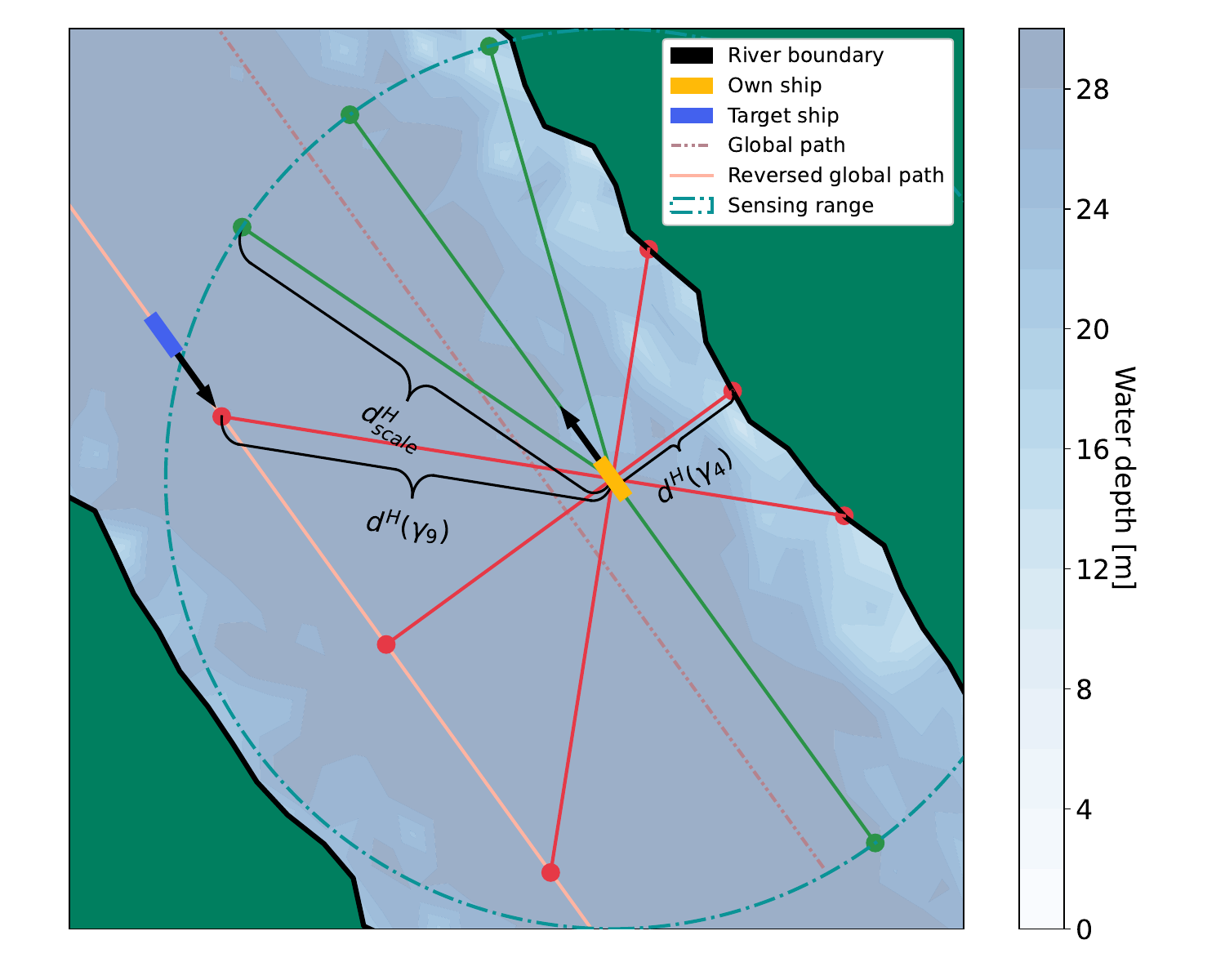}
    \caption{Visualization of $o_{\text{IW}, t}^{\rm LPP}$, representing the LPP units' awareness of the geometry of the waterway. The reversed global path is the path used to generate opposing traffic.}
    \label{fig:obs_lidar}
\end{figure}

\medskip
\noindent\textbf{Waterway observation.} To ensure the agent can plan without running aground, it requires information about the navigational area. We construct a vector, denoted as $o_{\text{IW}, t}^{\rm LPP}$, which contains proximities to the navigational boundary in 10 different directions\rd{; see Figure \ref{fig:obs_lidar}:}
\begin{equation}\label{eq:obs_coast}
o_{\text{IW},t}^{\rm LPP} = \left(1-\frac{d^{H}(\gamma_{1})}{d^{H}_{\rm scale}}, \ldots, 1-\frac{d^{H}(\gamma_{10})}{d^{H}_{\rm scale}} \right)^\top.
\end{equation}
The value $\gamma_i$ represents the $i$-th component, where $i = 1, \ldots, 10$, in radians of the vector with degree values $(0, 20, 45, 90, 135, 180, 225, 270, 315, 340)^\top$. We carefully selected these angles to provide the agent with a comprehensive view of its surroundings while avoiding the need for a high-dimensional feature vector that would require specifying angles, for example, for every degree around the clock. The function $d^H: [0, 2\pi) \rightarrow [0, d^H_{\text{scale}}]$ calculates the distance to either the coastline or the global path of the opposing traffic for a given angle.

During implementation, we examine for each angle up to 50 logarithmically scaled distances within a maximum range of $d^H_{\text{scale}} = \unit[1]{NM}$, leading to a normalization of the fractions in (\ref{eq:obs_coast}) to the interval $[0,1]$. We determine if a particular point is below the required water depth for the tanker or if it lies on the opposite side of the opposing traffic's path. Additionally, we follow the approach in \cite{heiberg2022risk} by reversing the signs of the fractions to provide the agent with information about the proximity rather than the distance to either the coastline or the opposing path.

\medskip
\noindent\textbf{Target ship observation.} For the observation of the $i$-th target ship at time $t$, denoted as $o_{\text{TS},i,t}$, the following features are considered:
\begin{equation}\label{eq:TS_obs}
\resizebox{1.\hsize}{!}{$
o_{\text{TS},i,t} = \left(
\frac{d^{i}_{\text{OS},t}-D(\alpha^{i}_{\text{OS},t})}{d_{\rm scale}},
\frac{[\alpha^{i}_{\text{OS},t}]_{-\pi}^{\pi}}{\pi},
\frac{[\psi_{i,t}-{\chi_{P_k, t}}]_{-\pi}^{\pi}}{\pi},
\frac{U_{i,t}- U_{\text{OS},t}}{U_{\rm scale}},
\sigma_{i,t},
\frac{t^{\rm cpa}_{i,t}}{t_{\rm norm}},
\frac{d^{\rm cpa,*}_{i,t}}{d_{\rm norm}}
\right)^{\top}.$}
\end{equation}
Here, $d^{i}_{\text{OS},t}$ represents the Euclidean distance between the own ship and the $i$-th target ship, $\alpha^{i}_{\text{OS},t}$ is the relative bearing of ship $i$ from the perspective of the own ship, $\psi_{i,t}$ and $U_{i,t}$ are the heading and speed of the target ship, respectively. The function $D: [0,2\pi) \rightarrow \mathbb{R}$ computes the ship domain around the own ship for a certain angle. Thus, the distance between the target ship's midpoint and the own ship's ship domain is provided to the agent, directly reflecting our definition of a collision event. The binary variable $\sigma_{i,t}$ indicates whether the other vessel is traveling in the same direction as the own ship and is defined as follows:

\begin{equation}
\sigma_{i,t} = 
    \begin{cases} 
      -1 & \text{if} \quad  \vert[\psi_{i,t}-\psi_{\text{OS},t}]_{-\pi}^{\pi}\vert \geq \pi/2, \\
      1 & \rm else. 
   \end{cases}
\end{equation}
The collision risk with the other ship is captured by $t_{i,t}^{\rm cpa}$, the TCPA between the own ship and ship $i$, and the variable $d_{i,t}^{\rm cpa, *}$. The latter follows \cite{waltz2022spatial} and is defined as $d_{i,t}^{\rm cpa, *} = \max{[0, d_{i,t}^{\rm cpa} - D(\alpha^{i, \rm cpa}_{\text{OS},t})]}$, where $d_{i,t}^{\rm cpa}$ is the regular DCPA and $\alpha^{i, \rm cpa}_{\text{OS},t}$ is the relative bearing of the ship $i$ from the perspective of the own ship at the \rdd{CPA}. Thus, we consider the ship domain in the computation of the DCPA metric similar to as we do it for the present distance to the target ship. The scaling parameters $t_{\rm norm} = \unit[300]{s}$ and $d_{\rm norm} = \unit[0.25]{NM}$ are used.

The complete target ship observation vector is constructed as: 
\begin{equation}
    o_{\text{TS}, t}^{\rm LPP} = \left(\left(o_{\text{TS},1,t}\right)^\top, \ldots, \left(o_{\text{TS},N_t,t}\right)^\top \right)^\top,
\end{equation}
where $N_t$ is the number of present target ships at time step $t$. The target ships within this vector are sorted in descending order based on their distance to the own ship. Furthermore, a target ship is only included if its distance to the own ship is less than $d_{\rm scale} = \unit[0.5]{NM}$, which is considered a reasonable range for practical operations on IWs.

\subsubsection{\rd{Neural network architecture}}
\begin{figure}
    \centering
    \includegraphics[width=01.\textwidth]{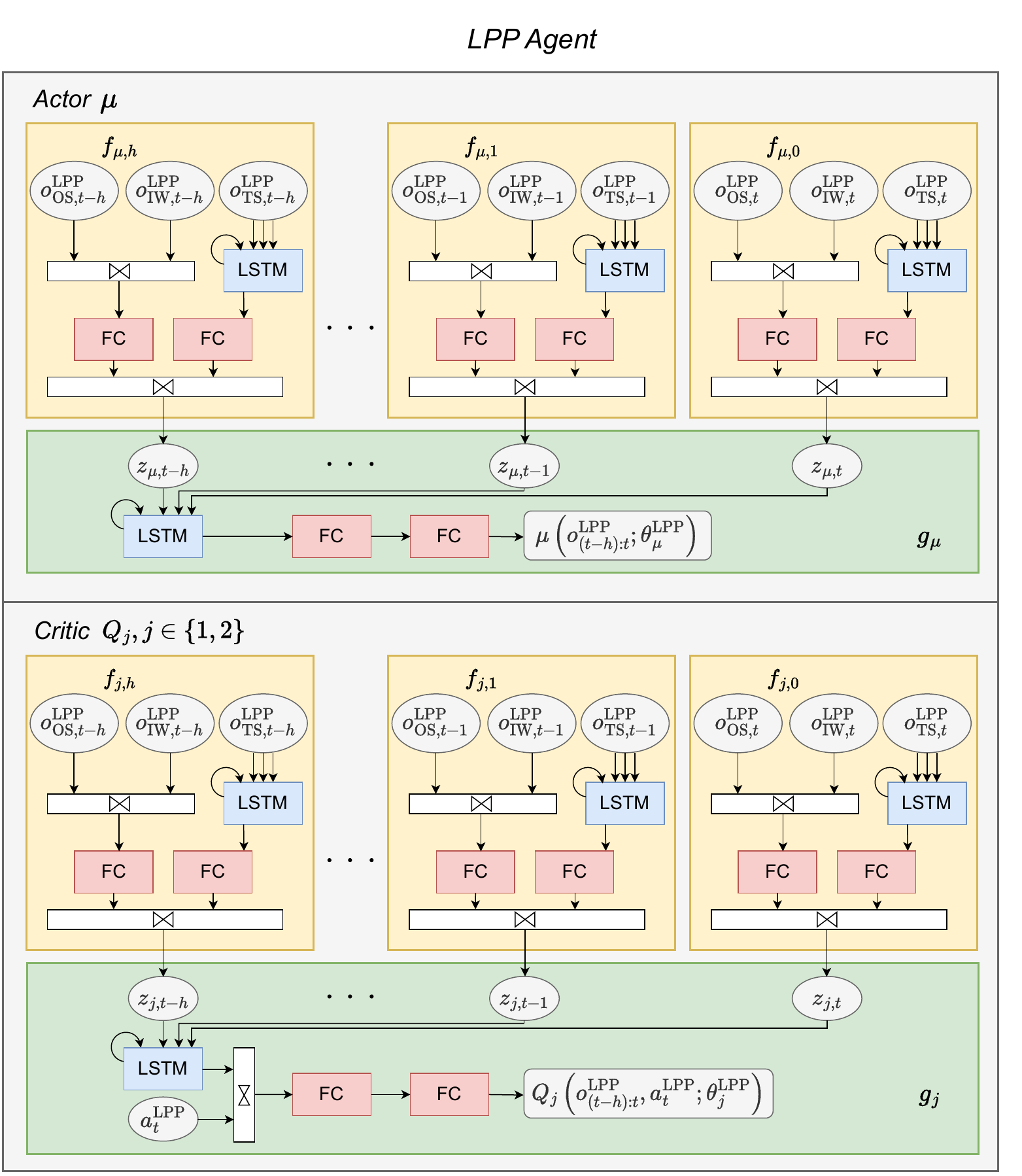}
    \caption{\rd{Neural network architecture for the LPP agent; adapted from \cite{waltz2022spatial}. We apply a ReLU activation after each FC layer, except for the last one of the actor and the critic, which use a tanh and linear activation, respectively.}}
    \label{fig:LPP_network}
\end{figure}

\rd{The varying number of surrounding ships, denoted as $N_t$, introduces a challenge for conventional feed-forward neural networks, which typically rely on a fixed input size. To overcome this challenge, we leverage the spatial-temporal recurrent neural network architecture proposed by \cite{waltz2022spatial} and integrate it into actor-critic methods. This adaptation empowers the model to effectively handle continuous action spaces. The schematic representation of our proposed architecture is illustrated in Figure \ref{fig:LPP_network}. In the figure, the concatenation operator is denoted as $\bowtie$, the fully connected (FC) layers have 64 neurons, and the number of hidden units in the LSTM layers is 64.}

\rd{In particular, we incorporate the spatial-temporal recurrent structure into both the actor and the two critics of the LSTM-TD3 algorithm. The spatial recurrent component loops over surrounding vessels, while the temporal recurrent component loops over time steps. This design choice equips our LPP agent with resilience against partial observability, while allowing it to adapt seamlessly to scenarios involving varying numbers of target ships.}

\rd{The actor's architecture closely mirrors that presented in \cite{waltz2022spatial}, except that we apply a hyperbolic tangent activation at the last layer to generate actions within the range $[-1,1]^{\vert \mathcal{A}\vert}$, where $\vert \mathcal{A}\vert$ denotes the cardinality of the action space. Further details on the action space \rdd{are provided} in Section \ref{subsubsec:action_space_LPP}. Formally, the actor network can be described as follows:
\begin{align}
    z_{\mu, t-l} &= f_{\mu,l}\left(o_{\text{OS},t-l}^{\rm LPP}, o_{\text{IW},t-l}^{\rm LPP}, o_{\text{TS},t-l}^{\rm LPP}; \theta_{f_{\mu,l}}\right) \quad \text{for} \quad l = 0, \ldots, h, \nonumber\\
    \mu\left(o_{(t-h):t}^{\rm LPP}; \theta_{\mu}^{\rm LPP}\right) &= g_{\mu}\left(z_{\mu,t-h}, \ldots, z_{\mu,t-1}, z_{\mu,t}; \theta_{g_{\mu}}\right),
\end{align}
where \rdd{$h$ is the history length,} the functions $f_{\mu,l}$ with parameter sets $\theta_{f_{\mu,l}}$ for $l=0, \ldots,h$ are the spatial recurrent components, and the function $g_{\mu}$ with parameter set $\theta_{g_{\mu}}$ represents the temporal recurrency. The notation $o_{(t-h):t}^{\rm LPP} = \cup_{l=0}^{h}o_{t-l}^{\rm LPP}$ is used to indicate that the actor network is a function of the past observations, while the complete parameter set of the actor is $\theta_{\mu}^{\rm LPP} = \left(\cup_{l=0}^{h} \theta_{f_{\mu,l}}\right) \cup \theta_{g_{\mu}}$.}

\rd{Similarly, the critics' architecture follows the principles outlined in \cite{waltz2022spatial}, with the difference that we concatenate the output of the temporal LSTM with the actions supplied by the actor. Formally, we have for critic $Q_j$ with $j \in \{1,2\}$:
\begin{align}
    z_{j, t-l} &= f_{j,l}\left(o_{\text{OS},t-l}^{\rm LPP}, o_{\text{IW},t-l}^{\rm LPP}, o_{\text{TS},t-l}^{\rm LPP}; \theta_{f_{j,l}}\right) \quad \text{for} \quad l = 0, \ldots, h,\nonumber\\
    Q_j\left(o_{(t-h):t}^{\rm LPP}, a_{t}^{\rm LPP}; \theta_{j}^{\rm LPP}\right) &= g_{j}\left(z_{j,t-h}, \ldots, z_{j,t-1}, z_{j,t}, a_{t}^{\rm LPP}; \theta_{g_{j}}\right),
\end{align}
with the spatial recurrent functions $f_{j,l}$ for $l = 0, \ldots,h$, parametrized with sets $\theta_{f_{j,l}}$, and the function $g_j$ with parameter set $\theta_{g_j}$, which represents the temporal recurrent component. The critics evaluate the action $a_{t}^{\rm LPP}$ provided by the actor, and the complete parameter set of critic $Q_j$ is denoted $\theta_{j}^{\rm LPP} = \left(\cup_{l=0}^{h} \theta_{f_{j,l}}\right) \cup \theta_{g_j}$.}

\rd{I}f there are no target ships present, we artificially create a no-risk ship with $o_{\text{TS},1,t} = (1, -1, 1, -1, 1, -1, 1)^\top$ to fulfill the requirement of having at least one target ship for the recurrence loop in the network architecture.

\subsubsection{Action space}\label{subsubsec:action_space_LPP}
We define a one-dimensional action space ($\vert \mathcal{A}\vert = 1$) for the LPP unit, which represents a change in the own ship's heading. At time step $t$, the agent generates an action $a_{t}^{\rm LPP} \in [-1,1]$ that is used to update the own ship's heading $\psi_{\text{OS},t}$ according to the following formula:
\begin{equation}
\psi_{\text{OS},t+1} = \psi_{\text{OS},t} + a_{t}^{\rm LPP} \cdot \rdd{a_c^{\rm LPP}},    
\end{equation}
where \rdd{$a_c^{\rm LPP} = 10^\circ$}. To ensure realistic and feasible paths, we only apply the agent's action every four time steps. Given our simulation step size of 5 seconds, this means that the planner can adjust the heading by a maximum of ten degrees every 20 seconds.

\subsubsection{Reward function}\label{subsubsec:reward_planner}
Designing an appropriate reward function is crucial in RL applications as it provides feedback to the agent's actions. By incorporating insights from \cite{waltz2022spatial} and \cite{paulig2023robust}, we have identified five reward components that facilitate LPP on IWs. These components address global \rdd{PF}, \rdd{COLAV}, traffic rule adherence, and comfort considerations.

The first two components focus on following the global path. We introduce a cross-track error-based reward, denoted as $r^{\rm LPP}_{y_e, t}$, and a course error-based component, denoted as $r^{\rm LPP}_{\chi_e, t}$. They are defined as follows:
\begin{equation}
r^{\rm LPP}_{y_e, t} = \exp \left(-k^{\rm LPP}_{y_e} \cdot \frac{\vert y^{\rm global}_{e, t} \vert}{y_{e, \text{norm}}} \right), \quad
r^{\rm LPP}_{\chi_e, t} = \exp \left(-k^{\rm LPP}_{\chi_e} \cdot \vert [\chi^{\rm global}_{e, t}]_{-\pi}^{\pi} \vert \right),
\end{equation}
where the power weights are $k^{\rm LPP}_{y_e} = 2$ and $k^{\rm LPP}_{\chi_e} = 4$, and the normalization length is $y_{e, \text{norm}} = \unit[128]{m}$, which corresponds to $2 L_{pp}$ of the KVLCC2 replica ship. These values are chosen to appropriately weigh the importance of the respective error terms in the reward computation.

\rd{The third reward component, denoted as $r^{\rm LPP}_{\text{coll}, t}$, penalizes collision events with other ships and leaving the navigational area. We, therefore, define the binary variables $\sigma_{\text{ground}, t}$ and $\sigma_{\text{lane}, t}$, which take the value one if the ship runs aground or crosses the global path of the opposing traffic, respectively, and zero otherwise. Both situations indicate leaving the navigational area and are therefore considered collision events. Further, we define $\sigma_{\text{coll}, i,t}$ as a binary variable that takes the value one if the own ship has a collision with target ship $i$ at time step $t$, and zero otherwise.}

\rd{In addition, research on COLAV for ASVs on open waters has shown the benefit of including a distance-based collision reward component \citep{xu2022path}. We also include such a signal into $r^{\rm LPP}_{\text{coll}, t}$ since it has the advantage of permanently yielding feedback for the RL agent, also in non-collision events. In particular, we will impose an elliptical collision reward around a target ship, where we encourage the agent to maintain a larger longitudinal distance from the target ship while allowing for smaller lateral distances, as the latter is necessary for overtaking maneuvers. This behavior is specified via the function $f : [0,2\pi) \times \mathbb{R} \rightarrow \mathbb{R}$ as follows:}
\begin{equation}
    f(\alpha, d) = \exp\left\{ \frac{-\left[d \cdot \sin(\alpha)\right]^2}{e_{\rm norm}^2}\right\} \cdot \exp\left\{ \frac{-\left[d \cdot \cos(\alpha)\right]^2}{n_{\rm norm}^2}\right\},
\end{equation}
where $e_{\text{norm}} = 3B$ and $n_{\text{norm}} = 1L_{pp}$, with $B$ being the width of the ship. \rd{On this basis, we introduce $r^{\rm LPP}_{\text{coll}, t}$ as:}
\begin{equation}\label{eq:r_coll}
\resizebox{1.\hsize}{!}{$
    r^{\rm LPP}_{\rm coll, t} = k_{\rm coll} \cdot \left( \sigma_{\rm ground, t} + \sigma_{\rm lane, t} + \sum_{i=1}^{N_t} \sigma_{\rm coll, i,t}\right) - \underset{i=1,\ldots,N_t}{\text{max}} f\left[\alpha_{i,t}^{\rm OS}, d^{i}_{\text{OS},t}-D\left(\alpha^{i}_{\text{OS},t}\right)\right]
$}
\end{equation}
\rd{where the collision weight $k_{\rm coll}$ is set to $-10$, the variable $\alpha_{i,t}^{\text{OS}}$ represents the relative bearing of the own ship from the perspective of target ship $i$, and $\alpha_{\text{OS},t}^{i}$ is the relative bearing of target ship $i$ from the perspective of the own ship.}

The fourth reward component, $r^{\rm LPP}_{\rm rule, t}$, concerns the compliance with traffic rules\rd{, which is investigated for each target ship separately. We define the binary variable $\sigma_{\text{rule},i, t}$ that takes value one if at time step $t$ a traffic rule with respect to target ship $i$ is violated. A traffic rule violation consists of three components in this study, which must be fulfilled simultaneously. First, the target ship needs to be close enough to be relevant. Second, it needs to travel in the same direction as the own ship. Third, one of the two behavioral rules we enforce during simulation is violated. As described in Section \ref{subsec:traffic_rules}, we require that other ships should be overtaken on their port side, corresponding to §23 (1) of \cite{SeeSchStrO1998}, and that if the own ship is being overtaken by another vessel, it should facilitate the maneuver by making space for the target ship. The latter requirement represents §23 (2) of the regulation, and we introduce the variable $\sigma_{\text{spd}, t}$ to check for its fulfillment. The variable takes the value one if all target ships traveling in the same direction as the own ship are faster than the own ship, and zero otherwise.}

\rd{In summary, we define $r^{\rm LPP}_{\rm rule, t} = k_{\rm rule} \cdot \sum_{i=1}^{N_t} \sigma_{\rm rule,i, t}$, where we set the constant to $k_{\rm rule} = -2$, and define the binary violation variable $\sigma_{\rm rule,i, t}$ as follows:}
\begin{equation}\label{eq:sigma_rule}
\resizebox{1.\hsize}{!}{$
\sigma_{\rm rule,i, t} = 
\begin{cases} 
  1 & \text{if} \quad 
  \left\{d_{\rm OS,t}^{i} \leq g(\alpha_{i,t}^{\rm OS})\right\}
  \land
  \left\{\vert[\psi_{i,t}-\psi_{\text{OS},t}]_{-\pi}^{\pi}\vert < \frac{\pi}{2}\right\}\\
  & \land
  \left\{\left[\left(U_{\rm OS,t} > U_{i,t}\right)
  \land
  \left(\frac{\pi}{2} \leq \alpha_{i,t}^{\rm OS} \leq \pi \right)\right]
  \lor\left[ \sigma_{\rm spd, t} \land  \left( \frac{3}{2}\pi \leq \alpha_{i,t}^{\rm OS} < 2\pi  \right) \right]\right\}\\
  0 & \rm else.
\end{cases}$}
\end{equation}
\rd{T}he function $g: [0,2\pi) \rightarrow \mathbb{R}$ calculates a bearing-dependent distance to determine if the target ship is within a range that requires consideration of traffic rules. Specifically, we select a lateral distance of \unit[0.25]{NM} and a longitudinal distance of \unit[0.5]{NM} from the target ship, and $g(\cdot)$ is defined to linearly interpolate between the four resulting corner points.

The fifth component is a comfort reward, $r^{\rm LPP}_{\rm comf,t}$, and \rd{should prevent the agent from frequently selecting large heading changes, resulting in a stable and smooth local path. Thus, we define: $r^{\rm LPP}_{\rm comf,t} = -\left(a_{t}^{\rm LPP}\right)^2$.}~Finally, we aggregate all five reward components into a single scalar via:
\begin{equation}
r^{\rm LPP}_t = 
r^{\rm LPP}_{y_e, t} \omega^{\rm LPP}_{y_e} + 
r^{\rm LPP}_{\chi_e,t} \omega^{\rm LPP}_{\chi_e} + 
r^{\rm LPP}_{\rm coll,t} \omega^{\rm LPP}_{\rm coll} + 
r^{\rm LPP}_{\rm rule,t} \omega^{\rm LPP}_{\rm rule} +
r^{\rm LPP}_{\rm comf,t} \omega^{\rm LPP}_{\rm comf},
\end{equation}
where we experimentally determined the weights as follows: $\omega^{\rm LPP}_{y_e} = \frac{4}{19} \approx 0.211$, $\omega^{\rm LPP}_{\chi_e} = \frac{1}{19} \approx 0.053$, $\omega_{\rm coll} = \omega^{\rm LPP}_{\rm rule} = \frac{6}{19} \approx 0.316$, and $\omega^{\rm LPP}_{\rm comf} = \frac{2}{19} \approx 0.105$.

\subsection{Training environment}
\subsubsection{Waterway generation}\label{subsubsec:river_generation}
\rdd{In the following, we denote a real-valued uniform distribution with support $[a,b]$ as $\mathcal{U}(a,b)$, an integer-valued uniform distribution with support $[a,b]$ as $\mathcal{DU}(a,b)$, and an exponential distribution with expectation $\beta$ as $Exp(\beta)$. To establish the simulation environment, we first sample a global path by interchanging straight and curved segments following the method described in \citet[Chapter 12]{fossen2021handbook}.} The length of each straight river interval is generated according to $\unit[400]{m} + \mathcal{DU}(0, 32) \cdot \unit[50]{m}$, and the radii of the curves are sampled from $\unit[1000]{m} + \mathcal{DU}(0, 4000) \cdot \unit[1]{m}$. Additionally, the curve angles are generated from $60^\circ + \mathcal{DU}(0, 40) \cdot 1^\circ$. We randomly sample left or right curves, respectively. Furthermore, we construct another path by imposing a fixed offset of 200 meters to the global path. The second path is referred to as reversed global path and is responsible for generating the opposing traffic.

Once we have established the global path, we proceed to create an IW by sampling water depths around it, following \cite{paulig2023robust}. Initially, we sample a maximum water depth in meters from $\text{clip}\left[Exp(35), 20, 100\right]$ with the clipping operation: $\text{clip}(x, l, u) = \min\left[u, \max(x,l)\right]$. The generated waterway has a maximum width of 500 meters. To introduce some variability to the depth data, we incorporate noise by adding realizations from $\mathcal{U}(-2, 2)$ in meters. We resample a generated waterway at the beginning of every fifth episode to keep the computational effort low.

\subsubsection{Target ship generation}
To initialize the target ships, we incorporate insights from the recent work conducted by \cite{hart2022enhanced}. Our aim is to create training scenarios that pose significant challenges to the agent, increasing the risk of collisions with other ships. In each episode, we randomly sample $N \sim \mathcal{DU}(0,10)$ target ships, encompassing a range of speeds, including both slower and faster vessels than the own ship, as well as ships traveling in opposing directions. The base vessel speed in our simulation is $U_{\rm base} = \unit[3]{m/s}$, and at each episode, we randomly sample the own ship's speed from $\mathcal{U}(0.8, 1.2) \cdot U_{\rm base}$. 

To initiate target ship $i$, where $i = 1, \ldots, N$ if $N > 0$, there is a 15\% probability of sampling a faster ship with a speed of $U_i \sim \mathcal{U}(1.3, 1.5) \cdot U_{\rm base}$. In this case, the vessel $i$ is initiated behind the own ship and travels in the same direction. The initial distance to the own ship along the global path, denoted $d_i$, is generated via $d_i \sim \mathcal{U}(0.3, 0.7) \cdot \unit[1]{NM}$, ensuring the vessel will create a threat for the own ship in the future. In the remaining 85\% of cases, vessel $i$ is slower with $U_i \sim \mathcal{U}(0.4, 0.8) \cdot U_{\rm base}$ and is initiated in front of the own ship. In this case, the vessel $i$ is randomly assigned to travel in the same or opposing direction. When traveling in the opposing direction, we sample $d_i \sim \mathcal{U}(1.1, 1.9) \cdot \unit[1]{NM}$ and place vessel $i$ on the reversed global path. Otherwise, we have $d_i \sim \mathcal{U}(0.3, 0.7) \cdot \unit[1]{NM}$.

\begin{figure}[ht]
    \centering
    \includegraphics[width=0.8\textwidth]{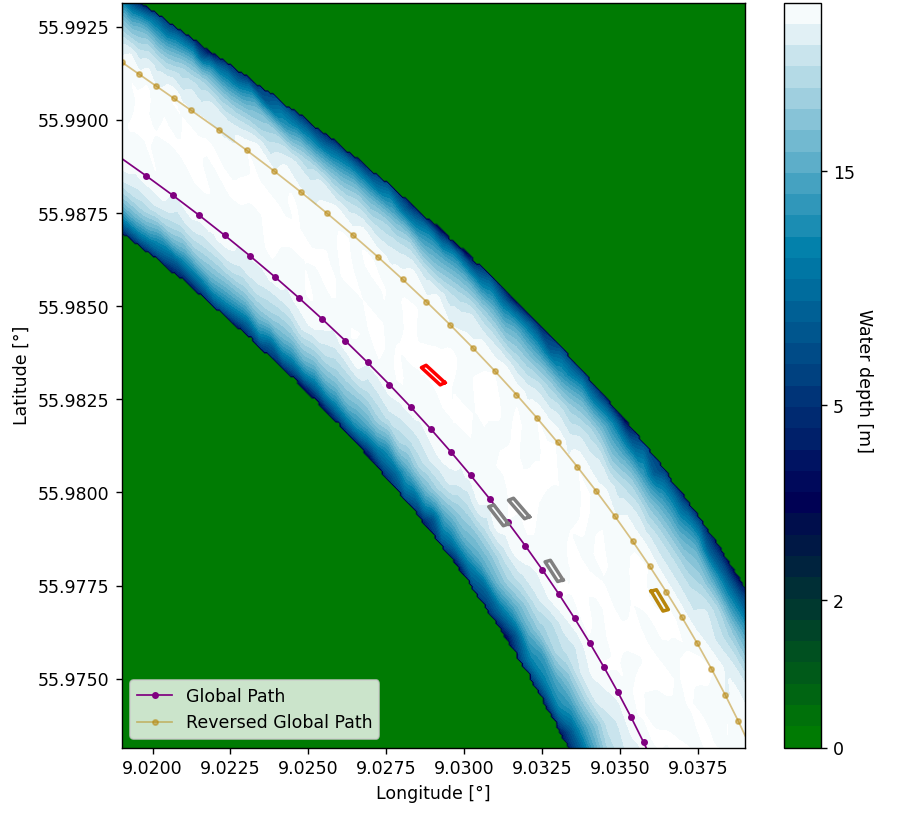}
    \caption{Screenshot of the simulation environment for the LPP agent (\rd{red}) with four target ships (\rd{grey in the agent's direction, golden otherwise}). \rd{T}he latitude and longitude values are artificial and serve as orientation.}
    \label{fig:LPP_Env_Screenshot}
\end{figure}

To introduce further variability and prevent the target ships from consistently spawning exactly on their respective paths, we introduce small positional noise to the target vessel's initial positions. An episode ends if a maximum of 150 steps has been reached or if the agent is more than $\unit[0.5]{NM}$ away from the global path. A screenshot of the full environment including waterway and target ships is shown in Figure \ref{fig:LPP_Env_Screenshot}. \rd{The own ship is depicted in red and travels in the south-east direction in this case, while the target ships are grey if they travel in the same direction as the own ship, and golden otherwise.}

\subsubsection{Target ship behavior modeling}\label{subsubsec:TS_behavior}
One of the key considerations in simulating AVSs is the control mechanism for the target ships \rdd{\citep{zhou2019review}}.~In our study, we focus on the non-communicative scenario \rdd{with} no explicit exchange of intentions or planned trajectories among vessels. In many existing works on ASVs, the assumption of linearly moving target ships with no course and speed changes \rdd{is adopted} \citep{guo2020autonomous, fan2022novel, xu2022path, xu2022colregs, sawada2021automatic}. However, while trained policies based on this assumption can generalize to non-linear target ship motions, as shown by \cite{waltz2022spatial}, the linearity and non-reactiveness assumptions are unrealistic in practical scenarios, particularly \rdd{on} IWs.

To address this limitation, we employ a simple rule-based controller for the target ships during the training of the LPP agent. The basis of this controller is the VFG method, as outlined in \rd{S}ection \ref{subsec:VFG}.~Additionally, the target ships are capable of performing basic overtaking maneuvers, in compliance with §23(1) of the collision regulations outlined in \cite{SeeSchStrO1998}. In head-on scenarios, they avoid collisions by executing a turn to starboard. For a detailed description of the complete controller, please refer to Algorithm \ref{algo:TS_control} in \ref{appendix:TS_control}. Importantly, considering the recent work of \cite{akdaug2022collaborative}, we incorporate a 20\% probability of generating a non-cooperative target vessel which does not perform any COLAV maneuvers.

\section{Path following module}\label{sec:PF_module}
\subsection{Configuration of the RL agent}
\subsubsection{Observation space}
\rd{Following the overall architecture in Figure \ref{fig:architecture},} the PF unit takes as input the local path generated from the LPP module and controls the rudder angle under consideration of the environmental disturbances. \rd{The PF module has no information about the IW geometry or the target ships and is solely responsible for tracking the given local path.}

The observation of the PF agent at time $t$ is denoted as $o_{t}^{\rm PF}$ and consists of a component describing the status of the own ship, denoted $o_{\text{OS},t}^{\rm PF}$, and a component regarding the environmental forces, denoted $o_{\text{Env},t}^{\rm PF}$. Thus, we have $o_{t}^{\rm PF} = \left(\left(o_{\text{OS}, t}^{\rm PF}\right)^\top, \left(o_{\text{Env}, t}^{\rm PF}\right)^\top\right)^\top$.

\medskip
\noindent\textbf{Own ship observation.} The component describing the status of the own ship is defined as:
\begin{equation}
    o_{\text{OS}, t}^{\rm PF} = \left( 
\frac{u_{\text{OS},t}}{u_{\rm scale}},
\frac{v_{\text{OS},t}}{v_{\rm scale}},
\frac{\Tilde{r}_{\text{OS},t}}{\Tilde{r}_{\rm scale}},
\frac{\Dot{\Tilde{r}}_{\text{OS},t}}{\Dot{\Tilde{r}}_{\rm scale}},
\frac{\delta_{\text{OS},t}}{\delta_{\rm max}},
\frac{y_{e,t}^{\rm local}}{y_{\rm scale}},
\frac{[\chi_{e,t}^{\rm local}]_{-\pi}^{\pi}}{\pi}
\right)^\top.
\end{equation}
Here, $u_{\text{OS},t}$, $v_{\text{OS},t}$, $\Tilde{r}_{\text{OS},t}$, $\Dot{\Tilde{r}}_{\text{OS},t}$, $\delta_{\text{OS},t}$ are the surge velocity, sway velocity, yaw rate, change in yaw rate, and rudder angle of the own ship, respectively. The scaling constants are: $u_{\rm scale} = \unit[3]{m/s}$, $v_{\rm scale} = \unit[0.2]{m/s}$, $\Tilde{r}_{\rm scale} = \unit[0.002]{rad/s}$, $\Dot{\Tilde{r}}_{\rm scale} = 8 \cdot \unit[10^{-5}]{rad/s^2}$, and $\delta_{\rm max} = 20^\circ$. The variables $y_{e,t}^{\rm local}$ and $\chi_{e,t}^{\rm local}$ are the cross-track error and the course error from the VFG method for the local path, which uses a gain parameter of $k^{\rm PF} = 0.01$.

\medskip
\noindent\textbf{Environmental force observation.} Further, we define the environmental observation component as follows:
\begin{equation}
\resizebox{1.\hsize}{!}{$
\begin{split}
    o_{\text{Env}, t}^{\rm PF} = \biggl( 
    &\frac{V_{c,t}}{V_{\rm c, norm}},
    \frac{[\beta_{c,t} -\psi_{\text{OS},t}]_{-\pi}^{\pi}}{\pi},
    \frac{V_{wi,t}}{V_{wi, \text{norm}}},
    \frac{[\beta_{wi,t} -\psi_{\text{OS},t}]_{-\pi}^{\pi}}{\pi},
    \frac{[\beta_{wa,t} -\psi_{\text{OS},t}]_{-\pi}^{\pi}}{\pi},\\
    &\frac{\zeta_{wa, t}}{\zeta_{wa, \text{norm}}},
    \frac{T_{wa, t}}{T_{wa, \text{norm}}},
    \frac{\lambda_{wa, t}}{\lambda_{wa, \text{norm}}},
    \frac{H_t}{H_{\rm norm}}\biggl)^\top,
\end{split}$}
\end{equation}
where $V_{c,t}$, $\beta_{c,t}$, $V_{wi,t}$, $\beta_{wi,t}$ are the current and wind's speed and angle of attack, respectively, at the position of the own ship at time $t$. Moreover, $\beta_{wa,t}$, $\zeta_{wa, t}$, $T_{wa, t}$, $\lambda_{wa, t}$, are the wave's angle, height, period, and length. Finally, we included the water depth $H_t$ at the own ship's position into the observation vector since it also affects the dynamics via the shallow water corrections outline in \rd{S}ection \ref{sec:vessel_dynamics}. The scaling parameters are set to: $V_{c, \text{norm}} = \unit[0.5]{m/s}$, $V_{wi, \text{norm}} = \unit[15]{m/s}$, $\zeta_{wa, \text{norm}} = \unit[2]{m}$, $T_{wa, \text{norm}} = \unit[7]{s^{-1}}$, and $\lambda_{wa, \text{norm}} = H_{\rm norm} = \unit[100]{m}$.

\subsubsection{\rd{Neural network architecture}}
\begin{figure}[!ht]
    \centering
    \includegraphics[width=0.7\textwidth]{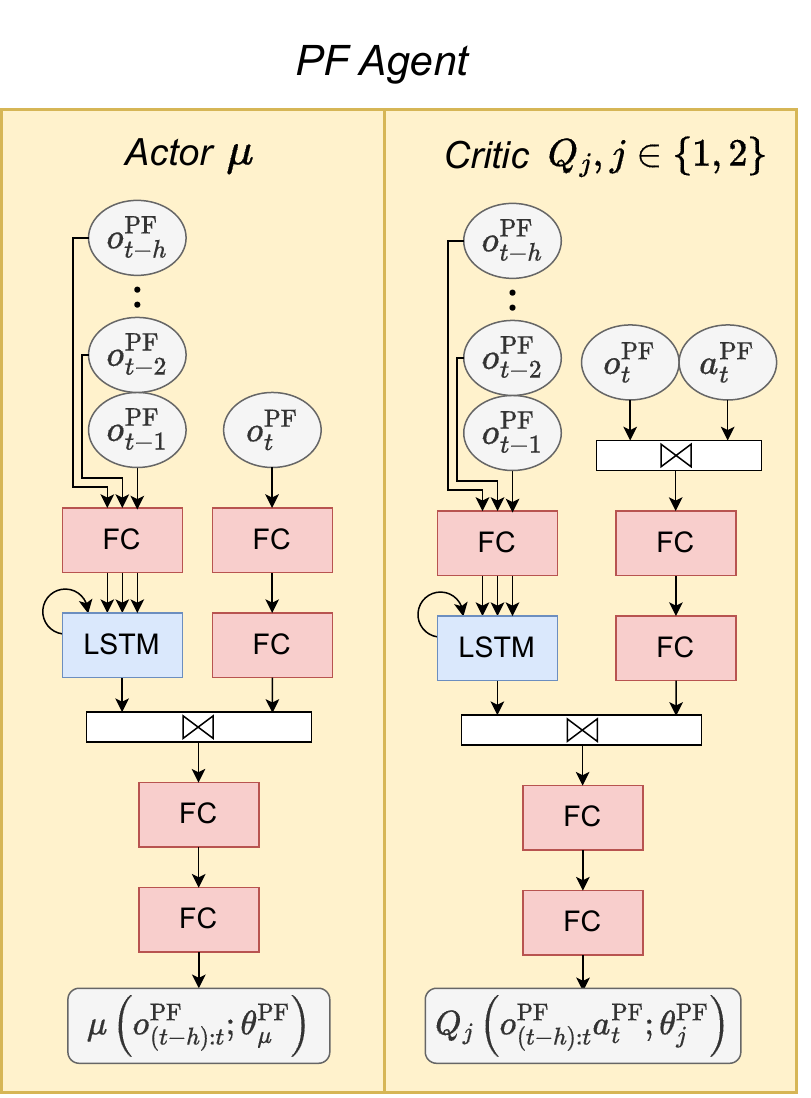}
    \caption{\rd{Neural network architecture used for the PF agent; adapted from \rdd{\cite{hart2023two}}. We apply a ReLU activation after each FC layer, except for the last one of the actor and the critic, which use a tanh and linear activation, respectively.}}
    \label{fig:PF_network}
\end{figure}

\rd{We use the network architecture outlined in \rdd{\cite{hart2023two}} for both actor $\mu$ and the critics $Q_j$, $j \in \{1,2\}$ of the PF unit, which is similar to the original proposal of \cite{meng2021memory}. The architecture is visualized in Figure \ref{fig:PF_network}, where the FC layers use 128 neurons and the number of hidden units of the LSTM layers is 128. In particular, the actor $\mu$, parametrized by $\theta_{\mu}^{\rm PF}$, processes the observations $o_{(t-h):t}^{\rm PF} = \cup_{l=0}^{h}o_{t-l}^{\rm PF}$ to yield an action $\mu\left(o_{(t-h):t}^{\rm PF}; \theta_{\mu}^{\rm PF}\right)$. The critic $Q_j$, $j \in \{1,2\}$, which is parametrized by the set $\theta_{j}^{\rm PF}$, evaluates a given action $a_{t}^{\rm PF}$ to produce an action-value estimate $Q_j\left(o_{(t-h):t}^{\rm PF}, a_{t}^{\rm PF}; \theta_{j}^{\rm PF}\right)$. Following \cite{meng2021memory}, the networks process the past observations $o_{t-h}^{\rm PF},\ldots,o_{t-1}^{\rm PF}$ through an LSTM layer, which constitutes a \emph{memory extraction} unit. The current observation $o_{t}^{\rm PF}$ is separately processed by \rdd{FC} layers, which \cite{meng2021memory} refer to as the \emph{current feature extraction} unit.}

\subsubsection{Action space}
The PF agent controls the rudder angle of the own ship. Building on the observation defined in the last subsection, the agent computes an action $a_{t}^{\rm PF} \in [-1,1]$ which adjusts the rudder angle as follows:
\begin{equation}
\delta_{\text{OS},t+1} = \text{clip}\left(\delta_{\text{OS},t} + a^{\rm PF}_{t} \cdot \rdd{a_c^{\rm PF}},  -\delta_{\rm max}, \delta_{\rm max}\right),
\end{equation}
where the clipping operation ensures the absolute value of the rudder angle does not exceed $\delta_{\rm max} = 20^\circ$. Further, we set $\rdd{a_c^{\rm PF}} = 5^\circ$, which results in combination with our \rdd{control frequency of $t_{\rm control} = \unit[5]{s}$} in realistic rudder changes.

\subsubsection{Reward function}
The reward function for the PF unit builds on \cite{paulig2023robust} and consists of three components: a cross-track error reward, $r^{\rm PF}_{y_e,t}$, a course error component, $r^{\rm PF}_{\chi_e,t}$, and a comfort reward, $r^{\rm PF}_{\text{comf},t}$. Similar to the LPP agent's reward function of \rd{S}ection \ref{subsubsec:reward_planner}, we define:
\begin{align}
    r^{\rm PF}_{y_e,t} &= \exp \left(-k_{y_e}^{\rm PF} \vert y_{e,t}^{\rm local}\vert \right), \quad r_{\text{comf}, t}^{\rm PF} = -\left(a^{\rm PF}_{t}\right)^2,\\
    r^{\rm PF}_{\chi_e,t} &= \begin{cases}
        k_{\rm turn} & \text{if}\;\; \vert \chi_{e,t}^{\rm local}\vert \geq \frac{\pi}{2}\\
        \exp \left(-k_{\chi_e}^{\rm PF} \vert \chi_{e,t}^{\rm local}\vert \right) & \text{else.}
    \end{cases}
\end{align}
We set the constants $k_{y_e}^{\rm PF} = 0.05$, $k_{\chi_e}^{\rm PF} = 5$, and $k_{\rm turn} = -10$. The condition with the large negative penalty for the course error is included to prevent the agent from completely turning around and following the path in the wrong direction. The reward components are aggregated to the reward for the PF agent at time $t$, denoted $r^{\rm PF}_{t}$, as follows:
\begin{equation}
    r^{\rm PF}_{t} = 
r^{\rm PF}_{y_e, t} \omega^{\rm PF}_{y_e} + 
r^{\rm PF}_{\chi_e,t} \omega^{\rm PF}_{\chi_e} + 
r^{\rm PF}_{\rm comf,t} \omega^{\rm PF}_{\rm comf},
\end{equation}
where a small grid search yielded the equal weights: $\omega^{\rm PF}_{y_e} = \omega^{\rm PF}_{\chi_e} = \omega^{\rm PF}_{\rm comf} = \frac{1}{3}$.

\subsection{Training environment}
The IW generation for the PF unit is identical to the one described in \rd{S}ection \ref{subsubsec:river_generation}. However, in this module, we need to sample the current, wind, and wave conditions since it is the PF unit's primary responsibility to account for these. The respective angles of attack in radians are separately sampled from $\mathcal{U}(0, 2\pi)$. Moreover, we sample a current speed from $\text{clip}\left[Exp(0.2), 0, 0.5\right] \cdot \unit[1]{m/s}$ and a wind speed from $\mathcal{U}(0, 15) \cdot \unit[1]{m/s}$. The wave height, length, and period are sampled from $\text{clip}\left[Exp(0.1), 0.01, 2\right] \cdot \unit[1]{m}$, $\text{clip}\left[Exp(20), 1, 100\right] \cdot \unit[1]{m}$, and $\text{clip}\left[Exp(1), 0.5, 7\right] \cdot \unit[1]{s^{-1}}$, respectively. Furthermore, we introduce zero-mean Gaussian noise to each value when queried by the own ship, enhancing the robustness of the policy.

The termination conditions for an episode are as follows: if more than 500 steps have elapsed, if the agent strays more than 400 meters away from the local path, or if the water depth at the own ship's position becomes insufficient, rendering the shallow water approximations infeasible.

\section{Results and validation}\label{sec:results_validation}
\subsection{Training details}
The LPP and PF agents are separately trained for $2 \cdot 10^6$ and $3 \cdot 10^6$ steps, respectively. The remaining hyperparameters are the same for both agents and are outlined in Table \ref{tbl:hyperparams}. \rdd{We thereby primarily followed the configuration of \cite{hart2023two}. However, an exception is the history length, which is set to $h=2$ for computational reasons, following the setup of \cite{waltz2022spatial}.} All experiments were conducted using Python 3.8.6 \citep{python3} and the PyTorch deep learning library \citep{paszke2019pytorch} version 1.10.0. The computational hardware employed for these experiments consisted of Intel(R) Xeon(R) CPUs E5-2680 v3 (12 cores) running at 2.50 GHz. We make the source code to this study publicly available in a GitHub repository in \cite{TUDRL} to ensure full reproducability. \rdd{In addition, we note that while the training of the DRL agents is computationally demanding due to the time-consuming optimization of the neural networks, the inference after training is extremely fast. Therefore, the proposed approach is highly practical, as the real-time computation of control commands during operations requires only a few milliseconds.}

\begin{table}[!htb]
\footnotesize
    \def\arraystretch{1.1}
    \centering
    \begin{tabular}{ll}
    Hyperparameter & Value\\
    \toprule   
    Batch size & 32\\
    Discount factor & 0.99\\
    History length ($h$) & 2\\
    Learning rate (actor) & 0.0001\\
    Learning rate (critic) & 0.0001 \\
    Loss function & Mean squared error\\
    Max. replay buffer size & $5 \cdot 10^5$\\
    Min. replay buffer size & 5,000\\
    Optimiser & Adam \citep{kingma2014adam}\\
    Policy update delay & 2\\
    Soft update rate & 0.001\\
    Target policy smoothing noise & 0.2 \\
    Target policy smoothing noise clip & 0.5 \\
    \end{tabular}
    \caption{List of hyperparameters. For a detailed description of each parameter we refer to \cite{fujimoto2018addressing} and \cite{meng2021memory}.}
    \label{tbl:hyperparams}
\end{table}

\begin{figure}[h]
\hfill
\subfigure[\rdd{LPP} unit]{\includegraphics[width=6.5cm]{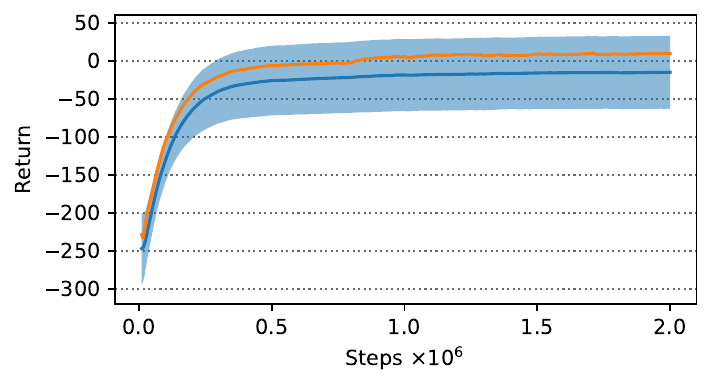}}
\hfill
\subfigure[\rdd{PF} unit]{\includegraphics[width=6.5cm]{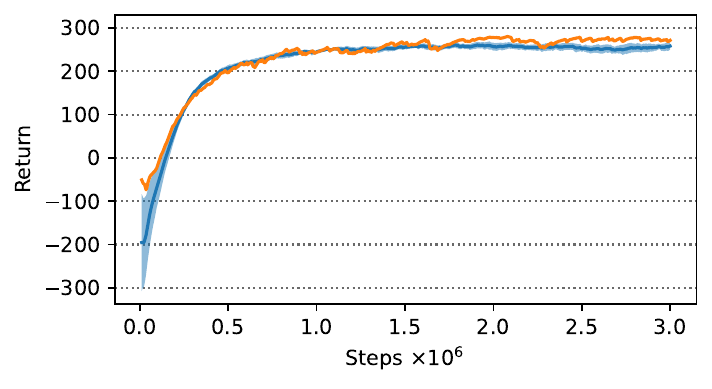}}
\hfill
\caption{The two plots show the test return's development during training for the two agents.}
\label{fig:training_plot}
\end{figure}

During the training process, we average every 5000 steps the test return, which is the sum of rewards, of 3 evaluation episodes. To enhance clarity, we apply exponential smoothing to these values. Additionally, we perform the experiment ten times under different random initialization of the neural networks. This enables us to compute confidence intervals that better represent the training performance, which are shown in Figure \ref{fig:training_plot}. \rd{The blue line in each of the plots is the respective mean over ten independent runs, and the light blue area around it represents the 95\% pointwise confidence interval. The orange line is the run whose final policy is validated in the upcoming section.} The LPP unit's training plot has a larger variance due to its dependence on five distinct reward components. In contrast, the PF unit's training is less variable since \rdd{PF} is a more straightforward control task, minimizing spatial and angular deviation from the local path.

In the upcoming sections, we separately validate the proposed LPP and PF modules on simulation examples.~Afterward, we employ the complete architecture described in \rd{S}ection \ref{sec:architecture} in validation scenarios on real AIS data.

\subsection{Validation: Local path planning module}\label{subsec:val_planning}
\subsubsection{\rd{Setup}}
We developed a comprehensive procedure to thoroughly evaluate the performance of the LPP unit. The procedure consists of six distinct setups, which are among the most challenging \rd{and safety-critical} scenarios encountered on \rdd{IWs}:
\begin{enumerate}
    \item overtaking a vessel train,
    \item overtaking an overtaker,
    \item overtaking under oncoming traffic,
    \item overtaking an overtaker under oncoming traffic,
    \item getting overtaken,
    \item and navigating along static obstacles. 
\end{enumerate}
The inclusion of static obstacles along the ship's path is particularly important to assess the LPP agent's generalization capabilities, as the agent has not been exposed to such obstacles during training. We analyze the behavior of the DRL agent separately in each of these six setups on a straight waterway segment, a left curve, and a right curve, respectively. This results in a total of 18 different scenarios being studied.

\rd{Moreover, we compare the performance of the DRL planning agent for the straight waterway segment with a state-of-the-art APF method, building on the recent proposals of \cite{liu2023colregs} and \cite{wang2019obstacle}. The detailed functionality of the APF approach is described in \ref{app:APF}. We choose three performance metrics to compare the two methods thoroughly. Following \cite{jadhav2023collision}, the first metric is the controller effort (CE) over a trajectory, which measures the average commanded heading change of the respective planning method. Formally, we define:
\begin{equation}
    \text{CE}_{\rm LPP} = \frac{1}{\Delta_{\psi} \cdot T}\sum_{t=1}^{T} \vert \psi_{\text{OS},t} - \psi_{\text{OS},t-1}\vert,
\end{equation}
where $T$ is the length of the trajectory. Note that the CE is normalized by $\Delta_\psi$, which is the maximum possible heading change between consecutive time steps. As described in Section \ref{subsubsec:action_space_LPP}, this value is $10^\circ$ for the DRL agent, while we reduced it to $2^\circ$ for the APF method, which heavily increased the latter method's performance. Generally, the CE should be as small as possible to generate smooth local paths that allow for fuel efficient operations.}

\rd{The second performance metric for the LPP task is the mean cross-track error (MCTE), which is defined by \cite{jadhav2023collision} as follows:
\begin{equation}
    \text{MCTE}_{\rm LPP} = \frac{1}{L_{pp} \cdot T}\sum_{t=0}^{T} \vert y_{e,t}^{\rm global}\vert,
\end{equation}
which includes a normalization by the length between perpendiculars $L_{pp} = \unit[64]{m}$ of the vessel. A smaller MCTE indicates a planned path close to the global path, reflecting strong tracking capabilities.}

\rd{Finally, incorporating safety considerations into our evaluation, we introduce a third performance metric named MinDist. This metric is the minimum distance encountered with respect to any target ship along the trajectory. Formally, we have:
\begin{equation}
   \text{MinDist} = \frac{1}{L_{pp}} \cdot \min_{t \in \{0,\ldots, T\}} \min_{i \in \{0, \ldots, N_t\}} \left[d_{\text{OS},t}^{i}-D\left(\alpha_{\text{OS},t}^{i}\right) \right], 
\end{equation}
where we consider the ship domain of the own ship similar to the target ship observation in (\ref{eq:TS_obs}). Generally, a sufficiently large MinDist indicates a safe operation.}

\subsubsection{\rd{Analysis}}
\begin{figure}[!htb]
    \centering
    \includegraphics[width=\textwidth]{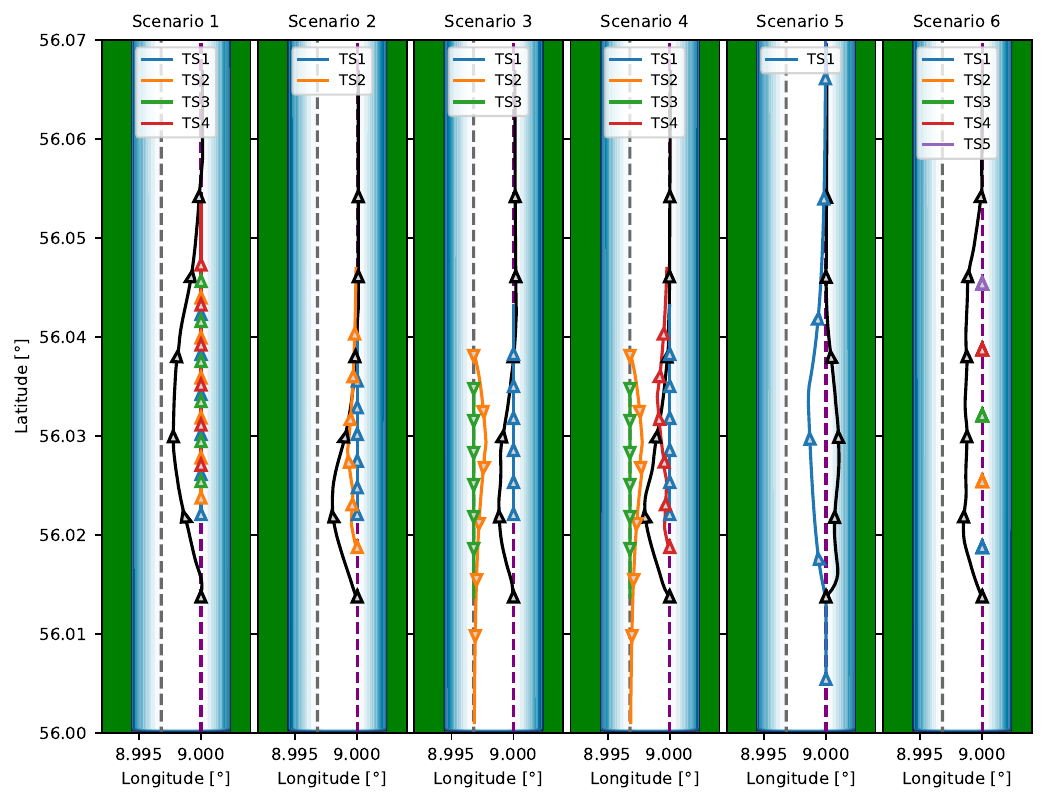}
    \caption{Trajectories of the \rd{LPP} validation scenarios of the \rd{DRL} agent on a straight river segment. Note that the latitude and longitude values are artificial and serve as orientation.}
    \label{fig:traj_val_plan_straight}
\end{figure}

\begin{figure}[!htb]
    \centering
    \includegraphics[width=\textwidth]{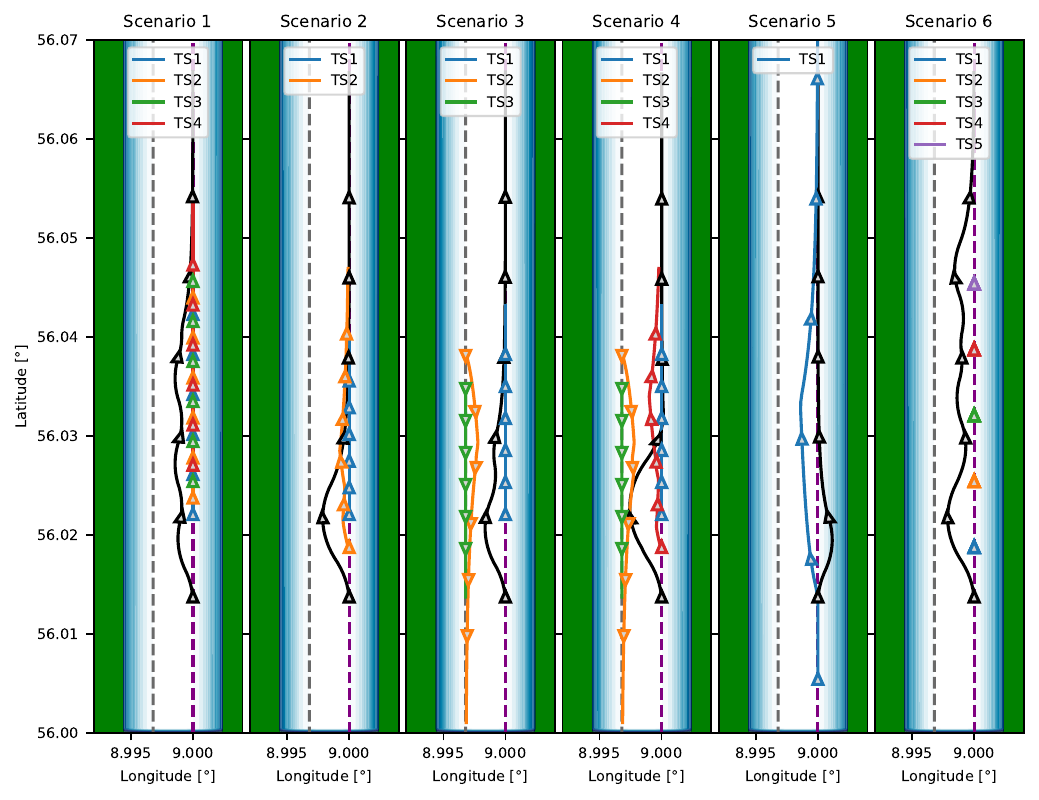}
    \caption{\rd{Trajectories of the LPP validation scenarios of the APF method on a straight river segment. Note that the latitude and longitude values are artificial and serve as orientation.}}
    \label{fig:APF_traj_val_plan_straight}
\end{figure}

\begin{figure}[!htb]
    \centering
    \includegraphics[width=\textwidth]{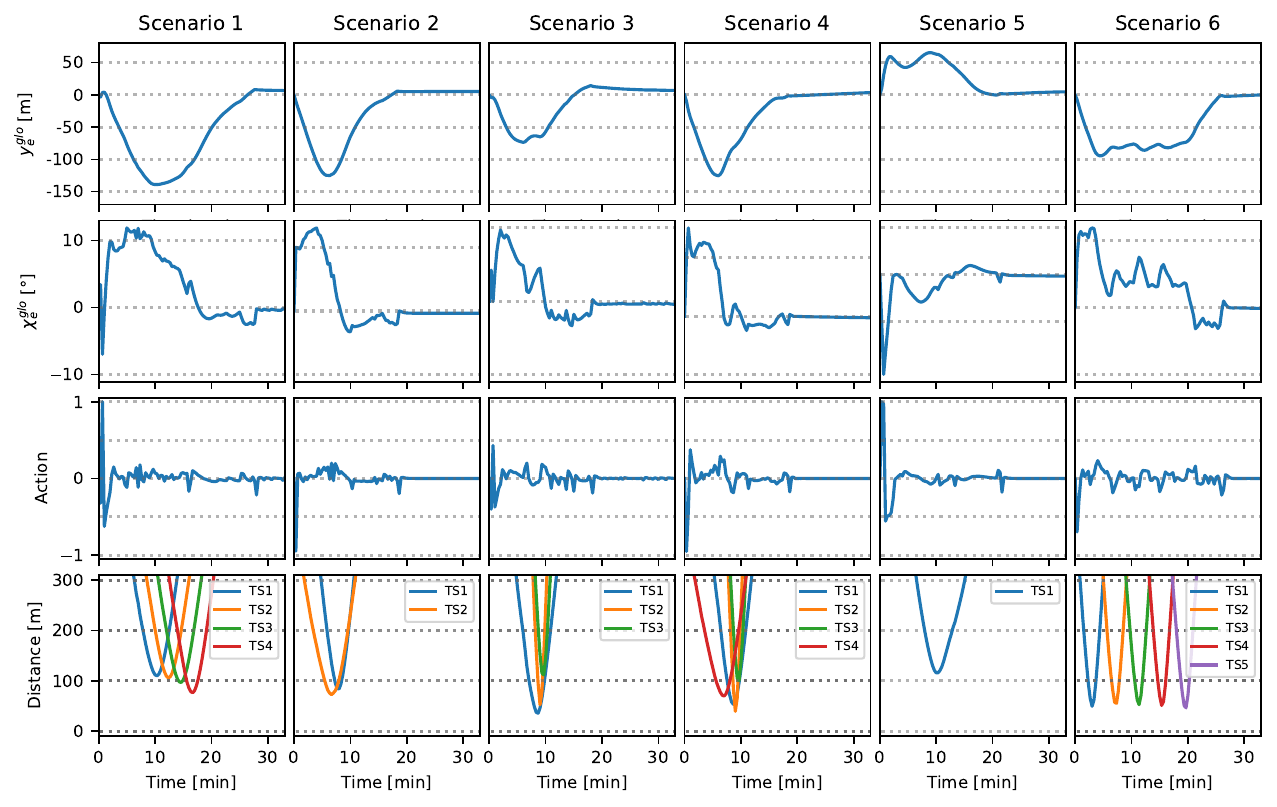}
    \caption{Global cross-track and course error, selected actions, and distances to the target ships during validation of the LPP agent on a straight waterway segment.}
    \label{fig:metric_val_plan_straight}
\end{figure}

\rd{The trajectories for the straight case of the DRL agent and the APF method are visualized in Figures \ref{fig:traj_val_plan_straight} and \ref{fig:APF_traj_val_plan_straight}, respectively. The black trajectories in these figures correspond to the respective planning agent, while the colorized ones are the target ships controlled by Algorithm \ref{algo:TS_control}. The purple and grey dotted lines are the global and reversed global paths. Further, Table \ref{tab:LPP_metrics} contains the corresponding performance metrics.} Additionally, Figure \ref{fig:metric_val_plan_straight} provides plots of the cross-track and course errors to the global path, the selected action by the \rd{DRL} agent, and the distance to the target ships. The results for the curved scenarios \rd{alongside the initial speed configurations of each vessel} can be found in \ref{app:val_planner}.

\begin{table}[htp!]
\color{red(ncs)}
\centering
\begin{tabular}{ccccccc}
\toprule[0.05em]
\multirow{2}{*}{Scenario} & \multicolumn{2}{c}{$\text{CE}_{\rm LPP}$} & \multicolumn{2}{c}{$\text{MCTE}_{\rm LPP}$} & \multicolumn{2}{c}{MinDist}\\ \cmidrule[0.05em]{2-7}
& DRL & APF & DRL & APF & DRL & APF \\ 
  \midrule
  1 & 0.067 & 0.565  & 0.954 & 0.576 & 1.197 & 0.434\\
  2 & 0.040 & 0.283 & 0.523 & 0.388 & 1.136 & 0.639 \\
  3 & 0.052 & 0.343 & 0.381 & 0.349 & 0.558 & 0.423 \\
  4 & 0.054 & 0.374 & 0.502 & 0.456  & 0.618 & 0.655 \\
  5 & 0.064 & 0.195 & 0.390 & 0.172 & 1.810 & 0.928 \\
  6 & 0.064 & 0.710 & 0.801 & 0.855 & 0.727 & 0.589 \\
  \midrule
  Average & 0.057 & 0.412 & 0.592 & 0.466 & 1.008 & 0.611\\
  \bottomrule[0.05em]
\end{tabular}
\caption{\rd{Performance comparison of the DRL agent with the APF method for the LPP task. All metrics are dimensionless.}}
\label{tab:LPP_metrics}
\end{table}

Focusing on Scenario 1 in Figures \ref{fig:traj_val_plan_straight} and \ref{fig:metric_val_plan_straight}, we can observe the successful overtaking maneuver performed by the LPP agent on a vessel train consisting of four target ships. Initially, the agent steers to the port side to avoid a collision with the nearest target ship and then overtakes each vessel one by one until the entire vessel train has been passed. Importantly, this behavior complies with §23(1) of the regulations specified in \cite{SeeSchStrO1998}, as the agent performs overtaking maneuvers on the target vessel's port side. Once the overtaking maneuver is completed safely, the agent returns to the global path until the cross-track and course errors are close to zero again. Moreover, the action selection during the maneuver is relatively moderate since the agent avoids large consecutive heading changes.

Similar successful results can be observed in the overtaking cases of Scenarios 2 to 4, demonstrating the LPP agent's proficiency in executing advanced maneuvers while maintaining appropriate safe distances from the target ship. Furthermore, the agent's action selection demonstrates a balanced approach in these scenarios, indicating the successful integration of the comfort reward component. In Scenario 5, we can observe the agent's compliance with §23(2), as it enables overtaking of the target ship by slightly moving towards starboard, which directly reflects the traffic rule reward component defined in (\ref{eq:sigma_rule}). Of particular note is Scenario 6, where the agent effectively navigates around static obstacles while maintaining a minimum distance of approximately 50 meters.

\rd{Comparing the DRL agent to the APF method, we observe that the APF trajectories are more unstable and clearly display the undesired change between stronger and weaker repulsive forces when passing target ships; see, for instance, Scenario 6 of Figure \ref{fig:APF_traj_val_plan_straight}. Remarkably, as shown in Table \ref{tab:LPP_metrics}, this unstable behavior leads to a slightly smaller average MCTE of the APF method in comparison to the DRL agent. However, this circumstance comes at the expense of the increased criticality of the APF trajectories. Generally, although the APF approach does not produce collisions, the MinDist of the DRL controller is on average 65\% larger. Further following Table \ref{tab:LPP_metrics}, the CE is massively reduced by the DRL approach, which is in line with the visually unstable behavior of the APF method.}

\rd{Lastly, we emphasize that the hyperparameters of the APF method have been carefully tuned to increase its overall performance during the six scenarios on the straight waterway segment. However, deploying this APF configuration without further tuning on the curved waterway segments is not feasible, so we limit comparing the DRL and APF methods to the straight waterway case. On the contrary, our DRL agent can seemingly handle different waterway curvatures without further adaptations, as shown in \ref{app:val_planner}.}

\subsection{Validation: Path following module}\label{subsec:val_PF}
\subsubsection{\rd{Setup}}
We evaluate the performance of our PF agent by subjecting it to various environmental conditions, \rd{including the three major forces: currents, winds, and waves.} Each force is tested separately in both moderate and extreme scenarios, resulting in a total of six validation scenarios. In each scenario, the agent is tasked with following a straight path initially unaffected by environmental forces. Subsequently, a force field perpendicular to the path is introduced, followed by another segment with zero forces. Finally, the force direction is reversed in the last segment. This testing procedure enables us to assess the agent's adaptability to different environmental conditions.~\rdd{Moderate scenarios are defined as those with a current speed of $V_c = \unit[0.25]{m/s}$, a wind speed of $V_{wi} = \unit[5]{m/s}$, and wave heights of $\zeta_{wa} = \unit[0.5]{m}$. Extreme scenarios are these with $V_c = \unit[1]{m/s}$, $V_{wi} = \unit[20]{m/s}$, and $\zeta_{wa} = \unit[1.5]{m}$, reflecting extremely challenging navigation conditions.}

For comparison purposes, we follow \cite{paramesh2021unified} and \cite{paulig2023robust} and include a PID controller for the rudder angle, which was optimized using the PSO approach outlined in \cite{eberhart2000comparing}; see \ref{app:PID} for details. It should be emphasized that the PID controller is specifically optimized for the validation scenarios, while the RL agent's training encompasses a broader range of scenarios.

\rd{Furthermore, we follow \cite{jadhav2023collision} and consider the CE and the MCTE to quantitatively compare the performance of the two methods. However, contrary to Section \ref{subsec:val_planning}, the CE for PF measures the average absolute rudder angle, while the MCTE in this task is computed with respect to the local instead of the global path. Formally, we set:
\begin{equation}
    \text{CE}_{\rm PF} = \frac{1}{\delta_{\rm max} \cdot T}\sum_{t=0}^{T} \vert \delta_{\text{OS},t}\vert, \quad \quad
    \text{MCTE}_{\rm PF} = \frac{1}{B \cdot T}\sum_{t=0}^{T} \vert y_{e,t}^{\rm local}\vert,
\end{equation}
where $B = \unit[11.6]{m}$ is the width of the downscaled KVLCC2 tanker.}

\subsubsection{\rd{Analysis}}
\begin{figure}[!htb]
    \centering
    \includegraphics[width=\textwidth]{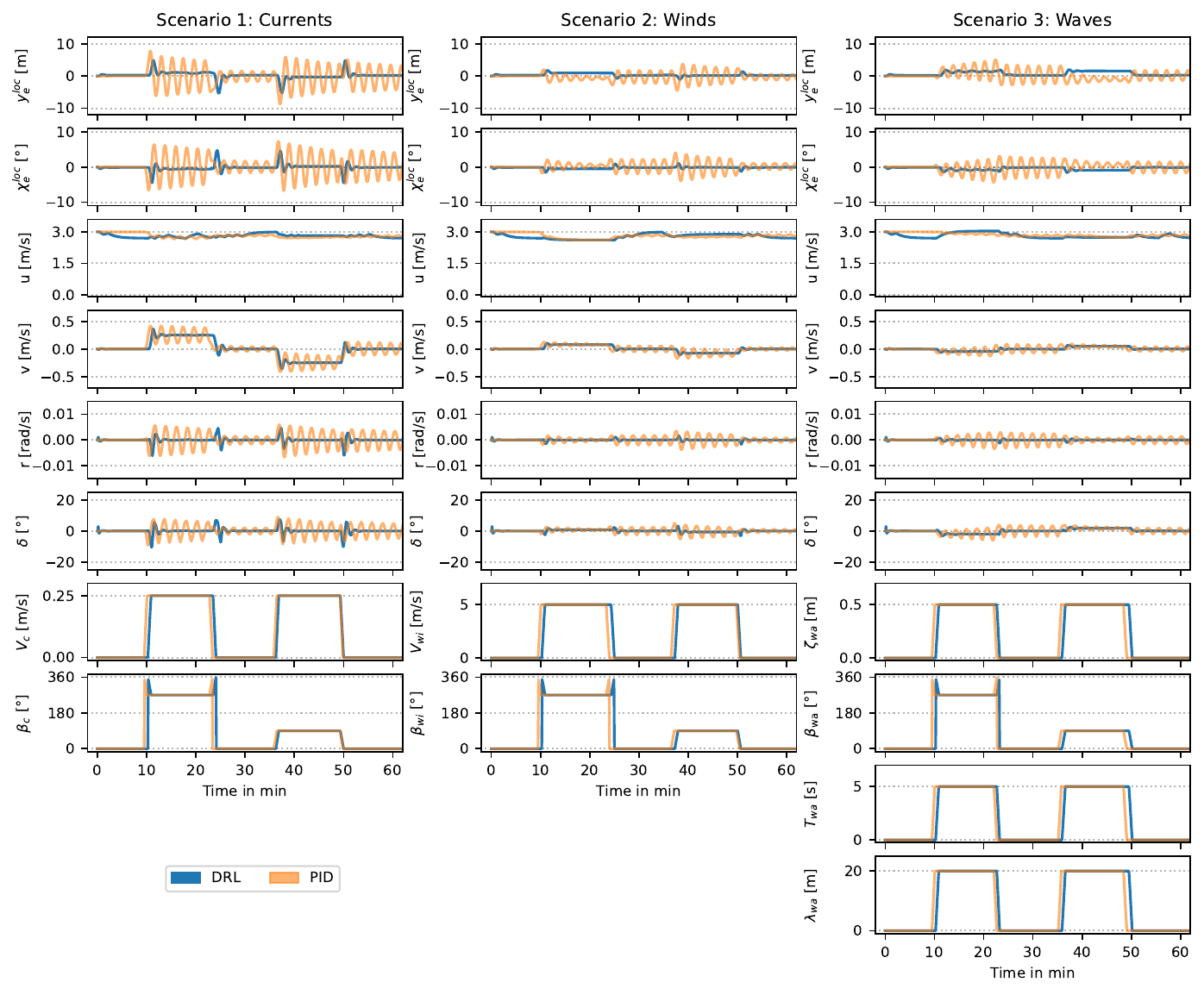}
    \caption{Validation results for \rdd{PF} under moderate environmental conditions.}
    \label{fig:follow_moderate}
\end{figure}

\begin{figure}[!htb]
    \centering
    \includegraphics[width=\textwidth]{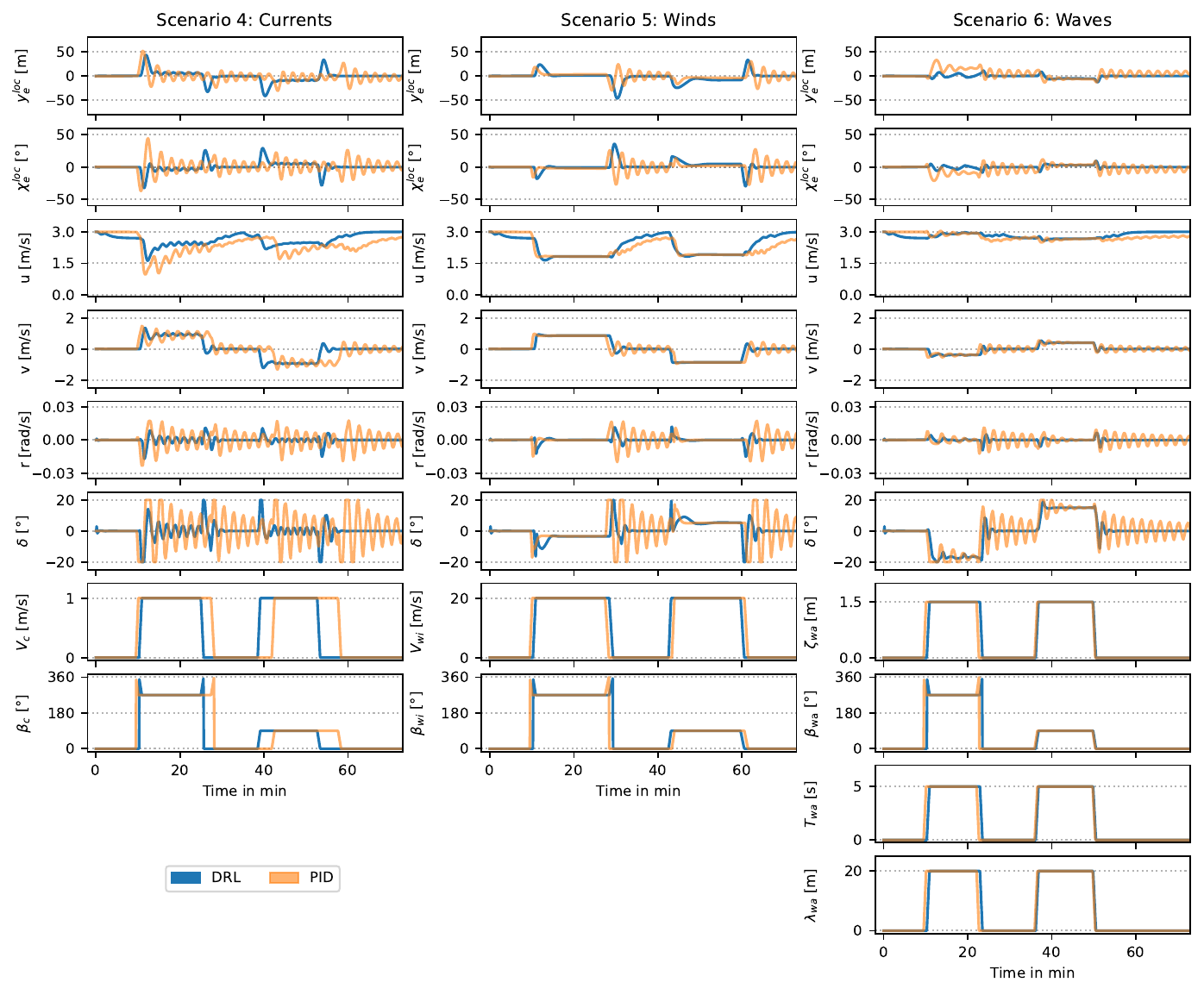}
    \caption{Validation results for \rdd{PF} under extreme environmental conditions.}
    \label{fig:follow_extreme}
\end{figure}

\rd{The results are presented in Figures \ref{fig:follow_moderate} and \ref{fig:follow_extreme}, where each column refers to a testing scenario for a separate environmental force, while the other two forces are set to zero. From top to bottom, the rows depict the local cross-track error, the local course error, the surge and sway velocities, the yaw rate, and the rudder angle, respectively. Afterward, the rows include the specific attributes for each force; for example, the speed and angle of attack of the currents. In addition, Table \ref{tab:PF_metrics} displays the CE and MCTE performance metrics.}

\begin{table}[htp]
\centering
\color{red(ncs)}
\begin{tabular}{ccccc}
\toprule[0.05em]
\multirow{2}{*}{Scenario} & \multicolumn{2}{c}{$\text{CE}_{\rm PF}$} & \multicolumn{2}{c}{$\text{MCTE}_{\rm PF}$}\\ \cmidrule[0.05em]{2-5}
& DRL & PID & DRL & PID\\ 
  \midrule
  1 & 0.032 & 0.137 & 0.056 & 0.202\\
  2 & 0.021 & 0.058 & 0.036 & 0.087\\
  3 & 0.041 & 0.078 & 0.067 & 0.105\\
  4 & 0.114 & 0.386 & 0.498 & 0.615\\
  5 & 0.165 & 0.311 & 0.460 & 0.488\\
  6 & 0.298 & 0.419 & 0.165 & 0.613\\
  \midrule
  Average & 0.112 & 0.232 & 0.214 & 0.352\\
  \bottomrule[0.05em]
\end{tabular}
\caption{\rd{Performance comparison of the DRL agent with the PID controller for the PF task. All metrics are dimensionless.}}
\label{tab:PF_metrics}
\end{table}

\rd{Generally, the DRL} agent demonstrates remarkable performance across all six scenarios, effectively adapting to force fields by promptly adjusting the rudder angle to steer back to the desired path. As a result, it successfully maintains minimal cross-track and course error. Notably, the inclusion of a comfort reward component has proven to be beneficial, as the agent generally exhibits smooth and moderate changes in the rudder angle. In contrast, the PID controller exhibits the typical undesired oscillation behavior, which persists even after applying the PSO technique. The challenging aspect lies in the heterogeneity of the scenarios, where the transition dynamics of the environment vary significantly under different force fields. This variability likely poses a difficulty for the PID controller in achieving stable and consistent performance. \rd{Following Table \ref{tab:PF_metrics}, the DRL agent strongly outperforms the PID controller both in terms of CE and MCTE. In particular, the CE of the DRL is, on average, approximately 48\% of the CE of the APF method, while the MCTE of the DRL method is simultaneously only circa 61\% of the MCTE of the APF approach.}

\subsection{Validation: Complete architecture}\label{subsec:val_complete_architecture}
\begin{figure}[!htb]
    \centering
    \includegraphics[width=\textwidth]{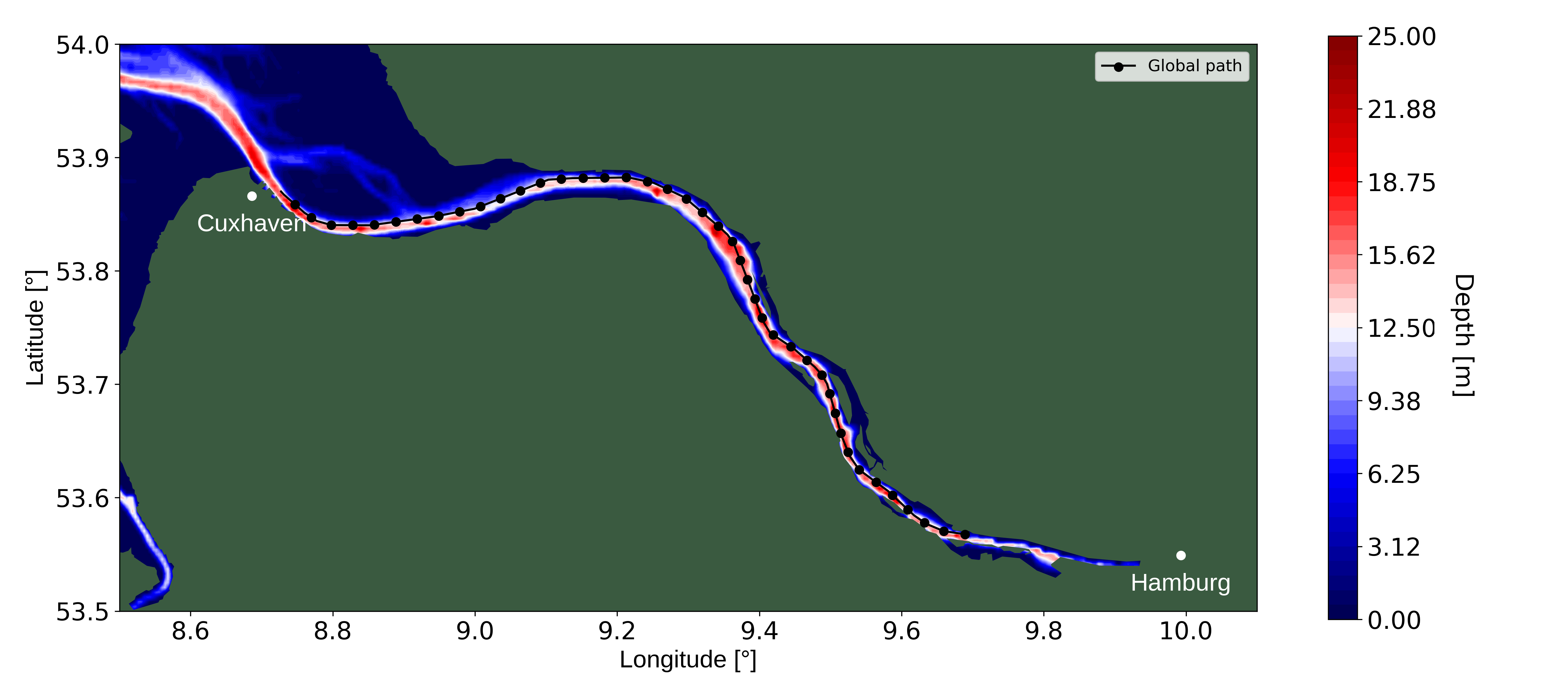}
    \caption{Global path for validation of the complete architecture based on AIS data.}
    \label{fig:AIS_global_path}
\end{figure}

After validating the LPP and PF units in separation, in this subsection, we aim to conduct a comprehensive evaluation of the complete architecture introduced in \rd{S}ection \ref{sec:architecture}. We deploy it in simulation on the lower part of the river Elbe in northern Germany. To ensure a realistic simulation, we specifically choose the date of January 29, 2022, and utilize the actual AIS vessel trajectories and environmental disturbances observed on that day. It is worth noting that the presence of the storm named \emph{Malik} in Middle Europe on that date adds an extra challenge to ASV. We manually specify a global path from Lighthouse Tinsdal to the Elbe estuary close to Cuxhaven, which is illustrated in Figure \ref{fig:AIS_global_path}. Further information regarding the data sources and the trajectory interpolation of the target ships are deferred to \ref{app:AIS_data_details}.

\begin{figure}[!htb]
    \centering
    \includegraphics[width=\textwidth]{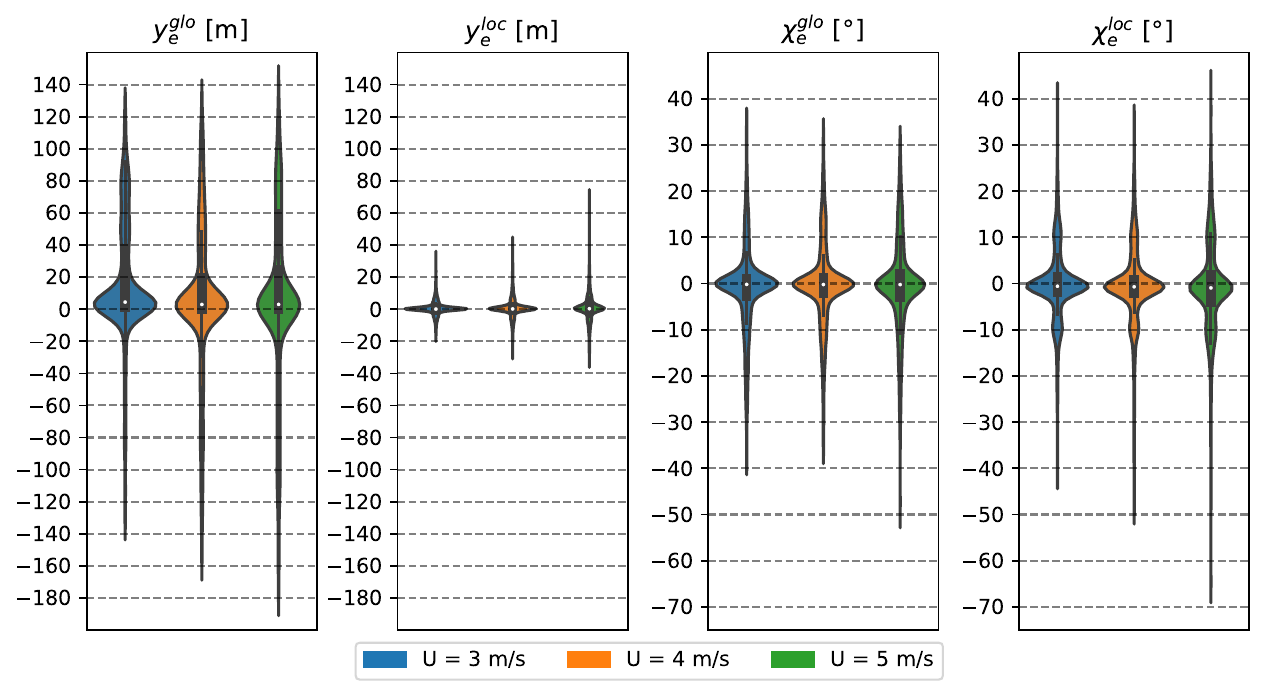}    \caption{Empirical distributions of the global and local cross-track and course errors over the complete journey from Lighthouse Tinsday to the Elbe estuary close to Cuxhaven.}
    \label{fig:pipeline_macro}
\end{figure}

\begin{figure}[!htb]
\centering
\subfigure[Head-On scenario at $U = 3 m/s$]{\includegraphics[width=6.5cm]{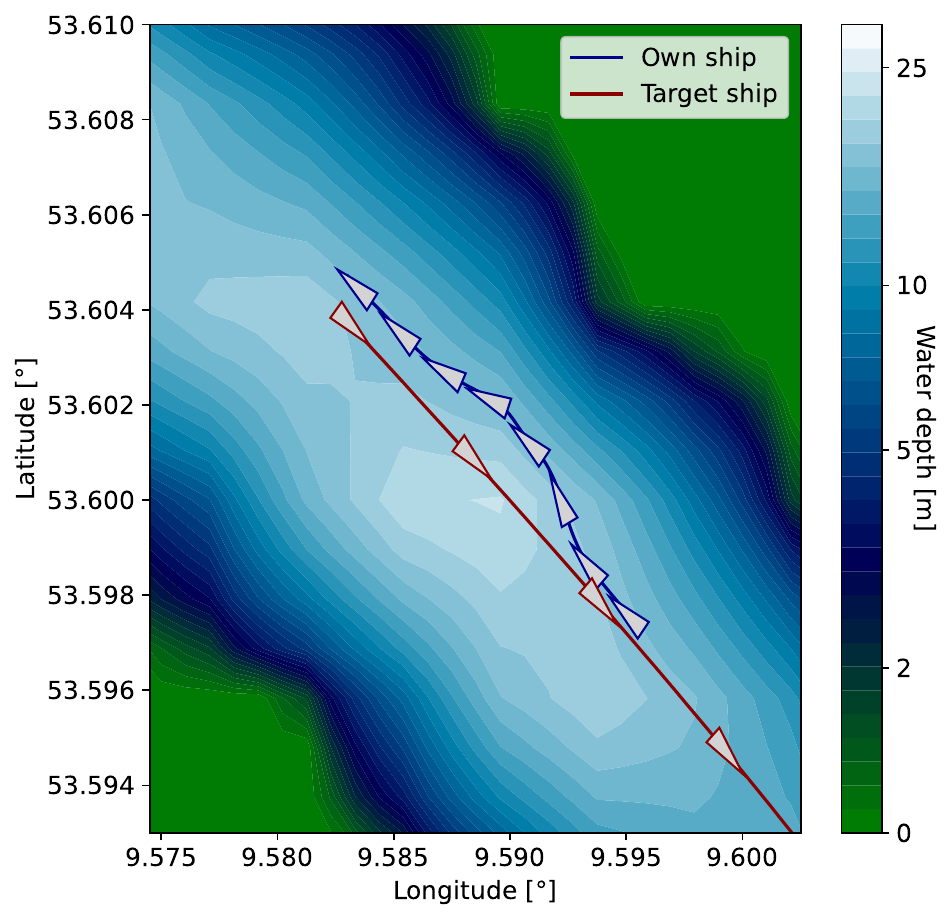}}
\hfill
\subfigure[Head-On scenario at $U = 5 m/s$]{\includegraphics[width=6.5cm]{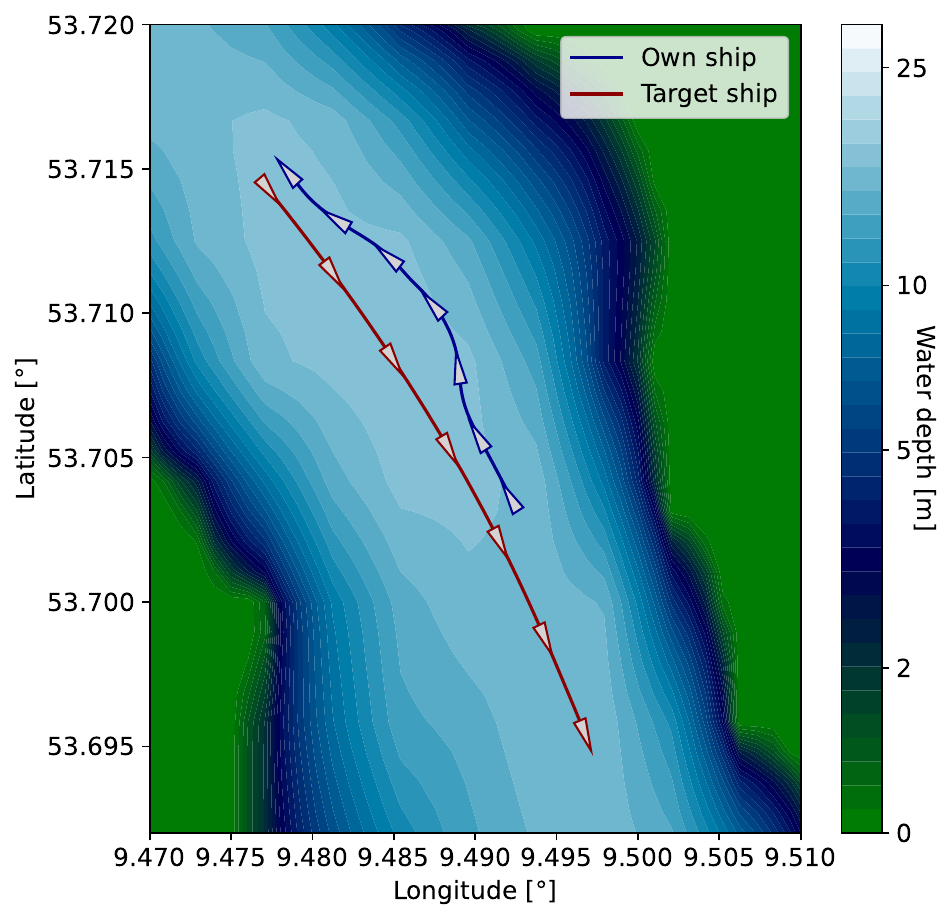}}
\hfill
\subfigure[Enabling overtaking at $U = 3 m/s$]{\includegraphics[width=6.5cm]{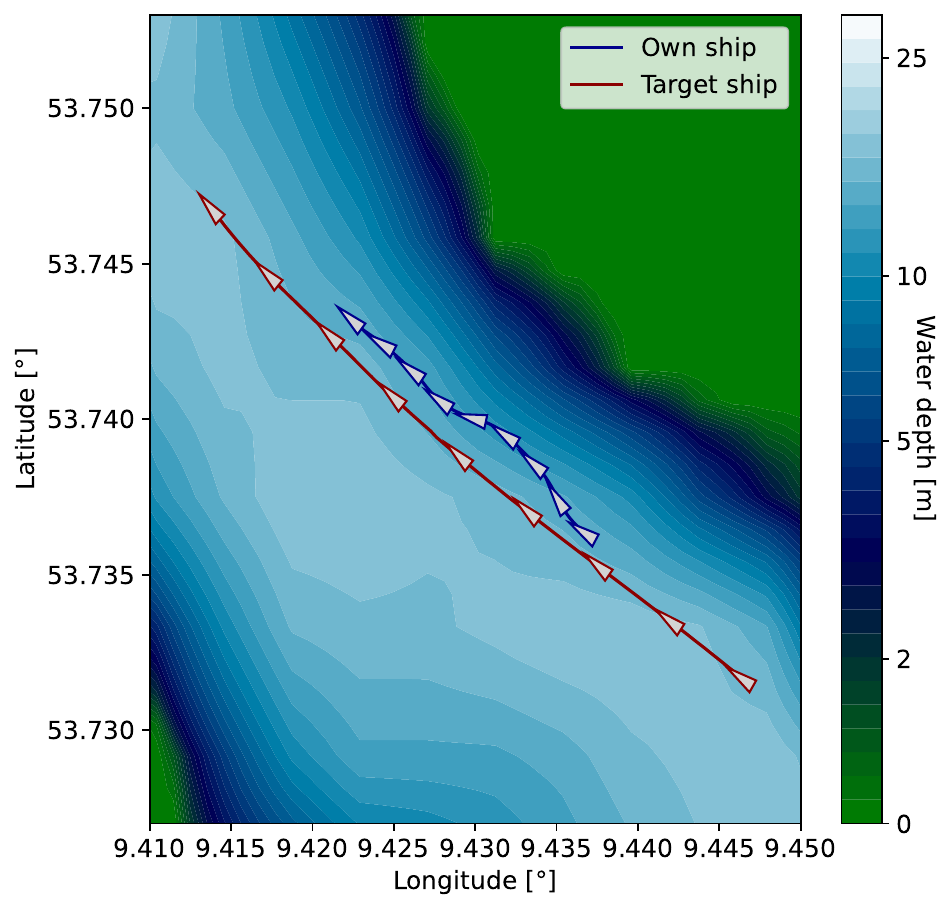}}
\hfill
\subfigure[Overtaking at $U = 4 m/s$]{\includegraphics[width=6.5cm]{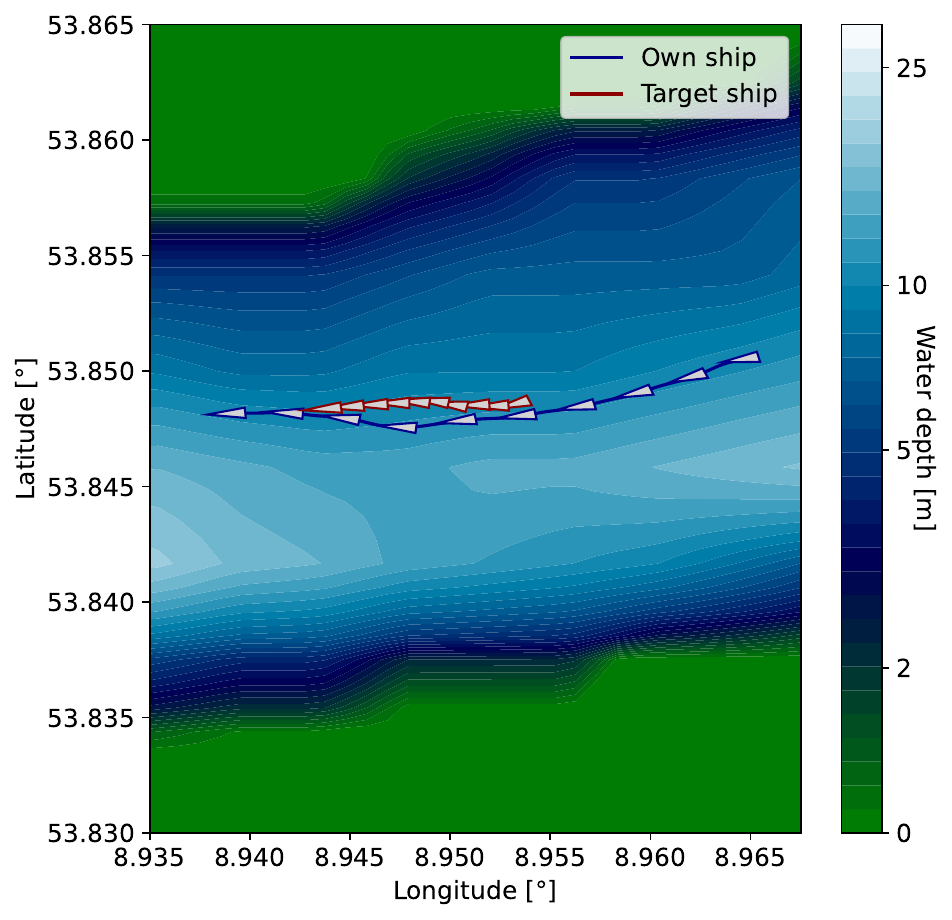}}
\hfill
\caption{Encounter scenarios based on real depth and AIS data using the complete architecture for ASV control.}
\label{fig:AIS_data_val}
\end{figure}

To test the architecture's generalization capabilities, we perform the same test for three different base speeds (in m/s) of the own ship: $U \in \{3, 4, 5\}$, even though both agents have only been trained on $U = \unit[3]{m/s}$. Figure \ref{fig:pipeline_macro} presents the angular and spatial deviations of the local and global paths, respectively, while selected COLAV maneuvers are displayed in Figure \ref{fig:AIS_data_val}.

Upon analyzing Figure \ref{fig:pipeline_macro}, we observe that the maximum spatial deviation from the global path reaches approximately 200 meters, while the vast majority of deviations are smaller than 40 meters in absolute value. These deviations are relatively low, considering the fairway width of the Elbe is typically between 600 and 2000 meters in the selected segment. The local cross-track error, which measures the deviation from the planned local path, is naturally of smaller magnitude and exceeds 20 meters only in a few selected cases. Additionally, we notice a slight increase in the spatial deviation from both the local and global paths as the speed increases. However, this increase remains moderate and demonstrates the successful generalization ability of both agents. Moreover, the angular deviations show resilience to changes in speed and, importantly, do not differ significantly between the global and local levels. This is likely attributed to the replanning of the local path, which presents the PF agent with a new path to react to.

Analyzing the maneuvers depicted in Figure \ref{fig:AIS_data_val}, it becomes evident that the framework possesses the capability to execute COLAV actions effectively. The model successfully performs starboard turns in head-on scenarios (Panels (a) and (b)), creates space for overtaking vessels (Panel (c)), and overtakes slower vessels on their portside (Panel (d)). These observations align with the validation results presented in \rd{S}ection \ref{subsec:val_planning}, further highlighting the architecture's suitability for real-world scenarios. Notably, the maneuvers prove successful for all three selected speeds.

While the architecture demonstrates strong performance in various validation scenarios, it is essential to note that collisions have occurred in rare cases during the AIS testing routine. However, these collisions arise due to the nature of the testing format, where the ASV is deployed into scenarios with static target ship trajectories. Consequently, the target ships remain unaware of the presence of the own ship, as illustrated in Figure \ref{fig:AIS_data_val}. Consider, for instance, Panel (c) of Figure \ref{fig:AIS_data_val}, where the own ship makes room for a faster target ship. If the target ship suddenly executes a starboard turn while being alongside the own ship, a collision will occur regardless of the own ship's reaction. As stated by \cite{lyu2019colregs}, it is impossible to avoid a collision if the target ship intentionally moves towards the own ship.

\subsection{\rd{Discussion and practical challenges}}\label{subsec:discussion_outlook}
\rd{The validation scenarios outlined in Sections \ref{subsec:val_planning}, \ref{subsec:val_PF}, and \ref{subsec:val_complete_architecture} have thoroughly assessed the effectiveness of the LPP and PF agents, both individually and in tandem, establishing a comprehensive evaluation of our autonomous system. Notably, the validation using AIS data in Section \ref{subsec:val_complete_architecture} delves into the behavior of our two-level system within the context of human vessel trajectories, demonstrating successful handling of various real-world encounter situations. However, we recognize the practical challenges associated with the interaction between autonomous systems and traditional, human-controlled vessels in \rdd{practical} operations \citep{kim2022safety}. For instance, an autonomous vessel needs to accurately estimate the intention of the conventional ship, while the conventional similarly needs to predict the future trajectory of the ASV \citep{porathe2019simplifying}.}

Recently, \cite{rodseth2023improving} presented several options for enhancing safety in encounters between conventional ships and ASVs, including the consideration of broadcasting intentions as a form of communication. We emphasize that such a broadcasting mechanism can be easily integrated into our architecture since the acquired information could be used during the planning iterations of the LPP agent. Generally, as emphasized by \cite{akdaug2022collaborative}, collaboration through communication between vessels is a critical component of \rdd{practical} operations\rdd{.}

\rdd{Finally, real-time applications can present unseen and unexpected uncertainties, such as abrupt alterations in weather patterns, unanticipated vessel maneuvers due to emergencies or navigational errors, or technical malfunctions in sensor equipment. Although our agents are trained in diverse scenarios and DRL offers strong generalization abilities, the sim-to-real transfer of autonomous systems should not be underestimated \citep{zhao2020sim}.}

\section{Conclusion}\label{sec:conclusion}
The use of \rdd{ASVs} for \rdd{IW} transportation shows promise in creating a sustainable and economically attractive transportation system. Our study introduces a two-level architecture for ASVs operating on \rdd{IWs} based on DRL, employing separate agents for \rdd{LPP} and \rdd{PF}. We consider relevant environmental disturbances, adhere to traffic rules, and validate our approach using simulated and real AIS trajectories.

We acknowledge the limitations of our work, which highlight avenues for future research. Firstly, we do not address limited visibility or sensor faults, which are critical high-risk scenarios in actual operations. Secondly, \rd{as emphasized in Section \ref{subsec:discussion_outlook}}, seafarers usually exchange communication signals when operating on \rdd{IWs}. Such information can be incorporated into our \rd{LPP} unit, thereby removing the simplifying assumption of linear target ship movement during the planning iterations. Lastly, our focus on simulation experiments calls for validating the proposed architecture in real-world waterways. Addressing these limitations will enable further advancements in developing ASVs for \rdd{IW} transportation, enhancing safety and effectiveness while paving the way for a sustainable transport system.

\section*{Acknowledgments}
The authors thank the members of the RL Dresden Group, especially Martin Treiber and Fabian Hart, for their constructive feedback on this work. Furthermore, the authors thank Thor Fossen for his insightful answers regarding the consideration of environmental forces in the simulation. The authors also acknowledge the Center for Information Services and High Performance Computing at TU Dresden for providing the resources for high-throughput calculations. Further, the authors thank the European Maritime Safety Agency for providing the AIS data for the validation scenarios. Finally, the authors express their gratitude to the Weiße Flotte Dresden for sharing their extensive experience in vessel operations. Their detailed explanations of the functionalities of the inland vessel 'Gräfin Kosel' have greatly enriched this research. Niklas Paulig was funded by BAW - Bundesanstalt für Wasserbau (Mikrosimulation des Schiffsverkehrs auf dem Niederrhein), Germany.

\bibliographystyle{apa}
\bibliography{bib}

\begin{thebibliography}{}

\bibitem[\protect\astroncite{Akda{\u{g}} et~al.}{2022}]{akdaug2022collaborative}
Akda{\u{g}}, M., Soln{\o}r, P., \& Johansen, T.~A. (2022).
\newblock Collaborative collision avoidance for maritime autonomous surface ships: A review.
\newblock {\em Ocean Engineering}, 250:110920.

\bibitem[\protect\astroncite{Al~Enezy et~al.}{2017}]{al2017developing}
Al~Enezy, O., van Hassel, E., Sys, C., \& Vanelslander, T. (2017).
\newblock Developing a cost calculation model for inland navigation.
\newblock {\em Research in Transportation Business \& Management}, 23:64--74.

\bibitem[\protect\astroncite{Amin and Hasegawa}{2010}]{amin2010generalised}
Amin, O.~M. \& Hasegawa, K. (2010).
\newblock Generalised mathematical model for ship manoeuvrability considering shallow water effect.
\newblock In {\em Conference Proceedings of the Japan Society of Naval Architects and Ocean Engineers}, volume~10, pages 531--534.

\bibitem[\protect\astroncite{Ankudinov et~al.}{1990}]{ankudinov1990manoeuvring}
Ankudinov, V., Miller, E., Jakobsen, B., \& Daggett, L. (1990).
\newblock Manoeuvring performance of tug/barge assemblies in restricted waterways.
\newblock In {\em International Conference on Marine Simulation and Ship Maneuverability}, volume~90, pages 515--525.

\bibitem[\protect\astroncite{Annamalai et~al.}{2015}]{annamalai2015robust}
Annamalai, A.~S., Sutton, R., Yang, C., Culverhouse, P., \& Sharma, S. (2015).
\newblock Robust adaptive control of an uninhabited surface vehicle.
\newblock {\em Journal of Intelligent \& Robotic Systems}, 78:319--338.

\bibitem[\protect\astroncite{Ba{\v{c}}kalov et~al.}{2023}]{bavckalov2023lessons}
Ba{\v{c}}kalov, I., Vidi{\'c}, M., \& Rudakovi{\'c}, S. (2023).
\newblock Lessons learned from accidents on some major european inland waterways.
\newblock {\em Ocean Engineering}, 273:113918.

\bibitem[\protect\astroncite{Bellemare et~al.}{2020}]{bellemare2020autonomous}
Bellemare, M.~G., Candido, S., Castro, P.~S., Gong, J., Machado, M.~C., Moitra, S., Ponda, S.~S., \& Wang, Z. (2020).
\newblock Autonomous navigation of stratospheric balloons using reinforcement learning.
\newblock {\em Nature}, 588(7836):77--82.

\bibitem[\protect\astroncite{Blindheim and Johansen}{2021}]{blindheim2021electronic}
Blindheim, S. \& Johansen, T.~A. (2021).
\newblock Electronic navigational charts for visualization, simulation, and autonomous ship control.
\newblock {\em IEEE Access}, 10:3716--3737.

\bibitem[\protect\astroncite{Breivik and Fossen}{2009}]{breivik2009guidance}
Breivik, M. \& Fossen, T.~I. (2009).
\newblock Guidance laws for autonomous underwater vehicles.
\newblock {\em Underwater vehicles}, 4:51--76.

\bibitem[\protect\astroncite{Breu and Fossen}{2011}]{breu2011}
Breu, D.~A. \& Fossen, T.~I. (2011).
\newblock L1 adaptive and extremum seeking control applied to roll parametric resonance in ships.
\newblock In {\em 2011 9th IEEE International Conference on Control and Automation (ICCA)}, pages 871--876. IEEE.

\bibitem[\protect\astroncite{{Bundesministerium für Digitales und Verkehr}}{1998}]{SeeSchStrO1998}
{Bundesministerium für Digitales und Verkehr} (1998).
\newblock {\em Seeschiffahrtsstraßen-Ordnung}.

\bibitem[\protect\astroncite{Cao et~al.}{2022}]{cao2022inland}
Cao, S., Fan, P., Yan, T., Xie, C., Deng, J., Xu, F., \& Shu, Y. (2022).
\newblock Inland waterway ship path planning based on improved {RRT} algorithm.
\newblock {\em Journal of Marine Science and Engineering}, 10(10):1460.

\bibitem[\protect\astroncite{Casta{\~n}eda et~al.}{2021}]{castaneda2021continuous}
Casta{\~n}eda, H., Rodriguez, J., \& Gordillo, J.~L. (2021).
\newblock Continuous and smooth differentiator based on adaptive sliding mode control for a quad-rotor mav.
\newblock {\em Asian Journal of Control}, 23(2):661--672.

\bibitem[\protect\astroncite{Chen et~al.}{2016}]{chen2016path}
Chen, L., Negenborn, R.~R., \& Lodewijks, G. (2016).
\newblock Path planning for autonomous inland vessels using {A* BG}.
\newblock In {\em International Conference on Computational Logistics}, pages 65--79. Springer.

\bibitem[\protect\astroncite{Cheng and Zhang}{2018}]{cheng2018concise}
Cheng, Y. \& Zhang, W. (2018).
\newblock Concise deep reinforcement learning obstacle avoidance for underactuated unmanned marine vessels.
\newblock {\em Neurocomputing}, 272:63--73.

\bibitem[\protect\astroncite{Chun et~al.}{2021}]{chun2021deep}
Chun, D.-H., Roh, M.-I., Lee, H.-W., Ha, J., \& Yu, D. (2021).
\newblock Deep reinforcement learning-based collision avoidance for an autonomous ship.
\newblock {\em Ocean Engineering}, 234:109216.

\bibitem[\protect\astroncite{de~Barros et~al.}{2022}]{de2022inland}
de~Barros, B. R.~C., de~Carvalho, E.~B., \& Junior, A. C. P.~B. (2022).
\newblock Inland waterway transport and the 2030 agenda: Taxonomy of sustainability issues.
\newblock {\em Cleaner Engineering and Technology}, page 100462.

\bibitem[\protect\astroncite{Ding et~al.}{2018}]{ding2018energy}
Ding, F., Zhang, Z., Fu, M., Wang, Y., \& Wang, C. (2018).
\newblock Energy-efficient path planning and control approach of {USV} based on particle swarm optimization.
\newblock In {\em OCEANS 2018 MTS/IEEE Charleston}, pages 1--6.

\bibitem[\protect\astroncite{Donha et~al.}{1998}]{donha1998}
Donha, D.~C., Desanj, D., Katebi, M., \& Grimble, M. (1998).
\newblock H$_\infty$ adaptive controllers for auto-pilot applications.
\newblock {\em International Journal of Adaptive Control and Signal Processing}, 12(8):623--648.

\bibitem[\protect\astroncite{Eberhart and Shi}{2000}]{eberhart2000comparing}
Eberhart, R.~C. \& Shi, Y. (2000).
\newblock Comparing inertia weights and constriction factors in particle swarm optimization.
\newblock In {\em Proceedings of the 2000 Congress on Evolutionary Computation}, volume~1, pages 84--88. IEEE.

\bibitem[\protect\astroncite{{E.U. Copernicus Marine Service}}{2023a}]{currentData}
{E.U. Copernicus Marine Service} (2023a).
\newblock Global ocean 1/12° physics analysis and forecast updated daily.
\newblock \url{https://doi.org/10.48670/moi-00016}.
\newblock Accessed: 2023-04-17.

\bibitem[\protect\astroncite{{E.U. Copernicus Marine Service}}{2023b}]{windData}
{E.U. Copernicus Marine Service} (2023b).
\newblock Global ocean daily gridded sea surface winds from scatterometer.
\newblock \url{https://doi.org/10.48670/moi-00182}.
\newblock Accessed: 2023-04-17.

\bibitem[\protect\astroncite{{E.U. Copernicus Marine Service}}{2023c}]{waveData}
{E.U. Copernicus Marine Service} (2023c).
\newblock Global ocean waves analysis and forecast.
\newblock \url{https://doi.org/10.48670/moi-00017}.
\newblock Accessed: 2023-04-17.

\bibitem[\protect\astroncite{Fan et~al.}{2022}]{fan2022novel}
Fan, Y., Sun, Z., \& Wang, G. (2022).
\newblock A novel reinforcement learning collision avoidance algorithm for {USVs} based on maneuvering characteristics and {COLREGs}.
\newblock {\em Sensors}, 22(6):2099.

\bibitem[\protect\astroncite{Fossen}{2021}]{fossen2021handbook}
Fossen, T.~I. (2021).
\newblock {\em Handbook of Marine Craft Hydrodynamics and Motion Control, 2nd Edition}.
\newblock John Wiley \& Sons.

\bibitem[\protect\astroncite{Fossen and Pettersen}{2014}]{fossen2014uniform}
Fossen, T.~I. \& Pettersen, K.~Y. (2014).
\newblock On uniform semiglobal exponential stability ({USGES}) of proportional line-of-sight guidance laws.
\newblock {\em Automatica}, 50(11):2912--2917.

\bibitem[\protect\astroncite{Fujimoto et~al.}{2018}]{fujimoto2018addressing}
Fujimoto, S., Hoof, H., \& Meger, D. (2018).
\newblock Addressing function approximation error in actor-critic methods.
\newblock In {\em International Conference on Machine Learning}, pages 1587--1596. PMLR.

\bibitem[\protect\astroncite{Gan et~al.}{2022}]{gan2022ship}
Gan, L., Yan, Z., Zhang, L., Liu, K., Zheng, Y., Zhou, C., \& Shu, Y. (2022).
\newblock Ship path planning based on safety potential field in inland rivers.
\newblock {\em Ocean Engineering}, 260:111928.

\bibitem[\protect\astroncite{Gonzalez-Garcia et~al.}{2020}]{gonzalez2020usv}
Gonzalez-Garcia, A., Casta{\~n}eda, H., \& Garrido, L. (2020).
\newblock Usv path-following control based on deep reinforcement learning and adaptive control.
\newblock In {\em Global Oceans 2020: Singapore--US Gulf Coast}, pages 1--7. IEEE.

\bibitem[\protect\astroncite{Goodfellow et~al.}{2016}]{goodfellow2016deep}
Goodfellow, I., Bengio, Y., \& Courville, A. (2016).
\newblock {\em Deep learning}.
\newblock MIT press.

\bibitem[\protect\astroncite{Guo et~al.}{2020}]{guo2020autonomous}
Guo, S., Zhang, X., Zheng, Y., \& Du, Y. (2020).
\newblock An autonomous path planning model for unmanned ships based on deep reinforcement learning.
\newblock {\em Sensors}, 20(2):426.

\bibitem[\protect\astroncite{Ha et~al.}{2021}]{ha2021quantitative}
Ha, J., Roh, M.-I., \& Lee, H.-W. (2021).
\newblock Quantitative calculation method of the collision risk for collision avoidance in ship navigation using the cpa and ship domain.
\newblock {\em Journal of Computational Design and Engineering}, 8(3):894--909.

\bibitem[\protect\astroncite{Han et~al.}{2022}]{han2022multi}
Han, Q., Yang, X., Song, H., \& Du, W. (2022).
\newblock Multi-objective ship path planning using non-dominant relationship-based {WOA} in marine meteorological environment.
\newblock {\em Ocean Engineering}, 266:112862.

\bibitem[\protect\astroncite{Hart and Okhrin}{2024}]{hart2022enhanced}
Hart, F. \& Okhrin, O. (2024).
\newblock Enhanced method for reinforcement learning based dynamic obstacle avoidance by assessment of collision risk.
\newblock {\em Neurocomputing}, 568:127097.

\bibitem[\protect\astroncite{Hart et~al.}{2023a}]{hart2023vessel}
Hart, F., Okhrin, O., \& Treiber, M. (2023a).
\newblock Vessel-following model for inland waterways based on deep reinforcement learning.
\newblock {\em Ocean Engineering}, 281:114679.

\bibitem[\protect\astroncite{Hart et~al.}{2023b}]{hart2023two}
Hart, F., Waltz, M., \& Okhrin, O. (2023b).
\newblock Two-step dynamic obstacle avoidance.
\newblock {\em arXiv preprint arXiv:2311.16841}.

\bibitem[\protect\astroncite{Hart et~al.}{1968}]{hart1968formal}
Hart, P.~E., Nilsson, N.~J., \& Raphael, B. (1968).
\newblock A formal basis for the heuristic determination of minimum cost paths.
\newblock {\em IEEE Transactions on Systems Science and Cybernetics}, 4(2):100--107.

\bibitem[\protect\astroncite{Heiberg et~al.}{2022}]{heiberg2022risk}
Heiberg, A., Larsen, T.~N., Meyer, E., Rasheed, A., San, O., \& Varagnolo, D. (2022).
\newblock Risk-based implementation of {COLREGs} for autonomous surface vehicles using deep reinforcement learning.
\newblock {\em Neural Networks}, 152:17--33.

\bibitem[\protect\astroncite{Hochreiter and Schmidhuber}{1997}]{hochreiter1997long}
Hochreiter, S. \& Schmidhuber, J. (1997).
\newblock Long short-term memory.
\newblock {\em Neural computation}, 9(8):1735--1780.

\bibitem[\protect\astroncite{Hofbauer and Putz}{2020}]{hofbauer2020external}
Hofbauer, F. \& Putz, L.-M. (2020).
\newblock External costs in inland waterway transport: An analysis of external cost categories and calculation methods.
\newblock {\em Sustainability}, 12(14):5874.

\bibitem[\protect\astroncite{Huang et~al.}{2020}]{huang2020ship}
Huang, Y., Chen, L., Chen, P., Negenborn, R.~R., \& Van~Gelder, P. (2020).
\newblock Ship collision avoidance methods: State-of-the-art.
\newblock {\em Safety science}, 121:451--473.

\bibitem[\protect\astroncite{Ibarz et~al.}{2021}]{ibarz2021train}
Ibarz, J., Tan, J., Finn, C., Kalakrishnan, M., Pastor, P., \& Levine, S. (2021).
\newblock How to train your robot with deep reinforcement learning: lessons we have learned.
\newblock {\em The International Journal of Robotics Research}, 40(4-5):698--721.

\bibitem[\protect\astroncite{{International Maritime Organization}}{1972}]{COLREGs1972}
{International Maritime Organization} (1972).
\newblock {\em {COLREG}: Convention on the International Regulations for Preventing Collisions at Sea}.

\bibitem[\protect\astroncite{{International Maritime Organization}}{2023}]{IMO_AIS}
{International Maritime Organization} (2023).
\newblock {AIS} transponders.
\newblock \url{https://www.imo.org/en/OurWork/Safety/Pages/AIS.aspx}.
\newblock Accessed June 2, 2023.

\bibitem[\protect\astroncite{{International Telecommunication Union}}{2014}]{series2014technical}
{International Telecommunication Union} (2014).
\newblock Technical characteristics for an automatic identification system using time-division multiple access in the vhf maritime mobile band.
\newblock {\em Recommendation ITU: Geneva, Switzerland}.

\bibitem[\protect\astroncite{Jadhav et~al.}{2023}]{jadhav2023collision}
Jadhav, A.~K., Pandi, A.~R., \& Somayajula, A. (2023).
\newblock Collision avoidance for autonomous surface vessels using novel artificial potential fields.
\newblock {\em Ocean Engineering}, 288:116011.

\bibitem[\protect\astroncite{Johansen et~al.}{2016}]{johansen2016ship}
Johansen, T.~A., Perez, T., \& Cristofaro, A. (2016).
\newblock Ship collision avoidance and {COLREGS} compliance using simulation-based control behavior selection with predictive hazard assessment.
\newblock {\em IEEE Transactions on Intelligent Transportation Systems}, 17(12):3407--3422.

\bibitem[\protect\astroncite{Kaelbling et~al.}{1998}]{kaelbling1998planning}
Kaelbling, L.~P., Littman, M.~L., \& Cassandra, A.~R. (1998).
\newblock Planning and acting in partially observable stochastic domains.
\newblock {\em Artificial Intelligence}, 101(1-2):99--134.

\bibitem[\protect\astroncite{Kang et~al.}{2019}]{kang2019generalization}
Kang, K., Belkhale, S., Kahn, G., Abbeel, P., \& Levine, S. (2019).
\newblock Generalization through simulation: Integrating simulated and real data into deep reinforcement learning for vision-based autonomous flight.
\newblock In {\em International Conference on Robotics and Automation}, pages 6008--6014. IEEE.

\bibitem[\protect\astroncite{Khatib}{1985}]{APF_original}
Khatib, O. (1985).
\newblock Real-time obstacle avoidance for manipulators and mobile robots.
\newblock In {\em Proceedings. 1985 IEEE International Conference on Robotics and Automation}, volume~2, pages 500--505.

\bibitem[\protect\astroncite{Kijima and Nakiri}{1990}]{kijima1990prediction}
Kijima, K. \& Nakiri, Y. (1990).
\newblock Prediction method of ship manoeuvrability in deep and shallow waters.
\newblock In {\em International Conference on Marine Simulation and Ship Maneuverability}, page 311.

\bibitem[\protect\astroncite{Kim et~al.}{2022}]{kim2022safety}
Kim, T.-e., Perera, L.~P., Sollid, M.-P., Batalden, B.-M., \& Sydnes, A.~K. (2022).
\newblock Safety challenges related to autonomous ships in mixed navigational environments.
\newblock {\em WMU Journal of Maritime Affairs}, 21(2):141--159.

\bibitem[\protect\astroncite{Kingma and Ba}{2014}]{kingma2014adam}
Kingma, D.~P. \& Ba, J. (2014).
\newblock Adam: A method for stochastic optimization.
\newblock {\em arXiv preprint arXiv:1412.6980}.

\bibitem[\protect\astroncite{Kitchat et~al.}{2024}]{kitchat2024deep}
Kitchat, K., Lin, M.-H., Chen, H.-S., Sun, M.-T., Sakai, K., Ku, W.-S., \& Surasak, T. (2024).
\newblock A deep reinforcement learning system for the allocation of epidemic prevention materials based on {DDPG}.
\newblock {\em Expert Systems with Applications}, 242:122763.

\bibitem[\protect\astroncite{Kuwata et~al.}{2013}]{kuwata2013safe}
Kuwata, Y., Wolf, M.~T., Zarzhitsky, D., \& Huntsberger, T.~L. (2013).
\newblock Safe maritime autonomous navigation with {COLREGS}, using velocity obstacles.
\newblock {\em IEEE Journal of Oceanic Engineering}, 39(1):110--119.

\bibitem[\protect\astroncite{Last et~al.}{2014}]{last2014comprehensive}
Last, P., Bahlke, C., Hering-Bertram, M., \& Linsen, L. (2014).
\newblock Comprehensive analysis of automatic identification system ({AIS}) data in regard to vessel movement prediction.
\newblock {\em The Journal of Navigation}, 67(5):791--809.

\bibitem[\protect\astroncite{Lefeber et~al.}{2003}]{lefeber2003tracking}
Lefeber, E., Pettersen, K.~Y., \& Nijmeijer, H. (2003).
\newblock Tracking control of an underactuated ship.
\newblock {\em IEEE Transactions on Control Systems Technology}, 11(1):52--61.

\bibitem[\protect\astroncite{Lenart}{1983}]{lenart1983collision}
Lenart, A.~S. (1983).
\newblock Collision threat parameters for a new radar display and plot technique.
\newblock {\em The Journal of Navigation}, 36(3):404--410.

\bibitem[\protect\astroncite{Li et~al.}{2021}]{li2021path}
Li, L., Wu, D., Huang, Y., \& Yuan, Z.-M. (2021).
\newblock A path planning strategy unified with a {COLREGs} collision avoidance function based on deep reinforcement learning and artificial potential field.
\newblock {\em Applied Ocean Research}, 113:102759.

\bibitem[\protect\astroncite{Lillicrap et~al.}{2015}]{lillicrap2015continuous}
Lillicrap, T.~P., Hunt, J.~J., Pritzel, A., Heess, N., Erez, T., Tassa, Y., Silver, D., \& Wierstra, D. (2015).
\newblock Continuous control with deep reinforcement learning.
\newblock {\em arXiv preprint arXiv:1509.02971}.

\bibitem[\protect\astroncite{Liu et~al.}{2023}]{liu2023colregs}
Liu, W., Qiu, K., Yang, X., Wang, R., Xiang, Z., Wang, Y., \& Xu, W. (2023).
\newblock {COLREGS}-based collision avoidance algorithm for unmanned surface vehicles using modified artificial potential fields.
\newblock {\em Physical Communication}, 57:101980.

\bibitem[\protect\astroncite{Liu et~al.}{2018}]{liu2018ship}
Liu, Y., Bu, R., \& Gao, X. (2018).
\newblock Ship trajectory tracking control system design based on sliding mode control algorithm.
\newblock {\em Polish Maritime Research}, 25(3):26--34.

\bibitem[\protect\astroncite{Liu and Bucknall}{2015}]{liu2015path}
Liu, Y. \& Bucknall, R. (2015).
\newblock Path planning algorithm for unmanned surface vehicle formations in a practical maritime environment.
\newblock {\em Ocean Engineering}, 97:126--144.

\bibitem[\protect\astroncite{Liu et~al.}{2016}]{liu2016unmanned}
Liu, Z., Zhang, Y., Yu, X., \& Yuan, C. (2016).
\newblock Unmanned surface vehicles: An overview of developments and challenges.
\newblock {\em Annual Reviews in Control}, 41:71--93.

\bibitem[\protect\astroncite{Lyu and Yin}{2018}]{lyu2018fast}
Lyu, H. \& Yin, Y. (2018).
\newblock Fast path planning for autonomous ships in restricted waters.
\newblock {\em Applied Sciences}, 8(12):2592.

\bibitem[\protect\astroncite{Lyu and Yin}{2019}]{lyu2019colregs}
Lyu, H. \& Yin, Y. (2019).
\newblock {COLREGS}-constrained real-time path planning for autonomous ships using modified artificial potential fields.
\newblock {\em The Journal of Navigation}, 72(3):588--608.

\bibitem[\protect\astroncite{Ma et~al.}{2021}]{ma2021intent}
Ma, J., Jia, C., Shu, Y., Liu, K., Zhang, Y., \& Hu, Y. (2021).
\newblock Intent prediction of vessels in intersection waterway based on learning vessel motion patterns with early observations.
\newblock {\em Ocean Engineering}, 232:109154.

\bibitem[\protect\astroncite{Martinsen and Lekkas}{2018}]{martinsen2018curved}
Martinsen, A.~B. \& Lekkas, A.~M. (2018).
\newblock Curved path following with deep reinforcement learning: Results from three vessel models.
\newblock In {\em OCEANS 2018 MTS/IEEE Charleston}, pages 1--8. IEEE.

\bibitem[\protect\astroncite{Matsuo et~al.}{2022}]{matsuo2022deep}
Matsuo, Y., LeCun, Y., Sahani, M., Precup, D., Silver, D., Sugiyama, M., Uchibe, E., \& Morimoto, J. (2022).
\newblock Deep learning, reinforcement learning, and world models.
\newblock {\em Neural Networks}.

\bibitem[\protect\astroncite{Meng et~al.}{2021}]{meng2021memory}
Meng, L., Gorbet, R., \& Kuli{\'c}, D. (2021).
\newblock Memory-based deep reinforcement learning for {POMDPS}.
\newblock In {\em International Conference on Intelligent Robots and Systems}, pages 5619--5626. IEEE.

\bibitem[\protect\astroncite{Meyer et~al.}{2020}]{meyer2020colreg}
Meyer, E., Heiberg, A., Rasheed, A., \& San, O. (2020).
\newblock {COLREG}-compliant collision avoidance for unmanned surface vehicle using deep reinforcement learning.
\newblock {\em IEEE Access}, 8:165344--165364.

\bibitem[\protect\astroncite{Mnih et~al.}{2015}]{mnih2015human}
Mnih, V., Kavukcuoglu, K., Silver, D., Rusu, A.~A., Veness, J., Bellemare, M.~G., Graves, A., Riedmiller, M., Fidjeland, A.~K., Ostrovski, G., et~al. (2015).
\newblock Human-level control through deep reinforcement learning.
\newblock {\em Nature}, 518(7540):529--533.

\bibitem[\protect\astroncite{Motwani et~al.}{2013}]{motwani2013interval}
Motwani, A., Sharma, S., Sutton, R., \& Culverhouse, P. (2013).
\newblock Interval kalman filtering in navigation system design for an uninhabited surface vehicle.
\newblock {\em The Journal of Navigation}, 66(5):639--652.

\bibitem[\protect\astroncite{Mou et~al.}{2010}]{mou2010study}
Mou, J.~M., Van Der~Tak, C., \& Ligteringen, H. (2010).
\newblock Study on collision avoidance in busy waterways by using {AIS} data.
\newblock {\em Ocean Engineering}, 37(5-6):483--490.

\bibitem[\protect\astroncite{Munim et~al.}{2020}]{munim2020big}
Munim, Z.~H., Dushenko, M., Jimenez, V.~J., Shakil, M.~H., \& Imset, M. (2020).
\newblock Big data and artificial intelligence in the maritime industry: a bibliometric review and future research directions.
\newblock {\em Maritime Policy \& Management}, 47(5):577--597.

\bibitem[\protect\astroncite{Negenborn et~al.}{2023}]{negenborn2023autonomous}
Negenborn, R.~R., Goerlandt, F., Johansen, T.~A., Slaets, P., Valdez~Banda, O.~A., Vanelslander, T., \& Ventikos, N.~P. (2023).
\newblock Autonomous ships are on the horizon: here’s what we need to know.
\newblock {\em Nature}, 615(7950):30--33.

\bibitem[\protect\astroncite{Nelson et~al.}{2007}]{nelson2006vector}
Nelson, D.~R., Barber, D.~B., McLain, T.~W., \& Beard, R.~W. (2007).
\newblock Vector field path following for miniature air vehicles.
\newblock {\em IEEE Transactions on Robotics}, 23(3):519--529.

\bibitem[\protect\astroncite{{\"O}zt{\"u}rk et~al.}{2022}]{ozturk2022review}
{\"O}zt{\"u}rk, {\"U}., Akda{\u{g}}, M., \& Ayabakan, T. (2022).
\newblock A review of path planning algorithms in maritime autonomous surface ships: Navigation safety perspective.
\newblock {\em Ocean Engineering}, 251:111010.

\bibitem[\protect\astroncite{{\"O}zt{\"u}rk and Cicek}{2019}]{ozturk2019individual}
{\"O}zt{\"u}rk, {\"U}. \& Cicek, K. (2019).
\newblock Individual collision risk assessment in ship navigation: A systematic literature review.
\newblock {\em Ocean Engineering}, 180:130--143.

\bibitem[\protect\astroncite{Paramesh and Rajendran}{2021}]{paramesh2021unified}
Paramesh, S. \& Rajendran, S. (2021).
\newblock A unified seakeeping and manoeuvring model with a {PID} controller for path following of a {KVLCC2} tanker in regular waves.
\newblock {\em Applied Ocean Research}, 116:102860.

\bibitem[\protect\astroncite{Paszke et~al.}{2019}]{paszke2019pytorch}
Paszke, A., Gross, S., Massa, F., Lerer, A., Bradbury, J., Chanan, G., Killeen, T., Lin, Z., Gimelshein, N., Antiga, L., Desmaison, A., Köpf, A., Yang, E., DeVito, Z., Raison, M., Tejani, A., Chilamkurthy, S., Steiner, B., Fang, L., Bai, J., \& Chintala, S. (2019).
\newblock Pytorch: An imperative style, high-performance deep learning library.
\newblock {\em Advances in Neural Information Processing Systems}, 32:8026--8037.

\bibitem[\protect\astroncite{Paulig}{2023}]{pytsa}
Paulig, N. (2023).
\newblock {pytsa: trajectories interpolation from {AIS} records}.
\newblock \url{https://github.com/nikpau/pytsa}.

\bibitem[\protect\astroncite{Paulig and Okhrin}{2024}]{paulig2023robust}
Paulig, N. \& Okhrin, O. (2024).
\newblock Robust path following on rivers using bootstrapped reinforcement learning.
\newblock {\em Ocean Engineering}, 298:117207.

\bibitem[\protect\astroncite{Peng et~al.}{2023}]{peng2023model}
Peng, Z., Liu, E., Pan, C., Wang, H., Wang, D., \& Liu, L. (2023).
\newblock Model-based deep reinforcement learning for data-driven motion control of an under-actuated unmanned surface vehicle: Path following and trajectory tracking.
\newblock {\em Journal of the Franklin Institute}, 360(6):4399--4426.

\bibitem[\protect\astroncite{Porathe and R{\o}dseth}{2019}]{porathe2019simplifying}
Porathe, T. \& R{\o}dseth, {\O}.~J. (2019).
\newblock Simplifying interactions between autonomous and conventional ships with e-navigation.
\newblock In {\em Journal of Physics: Conference Series}, volume 1357, page 012041. IOP Publishing.

\bibitem[\protect\astroncite{Puterman}{2014}]{puterman2014markov}
Puterman, M.~L. (2014).
\newblock {\em Markov decision processes: discrete stochastic dynamic programming}.
\newblock John Wiley \& Sons.

\bibitem[\protect\astroncite{Ringbom}{2019}]{ringbom2019regulating}
Ringbom, H. (2019).
\newblock Regulating autonomous ships—concepts, challenges and precedents.
\newblock {\em Ocean Development \& International Law}, 50(2-3):141--169.

\bibitem[\protect\astroncite{R{\o}dseth et~al.}{2023}]{rodseth2023improving}
R{\o}dseth, {\O}.~J., Wennersberg, L. A.~L., \& Nordahl, H. (2023).
\newblock Improving safety of interactions between conventional and autonomous ships.
\newblock {\em Ocean Engineering}, 284:115206.

\bibitem[\protect\astroncite{Roh{\'a}cs and Simongati}{2007}]{rohacs2007role}
Roh{\'a}cs, J. \& Simongati, G. (2007).
\newblock The role of inland waterway navigation in a sustainable transport system.
\newblock {\em Transport}, 22(3):148--153.

\bibitem[\protect\astroncite{Rong et~al.}{2022}]{rong2022ship}
Rong, H., Teixeira, A., \& Soares, C.~G. (2022).
\newblock Ship collision avoidance behaviour recognition and analysis based on {AIS} data.
\newblock {\em Ocean Engineering}, 245:110479.

\bibitem[\protect\astroncite{Sakamoto and Baba}{1986}]{sakamoto1986minimisation}
Sakamoto, T. \& Baba, E. (1986).
\newblock Minimisation of resistance of slowly moving full hull forms in short waves.
\newblock In {\em Proceedings of Sixteenth Symposium on Naval Hydrodynamics}, pages 598--613.

\bibitem[\protect\astroncite{Sawada et~al.}{2021}]{sawada2021automatic}
Sawada, R., Sato, K., \& Majima, T. (2021).
\newblock Automatic ship collision avoidance using deep reinforcement learning with {LSTM} in continuous action spaces.
\newblock {\em Journal of Marine Science and Technology}, 26(2):509--524.

\bibitem[\protect\astroncite{Schulman et~al.}{2017}]{schulman2017proximal}
Schulman, J., Wolski, F., Dhariwal, P., Radford, A., \& Klimov, O. (2017).
\newblock Proximal policy optimization algorithms.
\newblock {\em arXiv preprint arXiv:1707.06347}.

\bibitem[\protect\astroncite{Serigstad et~al.}{2018}]{serigstad2018hybrid}
Serigstad, E., Eriksen, B.-O.~H., \& Breivik, M. (2018).
\newblock Hybrid collision avoidance for autonomous surface vehicles.
\newblock {\em IFAC-PapersOnLine}, 51(29):1--7.

\bibitem[\protect\astroncite{Sharma et~al.}{2012}]{sharma2012autopilot}
Sharma, S., Naeem, W., \& Sutton, R. (2012).
\newblock An autopilot based on a local control network design for an unmanned surface vehicle.
\newblock {\em The Journal of Navigation}, 65(2):281--301.

\bibitem[\protect\astroncite{Siciliano et~al.}{2008}]{siciliano2008springer}
Siciliano, B., Khatib, O., \& Kr{\"o}ger, T. (2008).
\newblock {\em Springer handbook of robotics}, volume 200.
\newblock Springer.

\bibitem[\protect\astroncite{Siegwart et~al.}{2011}]{siegwart2011introduction}
Siegwart, R., Nourbakhsh, I.~R., \& Scaramuzza, D. (2011).
\newblock {\em Introduction to autonomous mobile robots}.
\newblock MIT press.

\bibitem[\protect\astroncite{Silver et~al.}{2018}]{silver2018general}
Silver, D., Hubert, T., Schrittwieser, J., Antonoglou, I., Lai, M., Guez, A., Lanctot, M., Sifre, L., Kumaran, D., Graepel, T., et~al. (2018).
\newblock A general reinforcement learning algorithm that masters chess, shogi, and go through self-play.
\newblock {\em Science}, 362(6419):1140--1144.

\bibitem[\protect\astroncite{Silver et~al.}{2014}]{silver2014deterministic}
Silver, D., Lever, G., Heess, N., Degris, T., Wierstra, D., \& Riedmiller, M. (2014).
\newblock Deterministic policy gradient algorithms.
\newblock In {\em International Conference on Machine Learning}, pages 387--395. Pmlr.

\bibitem[\protect\astroncite{Singh et~al.}{2018}]{singh2018constrained}
Singh, Y., Sharma, S., Sutton, R., Hatton, D., \& Khan, A. (2018).
\newblock A constrained {A}* approach towards optimal path planning for an unmanned surface vehicle in a maritime environment containing dynamic obstacles and ocean currents.
\newblock {\em Ocean Engineering}, 169:187--201.

\bibitem[\protect\astroncite{Soler et~al.}{2024}]{soler2024reinforcement}
Soler, D., Mari{\~n}o, O., Huergo, D., de~Frutos, M., \& Ferrer, E. (2024).
\newblock Reinforcement learning to maximize wind turbine energy generation.
\newblock {\em Expert Systems with Applications}, 249:123502.

\bibitem[\protect\astroncite{Sutton and Barto}{2018}]{sutton2018reinforcement}
Sutton, R.~S. \& Barto, A.~G. (2018).
\newblock {\em Reinforcement Learning: An Introduction}.
\newblock Cambridge: The MIT Press.

\bibitem[\protect\astroncite{Szepesv{\'a}ri}{2010}]{szepesvari2010algorithms}
Szepesv{\'a}ri, C. (2010).
\newblock Algorithms for reinforcement learning.
\newblock {\em Synthesis lectures on artificial intelligence and machine learning}, 4(1):1--103.

\bibitem[\protect\astroncite{Szlapczynski and Szlapczynska}{2017}]{szlapczynski2017review}
Szlapczynski, R. \& Szlapczynska, J. (2017).
\newblock Review of ship safety domains: Models and applications.
\newblock {\em Ocean Engineering}, 145:277--289.

\bibitem[\protect\astroncite{Taimuri et~al.}{2020}]{taimuri20206}
Taimuri, G., Matusiak, J., Mikkola, T., Kujala, P., \& Hirdaris, S. (2020).
\newblock A 6-{DoF} maneuvering model for the rapid estimation of hydrodynamic actions in deep and shallow waters.
\newblock {\em Ocean Engineering}, 218:108103.

\bibitem[\protect\astroncite{Tam and Bucknall}{2010}]{tam2010path}
Tam, C. \& Bucknall, R. (2010).
\newblock Path-planning algorithm for ships in close-range encounters.
\newblock {\em Journal of Marine Science and Technology}, 15:395--407.

\bibitem[\protect\astroncite{Tam et~al.}{2009}]{tam2009review}
Tam, C., Bucknall, R., \& Greig, A. (2009).
\newblock Review of collision avoidance and path planning methods for ships in close range encounters.
\newblock {\em The Journal of Navigation}, 62(3):455--476.

\bibitem[\protect\astroncite{Toyoda and Fujii}{1971}]{toyoda1971marine}
Toyoda, S. \& Fujii, Y. (1971).
\newblock Marine traffic engineering.
\newblock {\em The Journal of Navigation}, 24(1):24--34.

\bibitem[\protect\astroncite{Treiber and Kanagaraj}{2015}]{treiber2015comparing}
Treiber, M. \& Kanagaraj, V. (2015).
\newblock Comparing numerical integration schemes for time-continuous car-following models.
\newblock {\em Physica A: Statistical Mechanics and its Applications}, 419:183--195.

\bibitem[\protect\astroncite{Tu et~al.}{2017}]{tu2017exploiting}
Tu, E., Zhang, G., Rachmawati, L., Rajabally, E., \& Huang, G.-B. (2017).
\newblock Exploiting {AIS} data for intelligent maritime navigation: A comprehensive survey from data to methodology.
\newblock {\em IEEE Transactions on Intelligent Transportation Systems}, 19(5):1559--1582.

\bibitem[\protect\astroncite{Tutsoy et~al.}{2024}]{tutsoy2024minimum}
Tutsoy, O., Asadi, D., Ahmadi, K., Nabavi-Chashmi, S.~Y., \& Iqbal, J. (2024).
\newblock Minimum distance and minimum time optimal path planning with bioinspired machine learning algorithms for faulty unmanned air vehicles.
\newblock {\em IEEE Transactions on Intelligent Transportation Systems}.

\bibitem[\protect\astroncite{Vagale et~al.}{2021a}]{Vagale2021}
Vagale, A., Bye, R.~T., Oucheikh, R., Osen, O.~L., \& Fossen, T.~I. (2021a).
\newblock Path planning and collision avoidance for autonomous surface vehicles {II}: a comparative study of algorithms.
\newblock {\em Journal of Marine Science and Technology}, 26(4):1307--1323.

\bibitem[\protect\astroncite{Vagale et~al.}{2021b}]{vagale2021path}
Vagale, A., Oucheikh, R., Bye, R.~T., Osen, O.~L., \& Fossen, T.~I. (2021b).
\newblock Path planning and collision avoidance for autonomous surface vehicles {I}: a review.
\newblock {\em Journal of Marine Science and Technology}, 26:1292–1306.

\bibitem[\protect\astroncite{Van~Rossum and Drake}{2009}]{python3}
Van~Rossum, G. \& Drake, F.~L. (2009).
\newblock {\em Python 3 Reference Manual}.
\newblock CreateSpace, Scotts Valley, CA.

\bibitem[\protect\astroncite{Vanneste et~al.}{2022}]{vanneste2022safety}
Vanneste, A., Vanneste, S., Vasseur, O., Janssens, R., Billast, M., Anwar, A., Mets, K., De~Schepper, T., Mercelis, S., \& Hellinckx, P. (2022).
\newblock Safety aware autonomous path planning using model predictive reinforcement learning for inland waterways.
\newblock In {\em Annual Conference of the IEEE Industrial Electronics Society}, pages 1--6.

\bibitem[\protect\astroncite{Vinyals et~al.}{2019}]{vinyals2019grandmaster}
Vinyals, O., Babuschkin, I., Czarnecki, W.~M., Mathieu, M., Dudzik, A., Chung, J., Choi, D.~H., Powell, R., Ewalds, T., Georgiev, P., et~al. (2019).
\newblock Grandmaster level in starcraft {II} using multi-agent reinforcement learning.
\newblock {\em Nature}, 575(7782):350--354.

\bibitem[\protect\astroncite{Waltz and Okhrin}{2023}]{waltz2022spatial}
Waltz, M. \& Okhrin, O. (2023).
\newblock Spatial–temporal recurrent reinforcement learning for autonomous ships.
\newblock {\em Neural Networks}, 165:634--653.

\bibitem[\protect\astroncite{Waltz and Okhrin}{2024}]{waltz2022two}
Waltz, M. \& Okhrin, O. (2024).
\newblock Addressing maximization bias in reinforcement learning with two-sample testing.
\newblock {\em Artificial Intelligence}, 336:104204.

\bibitem[\protect\astroncite{Waltz and Paulig}{2022}]{TUDRL}
Waltz, M. \& Paulig, N. (2022).
\newblock {RL Dresden Algorithm Suite}.
\newblock \url{https://github.com/MarWaltz/TUD_RL}.

\bibitem[\protect\astroncite{Wan et~al.}{2020}]{wan2020improved}
Wan, L., Su, Y., Zhang, H., Shi, B., \& AbouOmar, M.~S. (2020).
\newblock An improved integral light-of-sight guidance law for path following of unmanned surface vehicles.
\newblock {\em Ocean Engineering}, 205:107302.

\bibitem[\protect\astroncite{Wang et~al.}{2019}]{wang2019obstacle}
Wang, P., Gao, S., Li, L., Sun, B., \& Cheng, S. (2019).
\newblock Obstacle avoidance path planning design for autonomous driving vehicles based on an improved artificial potential field algorithm.
\newblock {\em Energies}, 12(12):2342.

\bibitem[\protect\astroncite{Wang et~al.}{2017}]{wang2017ship}
Wang, X., Liu, Z., \& Cai, Y. (2017).
\newblock The ship maneuverability based collision avoidance dynamic support system in close-quarters situation.
\newblock {\em Ocean Engineering}, 146:486--497.

\bibitem[\protect\astroncite{Wang et~al.}{2023}]{wang2023path}
Wang, Y., Cao, J., Sun, J., Zou, X., \& Sun, C. (2023).
\newblock Path following control for unmanned surface vehicles: A reinforcement learning-based method with experimental validation.
\newblock {\em IEEE Transactions on Neural Networks and Learning Systems}.

\bibitem[\protect\astroncite{Woo et~al.}{2019}]{woo2019deep}
Woo, J., Yu, C., \& Kim, N. (2019).
\newblock Deep reinforcement learning-based controller for path following of an unmanned surface vehicle.
\newblock {\em Ocean Engineering}, 183:155--166.

\bibitem[\protect\astroncite{Xu and Guedes~Soares}{2023}]{xu2023review}
Xu, H. \& Guedes~Soares, C. (2023).
\newblock Review of path-following control systems for maritime autonomous surface ships.
\newblock {\em Journal of Marine Science and Application}, 22(2):153--171.

\bibitem[\protect\astroncite{Xu et~al.}{2022a}]{xu2022path}
Xu, X., Cai, P., Ahmed, Z., Yellapu, V.~S., \& Zhang, W. (2022a).
\newblock Path planning and dynamic collision avoidance algorithm under {COLREGs} via deep reinforcement learning.
\newblock {\em Neurocomputing}, 468:181--197.

\bibitem[\protect\astroncite{Xu et~al.}{2022b}]{xu2022colregs}
Xu, X., Lu, Y., Liu, G., Cai, P., \& Zhang, W. (2022b).
\newblock {COLREGs}-abiding hybrid collision avoidance algorithm based on deep reinforcement learning for {USVs}.
\newblock {\em Ocean Engineering}, 247:110749.

\bibitem[\protect\astroncite{Yasukawa and Yoshimura}{2015}]{yasukawa2015introduction}
Yasukawa, H. \& Yoshimura, Y. (2015).
\newblock Introduction of {MMG} standard method for ship maneuvering predictions.
\newblock {\em Journal of Marine Science and Technology}, 20(1):37--52.

\bibitem[\protect\astroncite{Yu et~al.}{2023}]{yu2023review}
Yu, H., Meng, Q., Fang, Z., Liu, J., \& Xu, L. (2023).
\newblock A review of ship collision risk assessment, hotspot detection and path planning for maritime traffic control in restricted waters.
\newblock {\em The Journal of Navigation}, pages 1--27.

\bibitem[\protect\astroncite{Zhai et~al.}{2022}]{zhai2022intelligent}
Zhai, P., Zhang, Y., \& Shaobo, W. (2022).
\newblock Intelligent ship collision avoidance algorithm based on {DDQN} with prioritized experience replay under {COLREGs}.
\newblock {\em Journal of Marine Science and Engineering}, 10(5):585.

\bibitem[\protect\astroncite{Zhang et~al.}{2017}]{zhang2017improved}
Zhang, X., Yang, G., Zhang, Q., Zhang, G., \& Zhang, Y. (2017).
\newblock Improved concise backstepping control of course keeping for ships using nonlinear feedback technique.
\newblock {\em The Journal of Navigation}, 70(6):1401--1414.

\bibitem[\protect\astroncite{Zhang et~al.}{2023}]{zhang2023path}
Zhang, Y., Chen, P., Chen, L., \& Mou, J. (2023).
\newblock A path planning method for the autonomous ship in restricted bridge area based on anisotropic fast marching algorithm.
\newblock {\em Ocean Engineering}, 269:113546.

\bibitem[\protect\astroncite{Zhang et~al.}{2019}]{zhang2019path}
Zhang, Z., Wu, D., Gu, J., \& Li, F. (2019).
\newblock A path-planning strategy for unmanned surface vehicles based on an adaptive hybrid dynamic stepsize and target attractive force-{RRT} algorithm.
\newblock {\em Journal of Marine Science and Engineering}, 7(5):132.

\bibitem[\protect\astroncite{Zhao et~al.}{2020a}]{zhao2020sim}
Zhao, W., Queralta, J.~P., \& Westerlund, T. (2020a).
\newblock Sim-to-real transfer in deep reinforcement learning for robotics: a survey.
\newblock In {\em IEEE Symposium Series on Computational Intelligence}, pages 737--744. IEEE.

\bibitem[\protect\astroncite{Zhao et~al.}{2021}]{zhao2021usv}
Zhao, Y., Ma, Y., \& Hu, S. (2021).
\newblock Usv formation and path-following control via deep reinforcement learning with random braking.
\newblock {\em IEEE Transactions on Neural Networks and Learning Systems}, 32(12):5468--5478.

\bibitem[\protect\astroncite{Zhao et~al.}{2020b}]{zhao2020path}
Zhao, Y., Qi, X., Ma, Y., Li, Z., Malekian, R., \& Sotelo, M.~A. (2020b).
\newblock Path following optimization for an underactuated usv using smoothly-convergent deep reinforcement learning.
\newblock {\em IEEE Transactions on Intelligent Transportation Systems}, 22(10):6208--6220.

\bibitem[\protect\astroncite{Zhou et~al.}{2019}]{zhou2019review}
Zhou, Y., Daamen, W., Vellinga, T., \& Hoogendoorn, S. (2019).
\newblock Review of maritime traffic models from vessel behavior modeling perspective.
\newblock {\em Transportation Research Part C: Emerging Technologies}, 105:323--345.

\bibitem[\protect\astroncite{Zhuge et~al.}{2023}]{ZHUGE2023}
Zhuge, D., Wang, S., Zhen, L., \& Psaraftis, H.~N. (2023).
\newblock Data-driven modeling of maritime transportation: Key issues, challenges, and solutions.
\newblock {\em Engineering}.
\newblock DOI: 10.1016/j.eng.2022.12.009.

\end{thebibliography}

\newpage
\appendix

\setcounter{table}{0}
\gdef\thesection{Appendix \Alph{section}}
\section{German \rdd{vessel} traffic rules}\label{app:traffic_rules}
The following paragraph is an excerpt from the German regulation for shipping lanes \citep{SeeSchStrO1998}, which we consider in our research. We first state the original legal text in German, followed by the official English translation.\\

\noindent\textbf{\S 23 Überholen}\\
\textit{(1) Grundsätzlich muß links überholt werden. Soweit die Umstände des Falles es erfordern, darf rechts überholt
werden.\\
(2) Das überholende Fahrzeug muß unter Beachtung von Regel 9 Buchstabe e und Regel 13 der Kollisionsverhütungsregeln die Fahrt so weit herabsetzen oder einen solchen seitlichen Abstand vom vorausfahrenden Fahrzeug einhalten, daß kein gefährlicher Sog entstehen kann und während des ganzen
Überholmanövers jede Gefährdung des Gegenverkehrs ausgeschlossen ist. Das vorausfahrende Fahrzeug muß das Überholen soweit wie möglich erleichtern.}\\

\noindent\textbf{\S 23 Overtaking}\\
\textit{(1) As a rule, an overtaking vessel shall pass the vessel being overtaken on the latter vessel’s port side. If the circumstances of the case so require, the overtaking vessel may pass the vessel being overtaken on the latter vessel’s starboard side.\\
(2) The overtaking vessel, acting in compliance with the provisions of Rule 9(e) and Rule 13 of the International Regulations for Preventing Collisions at Sea, 1972, as amended, shall slacken her speed so much, respectively, shall give the vessel being overtaken such a wide berth that no dangerous suction or wash can develop and that no vessel proceeding in the opposite direction will be put at any risk for the entire duration of the overtaking process. The vessel being overtaken shall facilitate the overtaking vessel’s action to the greatest possible extent.}\\

\setcounter{table}{0}
\setcounter{figure}{0}
\gdef\thesection{Appendix \Alph{section}}
\section{\rdd{Nomenclature}}
\label{app:Nomenclature}

\begin{center}
\begin{scriptsize}
\resizebox{\textwidth}{!}{
\begin{tabular}{|llll|}
    \hline
AIS  &  Automatic identification system  & $\delta$ &  Rudder angle \\ 
APF  &  Artificial potential field & $\delta_{max}$ &  Maximum rudder angle \\ 
ASV  &  Autonomous surface vehicle  & $\Delta_\psi$ &  Maximum heading change between time steps \\ 
CE  &  Controller effort  & $\Dot{\Tilde{r}}_{scale}$ & Normalization constant \\ 
COLAV  &  Collision avoidance & $\mathcal{DU}$ &  Discrete uniform distribution \\ 
COLREGs  &  The International Regulations for  & $e_{norm}$ &  Normalization constant \\ 
  & Preventing Collisions at Sea & $Exp$ &  Exponential distribution \\ 
CPA  &  Closest point of approach  & $F_{att,t}$ &  APF attractive force \\ 
DCPA  &  Distance to CPA  & $F_{e,t}$ &  East component of APF total forces \\ 
DQN  &  Deep $Q$-Network & $F_{n,t}$ &  North component of APF total forces \\ 
DRL  &  Deep reinforcement learning & $F_{rep,t}$ &  APF repulsive force \\ 
FC  &  Fully connected  & $F_t$ &  APF resulting total force \\ 
GPP  &  Global path planning  & $f_{j,l}(\cdot,\theta_{f_{j,l}})$ & Spatial recurrent network part for the  \\ 
IW  &  Inland waterway  &   & critic with parameter set $\theta_{f_{j,l}}$ \\ 
LPP  &  Local path planning  & $f_{\mu,l}(\cdot,\theta_{f_{\mu,l}})$ & Spatial recurrent network part for the  \\ 
LSTM  &  Long short-term memory  &   & actor with parameter set $\theta_{f_{\mu,l}}$ \\ 
MCTE  &  Mean cross-track error  & $\gamma$ &  Discount factor \\ 
MDP  &  Markov decision process  & $\gamma_i$ &  $i$-th lidar beam \\ 
PF  &  Path following  & $g_{j}(\cdot,\theta_{g_j})$ &  Temporal recurrent network part for the  \\ 
PID  &  Proportional-integral-derivative  &   & critic with parameter set theta $g_{j}$ \\ 
PSO  &  Particle swarm optimization  & $g_{\mu}(\cdot,\theta_{g_\mu})$ &  Temporal recurrent network part for the  \\ 
RL  &  Reinforcement learning  &   & actor with parameter set theta $g_{mu}$ \\ 
RRT  &  Rapidly exploring random tree  & $\eta$ & $=(x_n,y_n,\psi)^\top$ \\ 
TCPA  &  Time to CPA  & $h$ &  History length \\ 
VFG  &  Vector field guidance & $H$ &  Water depth \\ 
  &    & $H_{norm}$ & Normalization constant \\ 
$a$  &  Action & $I_{zG}$  &  Moment of inertia \\ 
$\mathcal{A}$ &  Action space & $J_z$  &  Added moment of inertia \\ 
$a_{\rm lat}$ &  Length of the minor axis of APF  & $k_{\rm coll}$ &  Collision reward weight \\ 
  & ellipse & $k_{\rm rule}$ &  Rule adherence reward weight \\ 
$a_{\rm lon}$ &  Length of the major axis of APF  & $k_{\rm turn}$ &  Maximum course error penalty \\ 
  & ellipse & $k^{\rm LPP},k^{\rm PF}$ &  VFG gain parameters \\ 
$a_c^{\rm LPP}, a_c^{\rm PF}$ & Action multiplier & $k^{\rm LPP}_{\chi_e}$ &  Power weight for  \\ 
$a^{\rm LPP}_t$ &  LPP action &   & course error reward for LPP\\ 
$a^{\rm PF}_t$ &  PF action & $k^{\rm LPP}_{y_e}$ &  Power weights for  \\ 
$\alpha^i_{\text{OS},t}$ & Relative bearing of ship $i$ from the  &   & cross-track error reward for LPP\\ 
  & perspective of the own ship & $k^{\rm PF}_{\chi_e}$ &  Power weight for  \\ 
$\alpha^{\rm OS}_{i,t}$ & Relative bearing of the own ship from  &   & course error reward for PF \\ 
  & the perspective of ship $i$ & $k^{\rm PF}_{y_e}$ &  Power weight for  \\ 
$\alpha_{\text{OS},t}^{i,\rm cpa}$ &  Relative bearing of ship $i$ at the CPA &   & cross-track error reward for PF\\ 
$\beta_{c}$ &  Current attack angle & $k_a, k_r$ &  APF force weights \\ 
$\beta_{wa}$ &  Wave attack angle & $K_p,K_i,K_d$ &  PID gains \\ 
$\beta_{wi}$ &  Wind attack angle & $\lambda_{wa}$ &  Wave length \\ 
$B$ &  Ship breadth & $\lambda_{wa,\rm norm}$ & Normalization constant \\ 
$\chi$  &  Course angle & $L_{pp}$ &  Ship length between perpendiculars \\ 
$\chi_d$ &  Desired course & $m$ &  Mass of ASV \\ 
$\chi_e$ &  Course error & $m_{x_b}$  &  Added mass in $x_b$ direction \\ 
$\chi^\infty$ & VFG parameter & $m_{y_b}$  &  Added mass in $y_b$ direction \\ 
$\chi_{P_k}$ &  Path segment angle & $\mu$ &  Actor network \\ 
$d_{\rm norm}$ & Scaling constant & $\mu(\cdot,\theta^{\rm LPP}_\mu)$ &  Actor network with combined  \\ 
$d^*,d_0,d_{l}$ & Distance constants &   & parameter set $\theta^{\rm LPP}_{\mu}$ \\ 
$d_{i,t}^{\rm cpa}$ &  DCPA to target ship $i$& $\{n\},\{b\}$ &  Earth and body frame \\ 
$d^i_{\text{OS},t}$ &  Distance from agent to ship $i$ & $N_H$ &  Yaw moment force on ship hull \\ 
$d^{\rm G}_{\text{OS},t}$ & Distance from own ship to goal point & $N_R$ &  Yaw moment force on ship rudder \\ 
$d_{\rm scale}$ &  Maximum considered distance from own to  & $N_m$  &  Yaw moment around midship \\ 
  & target ships & $N_{WA}$ &  Wave yaw forces \\ 
$d^{\rm cpa}_{i,t}$ &  DCPA to the $i$-th target ship & $N_{WI}$ &  Wind yaw forces \\ 
    \hline
\end{tabular}}
\newpage
\resizebox{\textwidth}{!}{
\begin{tabular}{|llll|}
    \hline
$\nu$ & $=(u,v,\Tilde{r})^\top$ & $t_q$ &  Query time for AIS data extraction \\ 
$n_{\rm norm}$ &  Longitudinal ship domain  & $t_{\rm replan}$  &  LPP replan interval \\ 
  & reward normalization constant & $T_{wa}$ &  Wave period \\ 
$\vec{n}_{\rm OG}$ &  APF normal vector from  & $T_{wa,\rm norm}$ &  Normalization constant \\ 
  & agent in direction of goal (OG) & $\theta^{\rm LPP}_{\mu}$ &  Actor network parameter set \\ 
$\vec{n}_{\rm OT}$ &  APF normal vector from  & $\theta^{\rm LPP}_j$ &  Critic network parameter set \\ 
  & agent in direction of target ship (OT) & $\theta_{f_{j,l}}$ &  Parameter set for spatial  \\ 
$\vec{n}_{\rm OT}$ &  APF normal vector from  &   & recurrent network part for the critic \\ 
  & target ship in direction of agent & $\theta_{f_{\mu,l}}$ &  Parameter set for spatial  \\ 
$\vec{n}_{\text{OT}\perp}$ &  APF unit vector perpendicular to portside unit  &   & recurrent network part for the actor \\ 
  & vector from the own ship to the target ship & $\theta_{g_j}$ &  Parameter set for temporal  \\ 
$o$  &  Observation &   & recurrent network part for the critic \\ 
$\mathcal{O}$ &  Observation space & $\theta_{g_\mu}$ &  Parameter set for temporal  \\ 
& &   & recurrent network part for the actor \\ 
$o^{LPP}_{\text{IW},t}$ &  Navigational area information & $u$  &  Surge \\ 
$o^{LPP}_{\text{OS},t}$ &  Own ship information & $\mathcal{U}$ &  Uniform distribution \\ 
$o^{LPP}_{\text{TS},t}$ &  Target ship information & $u_c$ &  Longitudinal component of current velocity \\ 
$o^{\rm LPP}_t$ &  Agent observation & $u_{\rm scale}$ &  Normalization constant \\ 
$o^{PF}_{\text{Env},t}$ &  Environmental observation & $U$ &  Vessel speed ($\sqrt{u^2+v^2}$) \\ 
$o^{\rm PF}_{\text{OS},t}$ &  Own ship observation & $U_{\rm base}$ &  Validation base speed of the agent \\ 
$o^{\rm PF}_{t}$ &  PF observation & $U_{i,t}$ &  Speed of target ship $i$ \\ 
$o_{\text{TS},i,t}$ &  Observation for target ship $i$ & $U_{\text{OS},t}$ & Own ship speed \\ 
$\mathcal{P}$ &  State transition probability function & $U_{\rm scale}$ &  Speed scaling constant \\ 
$P_k$ &  $k$-th waypoint of a path & $v$  &  Sway \\ 
$\psi$ &  Heading angle & $v_c$ &  Lateral component of current velocity \\ 
$\psi_{d,t}$ &  APF desired heading & $v_{\rm scale}$ &  Normalization constant \\ 
$\psi_{\text{OS},t}$ &  Own ship heading & $V_c$ &  Current velocity \\ 
$\pi$ &  Policy function & $V_{c,\rm norm}$ &  Normalization constant \\ 
$Q_j(\cdot,\cdot,\theta^{\rm LPP}_j)$ &  Critic network with combined  & $V_{wi}$ &  Wind speed \\ 
  & parameter set $\theta^{\rm LPP}_j$ & $V_{wi,\rm norm}$ &  Normalization constant \\ 
$Q^\pi, Q_1, Q_2$ &  Action-value function & $\omega^{\rm LPP}_{\chi_e},\omega^{\rm LPP}_{coll},\omega^{\rm LPP}_{\rm comf}$ &  Reward component weight \\ 
$\mathcal{R}$ &  Reward function & $\omega^{\rm LPP}_{y_e},\omega^{\rm LPP}_{\rm rule},\omega^{\rm LPP}_{\rm comf}$ &  Reward component weight \\ 
$\Tilde{r}$  &  yaw rate & $\omega^{\rm PF}_{\chi_e},\omega^{PF}_{y_e},\omega^{\rm PF}_{\rm comf}$ &  Reward component weight \\ 
$\Tilde{r}_{\rm scale}$  &  Normalization constant & $X$ &  Surge force \\ 
$\Dot{\Tilde{r}}$  &  First derivative of yaw rate & $X_H$ &  Forces on ship hull in surge direction \\ 
$\Dot{\Tilde{r}}_{\rm scale}$  &  Normalization constant & $X_P$ &  Forces on ship propeller in surge direction \\ 
$r^{\rm LPP}_{\chi_e,t}$ &  Course error reward & $X_R$ &  Forces on ship rudder in surge direction \\ 
$r^{\rm LPP}_{\text{coll},t}$ &  Collision penalty & $X_{WA}$ &  Wave forces on ship in surge direction \\ 
$r^{\rm LPP}_{\text{comf},t}$ &  LPP comfort reward & $X_{WI}$ &  Wind forces in surge direction \\ 
$r^{\rm LPP}_{\text{rule},t}$ &  Rule adherence reward & $x_b$ &  Ship-fixed longitudinal axis \\ 
$r^{\rm LPP}_{y_e,t}$ &  Cross-track error reward & $x_e$ &  Along-track error \\ 
$r^{\rm PF}_{\chi_e,t}$ &  Course error reward & $x_G$ &  Longitudinal coordinate of  \\ 
$r^{\rm PF}_{\text{comf},t}$ &  Comfort reward &   & ASV's center of gravity \\ 
$r^{\rm PF}_{y_e,t}$ &  Cross-track error reward & $x_n$ &  Northing \\ 
$\Tilde{r}_{\rm scale}$ &  Normalization constant & $y_b$ &  Ship-fixed transversal axis \\ 
$\sigma_{\text{ground},t}$ &  Binary indicator for grounding & $y_e$ &  Cross-track error \\ 
$\sigma_{i,t}$ &  Direction indicator & $y_{e,\rm norm}$ &  Cross-track error reward  \\ 
$\sigma_{\text{coll},i,t}$ &  Collision indicator &   & normalization constant \\ 
$\sigma_{\text{lane},t}$ &  Binary indicator for lane crossing & $Y$ &  Sway force \\ 
$\sigma_{\text{off},j}$ &  Binary indicator for leaving path & $Y_H$ &  Forces on ship hull in sway direction \\ 
$\sigma_{\text{rule},t}$ &  Rule adherence indicator & $Y_R$ &  Forces on ship rudder in sway direction \\ 
$\sigma_{\text{spd},t}$ &  Target ship speed check indicator & $Y_{WA}$ &  Wave forces in sway direction \\ 
$s$  &  State & $Y_{WI}$ &  Wind forces in sway direction \\ 
$\mathcal{S}$ &  State space & $y_n$ &  Easting \\ 
$T$ &  Length of trajectory & $y_{\rm scale}$ &  Normalization constant \\ 
$t^{\rm cpa}_{i,t}$ &  TCPA to the $i$-th target ship & $\mathcal{Z}$ &  Observation function \\ 
$t_{\rm control}$  &  PF activation interval & $\zeta_{wa}$ &  Wave amplitude \\ 
$t_{\rm norm}$ &  Scaling constant & $\zeta_{wa,\rm norm}$ & Normalization constant \\ 
    \hline
\end{tabular}}
\end{scriptsize}
\end{center}

\gdef\thesection{Appendix \Alph{section}}
\section{Target ship control}\label{appendix:TS_control}
In Algorithm \ref{algo:TS_control}, we specify the behavior of the target ships during the training and validation phases of the LPP agent. The returned angle from the algorithm is assigned as the new heading for the controlled target ship. The phrase \emph{closest ship} in the algorithm refers to the surrounding ship with the closest Euclidean distance between the vessel's midpoints. \\

\begin{algorithm}[ht]
\begin{small}
\setstretch{1.10}
\DontPrintSemicolon
\SetAlgoLined
\SetKwComment{Comment}{\# }{}
\textbf{Input:}\\
 $B, L_{pp}$: width and length between perpendiculars of the controlled ship\\
 $\chi_{d}$: desired course from VFG (with gain parameter $k = 0.001$)\\
 $\chi_{P_k}$: path angle between current waypoints\\
 $\Tilde{d}$: distance to closest ship\\
 $U_0, U_1$: speed of the controlled ship and its closest surrounding ship\\
 $t^{\rm CPA}, d^{\rm CPA}$: time and distance to CPA with the closest ship\\
 $\alpha$: relative bearing of the controlled ship from the perspective of its closest surrounding ship\\
 $\sigma$: boolean, True if closest ship travels in reversed direction, False otherwise\\
 \algrule
 \textbf{Procedure:}\\
 \Comment{Turn right in case of opposing traffic}
 \uIf{($t^{\rm CPA}$ $>$ 0) $\land$ ($d^{\rm CPA}$ $<$ 2B) $\land$ ($\Tilde{d} \leq$ 10Lpp) $\land$ $\sigma$}{
 \textbf{return} $\chi_{d} + 5^{\circ}$
  }
 \texttt{\\}
 \Comment{Overtake left}
  \uElseIf{($U_0 > U_1$) $\land$ ($\Tilde{d} \leq$ 10Lpp) $\land$ ($135^{\circ} \leq \alpha \leq 315^{\circ}$)}{
  \uIf{$y_e > 0$}{
 \textbf{return} $\psi_{dc} - 8^\circ$
  }\Else{
    \textbf{return} $\chi_{P_k} - 8^\circ\cdot\exp\left[(y_e/5B) \log(4)\right]$}
  }
  \texttt{\\}
  \Comment{Default VFG guidance}
  \Else{
  \textbf{return} $\chi_{d}$
  }
\caption{Rule-based heading control for target ships on \rdd{IWs}}
\label{algo:TS_control}
\end{small}
\end{algorithm}

\setcounter{figure}{0} 
\setcounter{table}{0}
\gdef\thesection{Appendix \Alph{section}}
\section{\rd{Baseline: Artificial potential field method}}\label{app:APF}
\begin{figure}[h]
    \centering
    \includegraphics[width=\textwidth]{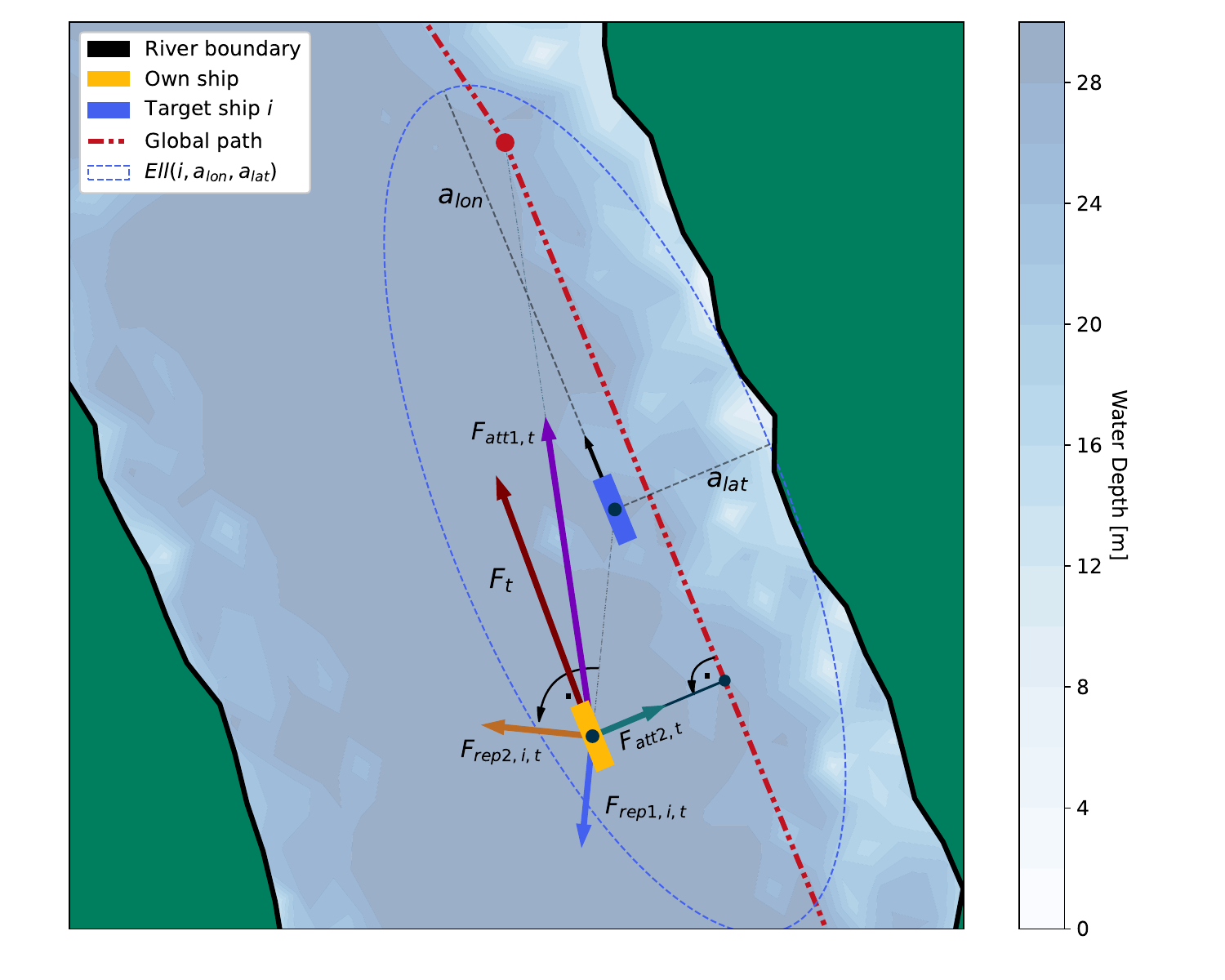}
    \caption{\rd{Visualization of the forces of the APF method building on \cite{liu2023colregs} and \cite{wang2019obstacle}.}}
    \label{fig:APF_forces}
\end{figure}

\rd{We select an APF approach as a baseline method for the \rdd{LPP} task. The approach is based on the recent proposals of \cite{liu2023colregs} and \cite{wang2019obstacle}, which we slightly adapt to improve the method's performance on \rdd{IW} scenarios. According to the original proposal from the robotics domain of \cite{APF_original}, an APF can be constructed by superpositioning attractive and repulsive forces. The attractive forces $F_{\text{att},t}$ at time step $t$ pull the planning agent toward the desired goal position, while the repulsive forces $F_{\text{rep},t}$ push the agent away from obstacles to avoid collisions. The resulting total force $F_t = F_{\text{att},t} + F_{\text{rep},t}$ is used to derive a desired heading as follows:
\begin{equation}
    \psi_{d,t} = \operatorname{arctan}\left(F_{e,t}/F_{n,t}\right),
\end{equation}
where $F_{n,t}$ and $F_{e,t}$ are the north and east components of $F_t$, respectively. Based on the desired heading $\psi_{d,t}$, the new heading can be set via:
\begin{equation}
    \psi_{\text{OS}, t+1} = \psi_{\text{OS}, t} + \operatorname{clip}\left(\psi_{d,t} - \psi_{\text{OS}, t}, -\Delta_{\psi}, \Delta_{\psi} \right),
\end{equation}
where $\psi_{\text{OS},t}$ is the heading of the own ship of time $t$ and $\Delta_{\psi}$ is the maximum possible heading change between time steps, considering the limited maneuverability of the vessel. Figure \ref{fig:APF_forces} visualizes the attractive and repulsive forces imposed in this study, which we describe in detail in the following.}

\rd{Following \cite{liu2023colregs}, we construct the attractive forces as a sum of two components: $F_{\text{att}, t}= F_{\text{att1}, t} + F_{\text{att2},t}$. The first component, $F_{\text{att1}, t}$, is the goal-oriented attractive force:
\begin{equation}
F_{\text{att1}, t} = 
\begin{cases} 
  k_{a1} \cdot d_{\text{OS},t}^{\rm G} \cdot \vec{n}_{\rm OG} & \text{if} \quad d_{\text{OS},t}^{\rm G} < d^*,\\
  k_{a1} \cdot d^* \cdot \vec{n}_{\rm OG} & \rm else,
\end{cases}
\end{equation}
where $k_{a1}$ and $d^*$ are constants, $d_{\text{OS},t}^{\rm G}$ is the Euclidean distance between the own ship and the goal point at time $t$, and $\vec{n}_{\rm OG}$ is the unit vector pointing from the own ship toward the goal. We thereby define the goal as a future waypoint of the global path, which is updated as the vessel progresses. The second component, $F_{\text{att2},t}$, is responsible for pulling the vessel back toward the path to the goal if the cross-track error is too large. Formally, we have:
\begin{equation}
F_{\text{att2}, t} = 
\begin{cases}
  k_{a2} \cdot \vert y_{e,t} \vert \cdot \vec{n}_{\rm OP} & \text{if} \quad y_{e,t} \geq d_l,\\
  0 & \rm else,
\end{cases}
\end{equation}
where $k_{\rm a2}$ and $d_l$ are constants, $y_{e,t}$ is the cross-track error of the own ship at time $t$, and $\vec{n}_{\rm OP}$ is the unit vector from the own ship perpendicular to the linear path between the initial vessel position and the goal; see \cite{liu2023colregs} and Figure \ref{fig:APF_forces}.}

\rd{Denoting the repulsive force with each target ship $i$ at time $t$ as $F_{\text{rep},i,t}$, the overall repulsive force is attained via summation: 
\begin{equation}
    F_{\text{rep},t} = \sum_i F_{\text{rep},i,t}.
\end{equation}
Similar to the attractive forces, we construct the repulsive force with relation to target ship $i$ as the sum of two components. The first component, $F_{\text{rep1},i, t}$, is the conventional repulsive force pushing the vessel away from the obstacle \citep{APF_original}, while the second component, $F_{\text{rep2},i, t}$, being orthogonal to $F_{\text{rep1},i, t}$, is inspired by \cite{liu2023colregs} and ensures that overtaking occurs on the portside of the target ships.}

\rd{Crucially, we impose two constraints that must be fulfilled to achieve a non-zero repulsive force from an obstacle. The first follows the proposal of \cite{wang2019obstacle} and requires the own ship to be in an ellipse around a target ship, allowing the repulsive force to differ in the lateral and longitudinal direction of the target vessel. Note that this approach is similar to how we specified the collision reward component of the LPP agent in Section \ref{subsubsec:reward_planner}. The second constraint is that the TCPA between the own ship and the target ship at time $t$, denoted $t_{i,t}^{\rm cpa}$, has to be larger or equal to zero. We found this second constraint crucial to improving the performance of the APF methods on our \rdd{IW} scenarios since it eliminates repulsive forces after a vessel has been overtaken. Aggregating these thoughts, we specify:
\begin{equation}
F_{\text{rep},i, t} = 
\begin{cases}
  F_{\text{rep1},i, t} + F_{\text{rep2},i, t} \quad \text{if} & \left[p_{\text{OS},t} \in \operatorname{Ell}(i, a_{\rm lon}, a_{\rm lat})\right] \land (t_{i,t}^{\rm cpa} \geq 0),\\
  0 & \rm else,
\end{cases}
\end{equation}
where $p_{\text{OS},t}$ is the current north-east position of the own ship, and $\operatorname{Ell}(i, a_{\rm lon}, a_{\rm lat})$ is the set of points inside the ellipse around target ship $i$. The semi-major axis of the ellipse has length $a_{\rm lon}$ and points in the direction of the heading of the target ship, while the semi-minor axis has length $a_{\rm lat}$.}

\rd{Following \cite{APF_original}, we thereby set:
\begin{equation}
    F_{\text{rep1},i, t} = k_{r1} \cdot \left(\frac{1}{d_{\text{OS},t}^{i}} - \frac{1}{d_0}\right) \cdot \frac{1}{\left(d_{\text{OS},t}^{i}\right)^2} \cdot \vec{n}_{\rm TO},
\end{equation}
where $k_{\rm r1}$ and $d_0$ are constant coefficients, and $d_{\text{OS},t}^{i}$ is the Euclidean distance between the own ship and target ship $i$ at time $t$. The unit vector $\vec{n}_{\rm TO}$ points from the target ship toward the own ship. Lastly, we define:
\begin{equation}\label{eq:F_rep2it}
F_{\text{rep2},i, t} = 
\begin{cases}
  k_{r2} \cdot \vec{n}_{\rm OT\perp} & \text{if} \quad 
  \vert[\psi_{i,t}-\psi_{\text{OS},t}]_{-\pi}^{\pi}\vert < \frac{\pi}{2},\\
  0 & \rm else,
\end{cases}
\end{equation}
where $k_{\rm  r2}$ is a constant, $\vec{n}_{\rm OT\perp}$ is the unit vector that is perpendicular to portside on the unit vector from the own ship to the target ship. The variables $\psi_{i,t}$ and $\psi_{\text{OS},t}$ are the heading of the target ship $i$ and the own ship, respectively. Hence, the condition in (\ref{eq:F_rep2it}) requires that the vessels travel in opposing directions. Otherwise, the overtaking-related force $F_{\text{rep2},i, t}$ is set to zero.}

\rd{We empirically determine all parameters via a grid search, leading to the values: $\Delta_\psi = 2^\circ$, $d^* = \unit[0.5]{NM}$, $d_l = \unit[50]{m}$, $k_{\rm a1} = 1$, $k_{\rm a2} = 0.1$, $k_{\rm r1} = 0.1$, $k_{\rm r2} = 0.1$, $a_{\rm lat} = d_0 = \unit[0.5]{NM}$, and $a_{\rm lon} = \unit[0.04]{NM}$.}

\setcounter{figure}{0} 
\setcounter{table}{0}
\gdef\thesection{Appendix \Alph{section}}
\section{Further validation: Local path planning agent}\label{app:val_planner}
\begin{figure}[htp]
    \centering
    \includegraphics[width=\textwidth]{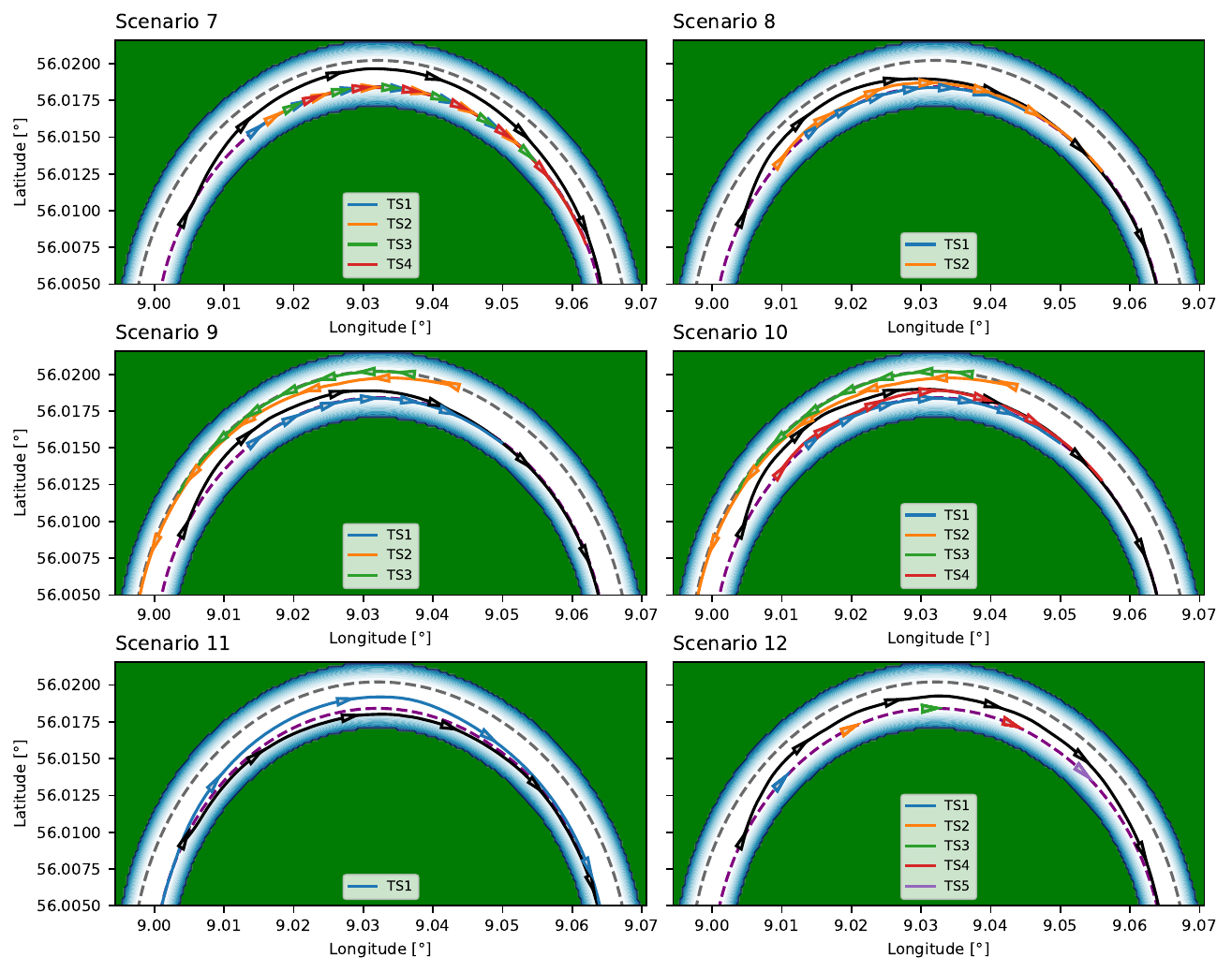}
    \caption{Trajectories of the validation scenarios of the LPP agent on a right curve. Note that the latitude and longitude values are artificial and serve as orientation.}
    \label{fig:val_plan_traj_right}
\end{figure}

\begin{figure}[htp]
    \centering
    \includegraphics[width=\textwidth]{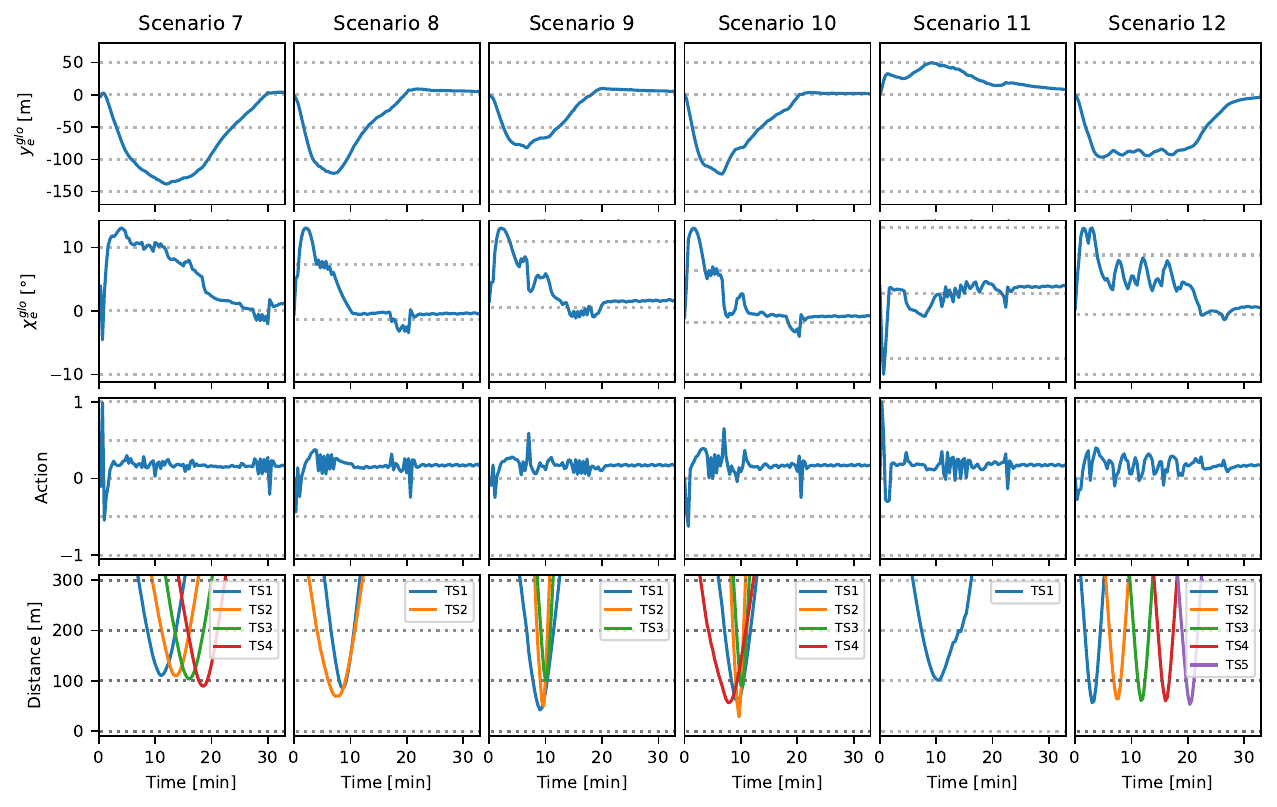}
    \caption{Global cross-track and course error, selected actions, and distances to the target ships during validation of the LPP agent on a right curve.}
    \label{fig:val_plan_metric_right}
\end{figure}

\begin{figure}[htp]
    \centering
    \includegraphics[width=\textwidth]{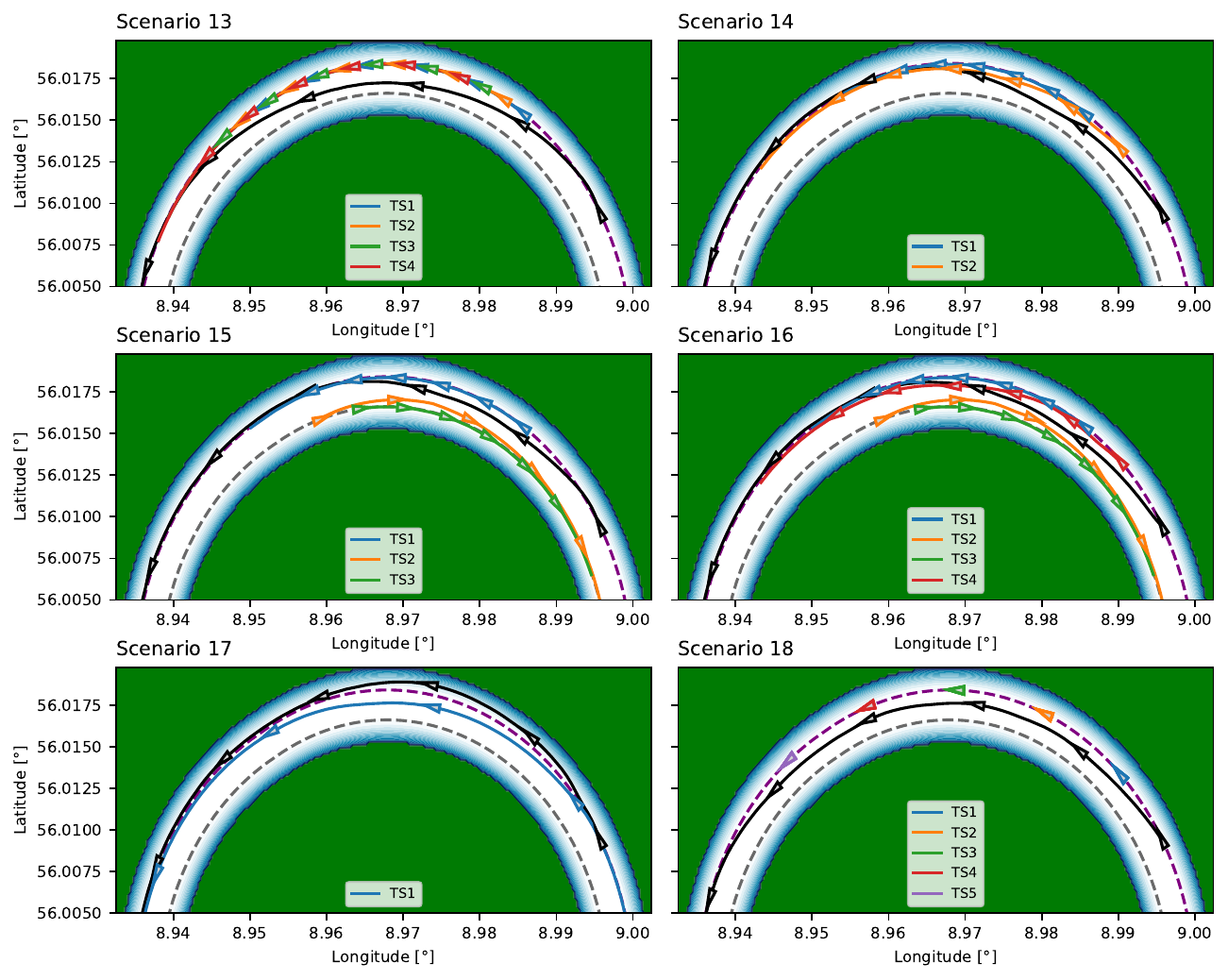}
    \caption{Trajectories of the validation scenarios of the LPP agent on a left curve. Note that the latitude and longitude values are artificial and serve as orientation.}
    \label{fig:val_plan_traj_left}
\end{figure}

\begin{figure}[htp]
    \centering
    \includegraphics[width=\textwidth]{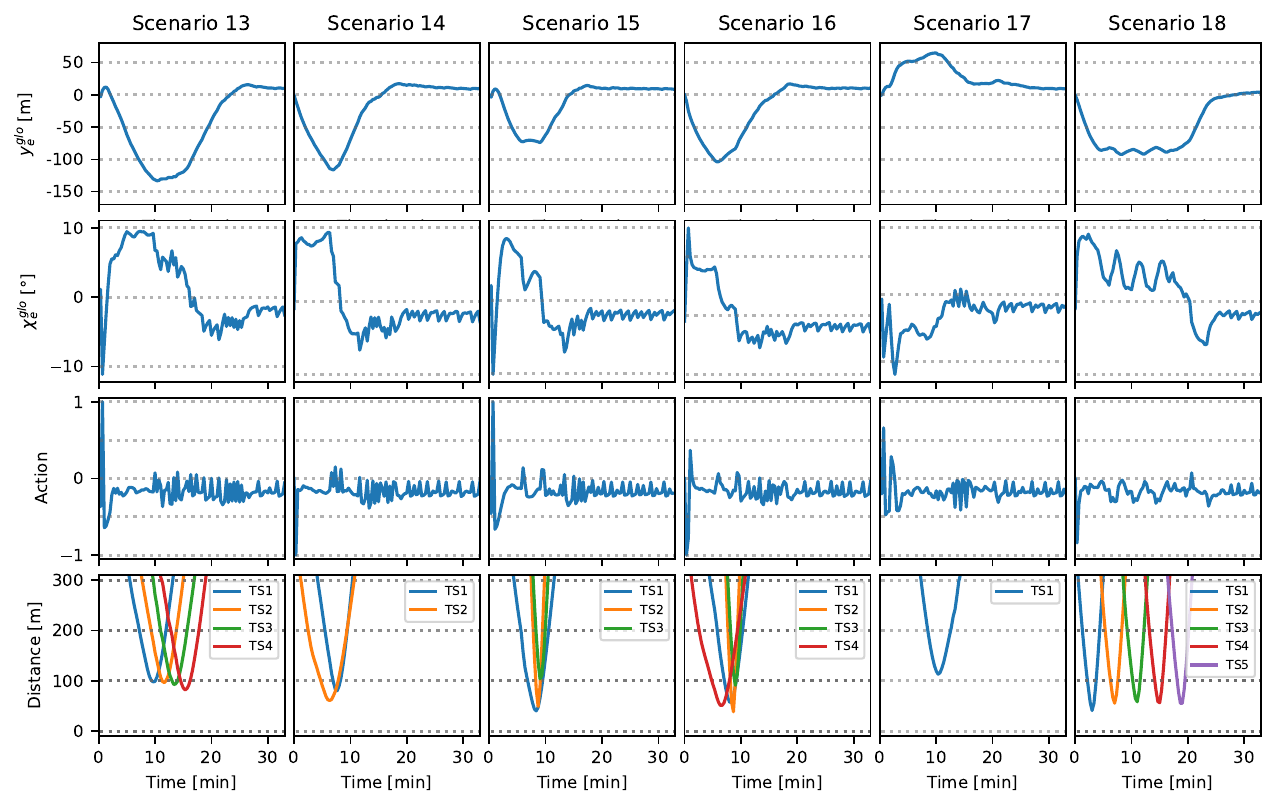}
    \caption{Global cross-track and course error, selected actions, and distances to the target ships during validation of the LPP agent on a left curve.}
    \label{fig:val_plan_left_metric}
\end{figure}

\rd{In the following, we provide further details on the discussed validation scenarios from \rd{S}ection \ref{subsec:val_planning}, including the results for curved waterway segments of the LPP agent. As emphasized in Section \ref{subsec:val_planning}, the APF method cannot be directly transferred to curved scenarios since its hyperparameters have been carefully tuned for the straight waterway case. Table \ref{tab:LPP_validate_scen_setup} includes the initial speed configuration of the target ships, while the speed of the own ship is set to $\unit[3]{m/s}$ in all cases. Furthermore, Figures \ref{fig:val_plan_traj_right} and \ref{fig:val_plan_traj_left} display the trajectories of the DRL planning agent for right and left curves, respectively. The black trajectories in these figures correspond to the DRL agent, while the colorized ones are the target ships controlled by Algorithm \ref{algo:TS_control}. The purple and grey dotted lines are the global and reversed global paths. Moreover, we display additional metrics during the curved scenarios in Figures \ref{fig:val_plan_metric_right} and \ref{fig:val_plan_left_metric}, respectively.}

Similar to the findings for the straight waterway segment, the LPP agent demonstrates the successful execution of all necessary maneuvers while effectively maintaining safe distances and avoiding consecutive large changes in heading. It is worth highlighting that the agent's need to navigate through curved paths in the scenarios depicted in Figures \ref{fig:val_plan_traj_right} and \ref{fig:val_plan_traj_left} is also reflected in the average action values \rd{as} shown in Figures \ref{fig:val_plan_metric_right} and \ref{fig:val_plan_left_metric}, respectively. In these figures, we observe a slightly positive average action in the right curve scenario, while a slightly negative average action is observed in the left curve scenario.

\begin{table}[htp]
    \centering
    \resizebox{\textwidth}{!}{%
    \begin{tabular}{ccccccc}
        \midrule
        Scenario & Waterway & $U_1$ [m/s]  & $U_2$ [m/s] & 
        $U_3$ [m/s] & $U_4$ [m/s] & $U_5$ [m/s]  \\
        \midrule
        1 & straight & 1.50 & 1.50 & 1.50 & 1.50 & 1.50\\
        2 & straight & 1.05 & 1.65 & - & - & -\\
        3 & straight & 1.20 & 2.10 & 1.20 & - & -\\
        4 & straight & 1.20 & 2.10 & 1.20 & 1.65 & -\\
        5 & straight & 4.50 & - & - & - & -\\
        6 & straight & 0.00 & 0.00 & 0.00 & 0.00 & 0.0\\
        \midrule
        7 & right curve & 1.50 & 1.50 & 1.50 & 1.50 & 1.50\\
        8 & right curve & 1.05 & 1.65 & - & - & -\\
        9 & right curve & 1.20 & 2.10 & 1.20 & - & -\\
        10 & right curve & 1.20 & 2.10 & 1.20 & 1.65 & -\\
        11 & right curve & 4.50 & - & - & - & -\\
        12 & right curve & 0.00 & 0.00 & 0.00 & 0.00 & 0.0\\
        \midrule
        13 & left curve & 1.50 & 1.50 & 1.50 & 1.50 & 1.50\\
        14 & left curve & 1.05 & 1.65 & - & - & -\\
        15 & left curve & 1.20 & 2.10 & 1.20 & - & -\\
        16 & left curve & 1.20 & 2.10 & 1.20 & 1.65 & -\\
        17 & left curve & 4.50 & - & - & - & -\\
        18 & left curve & 0.00 & 0.00 & 0.00 & 0.00 & 0.0\\
    \end{tabular}%
    }
    \caption{\rd{Initial target ship speeds for the validation scenarios in the LPP task. The variable $U_i$, $i=1,\ldots,5$, is the speed of target ship $i$.}}
    \label{tab:LPP_validate_scen_setup}
\end{table}

\setcounter{table}{0}
\setcounter{figure}{0}
\gdef\thesection{Appendix \Alph{section}}
\section{Optimization of the PID controller}\label{app:PID}
Following \cite{paramesh2021unified}, we select a PID controller as a benchmark for our PF agent. In general, the PID-controlled rudder angle is computed as follows:
\begin{align}
    \Tilde{\delta}_{t+1} &= K_p \cdot \chi_{e,t} + K_i \cdot \sum_{\Tilde{t}=0}^{t} \chi_{e,\Tilde{t}} + K_d \cdot \Tilde{r}_t, \\
    \delta_{t+1} &= \text{clip}\left[ \text{clip}\left(\Tilde{\delta}_{t+1}, \delta_{t} - a^{\rm PF}, \delta_{t} + a^{\rm PF} \right), - \delta_{\rm max}, \delta_{\rm max} \right],
\end{align}
where $\delta_0 = 0^\circ$ and $K_p$, $K_i$, and $K_d$ are the gain parameters. The two clipping operations ensure that the rudder angle does not change by more than $a^{\rm PF} = 5^\circ$ between time steps and does not exceed $\delta_{\rm max} = 20^\circ$ in absolute value. Note that these constraints also apply to our PF agent. On this basis, we use the PSO approach of \cite{eberhart2000comparing} to solve the following optimization problem:
\begin{equation}
    \min_{K_p, K_i, K_d} \sum_{j=1}^{6}\left(10^7 \cdot \sigma_{\text{off}, j} + \sum_{t=0}^{T_j} \chi_{j,e,t}^{2} \right),
\end{equation}
where the sum goes over the six validation scenarios shown in Figures \ref{fig:follow_moderate} and \ref{fig:follow_extreme}, respectively. The binary variable $\sigma_{\text{off}, j}$ is set to one if the controlled vessel deviates significantly from the specified path and actually leaves the river during the corresponding scenario. Otherwise, $\sigma_{\text{off}, j} = 0$. The episode length for scenario $j$ is denoted as $T_j$, and if the vessel remains within the river throughout the episode, we have $T_j = 750$. Additionally, $\chi_{j,e,t}$ represents the course-error with respect to the global path at time $t$ in scenario $j$. In summary, the optimization objective is to minimize the sum of squared course-errors while ensuring that deviations from the global path are not excessive enough to result in the vessel leaving the river.

Regarding the parametrization of the PSO method, we select a population size of 20 particles whose initial $K_p$, $K_d$, and $K_i$ values are sampled from $\mathcal{U}(0.25, 3.75)$, $\mathcal{U}(10, 30)$, and $\mathcal{U}(0.025, 0.075)$, respectively, while the velocities where set to 0.05, 1, and 0.05. The algorithm was run over 1000 iterations and the inertia weight was linearly decayed from 0.9 to 0.4, following the suggestions of \cite{eberhart2000comparing}.

\setcounter{table}{0}
\setcounter{figure}{0}
\gdef\thesection{Appendix \Alph{section}}
\section{Details on the validation data}\label{app:AIS_data_details}
We consider the segment of the river Elbe from Lighthouse Tinsdal to the Elbe estuary close to Cuxhaven. The environmental data was gathered from the E.U. Copernicus Marine Service. \rd{In particular, current data stems from \cite{currentData}, wind data from \cite{windData}, and wave data from \cite{waveData}.}

The AIS data was kindly provided by the European Maritime Safety Agency. The collected records can be categorized into two groups:
\begin{enumerate}
    \item \emph{Static and voyage-related information:} This includes details such as the vessels' call sign, IMO number, estimated time of arrival (ETA), maximum draught, vessel type, and cargo type.
    \item \emph{Dynamic information:} This includes real-time data such as the vessels' current position (latitude, longitude), navigational status (under way, at anchor, etc.), Maritime Mobile Service Identity (MMSI), speed over ground (SOG), course over ground (COG), and rate of turn.
\end{enumerate}
The static information is recorded every 6 minutes, while dynamic information is recorded at various intervals ranging from 2 to 180 seconds, depending on the vessels' dynamic conditions. For a detailed listing, please refer to \citet[p.~8]{series2014technical}.

AIS records may contain incomplete, erroneous, or duplicate information \citep{last2014comprehensive, tu2017exploiting}. Incomplete or faulty data can often be attributed to impaired measurement or transmission equipment onboard the vessels, while duplicate records occur when more than one receiving base station captures and records the vessel's message. To ensure accurate trajectory interpolation, we employ filtering techniques to eliminate duplicates. We compare the MMSI, timestamp, and receiving base station for each vessel, retaining only the first record if a vessel reports multiple times within a two-second window. Additionally, we remove records that have missing or incomplete information regarding position, SOG, COG, or MMSI. To facilitate the route extraction process, we sort the messages in ascending order based on time.

\begin{figure}
    \centering
    \includegraphics[width=0.7\textwidth]{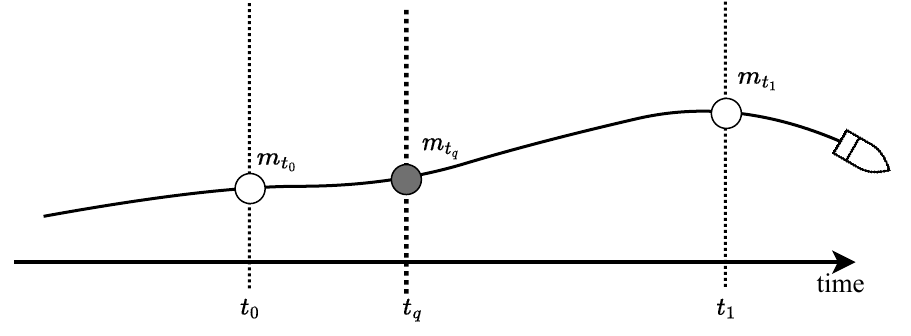}
    \caption{Interpolation between two AIS messages of two ships, to receive information at the query time $t_q$; inspired by \cite{rong2022ship}.}
    \label{aisalign}
\end{figure}

The LPP module relies on information about surrounding target ships at a specific time point, denoted as $t_q$. However, in cases where there is no exact record for $t_q$, interpolation between messages from different time points becomes necessary. This issue is illustrated in Figure \ref{aisalign}, which depicts two AIS messages, $m_{t_0}$ and $m_{t_1}$, recorded at times $t_0$ and $t_1$ respectively, with $t_0 < t_q < t_1$. Leveraging ideas from \cite{rong2022ship}, we estimate the message at the query time, $m_{t_q}$, using a cubic spline. Importantly, we fit separate univariate splines for the northing, easting, course over ground, and speed over ground of each vessel. In cases where there are insufficient data points, we resort to linear interpolation for the respective quantity. The complete trajectory extraction pipeline is publicly available \rd{in \cite{pytsa}}.

\newpage

\end{document}